\documentstyle[10pt,epsf,epsfig,hangcaption,xspace,amssymb,amsfonts,amsmath,amsthm,cite,
dp_delphititle,lineno]{dp_delphi}
\makeindex
\def\DpPaperGroup{EP}
\def\DpPaperRef{2003-061}
\def\DpDate{12 June 2003}
\def\DpAuthors{DELPHI Collaboration}
\def\DpSubmit{(Accepted by Eur. Phys. J. C)}
\def\DpTitle{{ Searches for Neutral Higgs Bosons \\
in Extended Models}}
\def\DpComment{ }
\def\DpEMail{ }

\newcommand{\epem}{$\mathrm{e^+e^-\,}$}
\newcommand{\mb}{$\mathrm{m_b}$}
\newcommand{\mh}{$\mathrm{m_h}$}
\newcommand{\mA}{$\mathrm{m_A}$}
\newcommand{\mZ}{$\mathrm{m_Z}$}
\newcommand{\ra}{\!\!$\rightarrow$}
\newcommand{\bb}{$\mathrm{b\bar{b}}$}
\newcommand{\ff}{$\mathrm{f\bar{f}}$}
\newcommand{\lplm}{$\mathrm{l^+l^-}$}
\newcommand{\qq}{$\mathrm{q\bar{q}}$}
\newcommand{\tautau}{$\mathrm{\tau^+\tau^-}$}

\newcommand{\mra}{\rightarrow}
\newcommand{\mbb}{\mathrm{b\bar{b}}}
\newcommand{\mff}{\mathrm{f\bar{f}}}

\newcommand{\mtautau}{\mathrm{\tau^+\tau^-}}

 \newcommand{ \nvg }{\mbox{$N_{vg} $}}
 \newcommand{ \nva }{\mbox{$N_{v} $}}
 \newcommand{ \xbi }{\mbox{$x_{b1} $}}
 \newcommand{ \xbii}{\mbox{$x_{b2} $}}
 \newcommand{ \xbiii}{\mbox{$x_{b3} $}}
 \newcommand{ \xbiv}{\mbox{$x_{b34} $}}
 \newcommand{ \xbtf}{\mbox{$x_{b34} $}}

\newcommand{\MeVcc}{\mbox{$\mathrm{MeV}/c^{2}$}}
\newcommand{\GeVcc}{\mbox{$\mathrm{GeV}/c^{2}$}}
\newcommand{\MeVc} {\mbox{$\mathrm{MeV}/c$}}
\newcommand{\GeVc} {\mbox{$\mathrm{GeV}/c$}}
\newfont{\scsl}{ecsc1200} 
\newcommand{\Zfitter}{{\scsl{%
\raisebox{-0.4ex}{z\kern-0.05em{}f}\kern-0.1em{}I\kern-0.15em%
\raisebox{0.8ex}{T}\kern-0.25em{}T\kern-0.25em%
\raisebox{-0.8ex}{E\kern-0.05em{}r}}}}
\begin{document}
\makeatletter
\newcount\@tempcntc
\def\@citex[#1]#2{\if@filesw\immediate\write\@auxout{\string\citation{#2}}\fi
  \@tempcnta\z@\@tempcntb\m@ne\def\@citea{}\@cite{\@for\@citeb:=#2\do
    {\@ifundefined
       {b@\@citeb}{\@citeo\@tempcntb\m@ne\@citea\def\@citea{,}{\bf ?}\@warning
       {Citation `\@citeb' on page \thepage \space undefined}}%
    {\setbox\z@\hbox{\global\@tempcntc0\csname b@\@citeb\endcsname\relax}%
     \ifnum\@tempcntc=\z@ \@citeo\@tempcntb\m@ne
       \@citea\def\@citea{,}\hbox{\csname b@\@citeb\endcsname}%
     \else
      \advance\@tempcntb\@ne
      \ifnum\@tempcntb=\@tempcntc
      \else\advance\@tempcntb\m@ne\@citeo
      \@tempcnta\@tempcntc\@tempcntb\@tempcntc\fi\fi}}\@citeo}{#1}}
\def\@citeo{\ifnum\@tempcnta>\@tempcntb\else\@citea\def\@citea{,}%
  \ifnum\@tempcnta=\@tempcntb\the\@tempcnta\else
   {\advance\@tempcnta\@ne\ifnum\@tempcnta=\@tempcntb \else \def\@citea{--}\fi
    \advance\@tempcnta\m@ne\the\@tempcnta\@citea\the\@tempcntb}\fi\fi}
 
\makeatother
\begin{titlepage}
\pagenumbering{roman}
\CERNpreprint{\DpPaperGroup}{\DpPaperRef} 
\date{{\small\DpDate}} 
\title{\DpTitle} 
\address{\DpAuthors} 
\begin{shortabs} 
\noindent
Searches for neutral Higgs bosons produced at LEP
in association with Z bosons, in pairs and in the Yukawa
process are presented in this paper. Higgs boson decays into b~quarks, $\tau$~leptons, or other Higgs
bosons are considered, giving rise to four-b, four-b+jets, six-b and four-$\tau$ final states,
as well as mixed modes with b~quarks and $\tau$~leptons. The whole mass domain kinematically accessible at
LEP in these topologies is searched. The analysed data set covers both the LEP1 and LEP2 energy ranges and
exploits most of the luminosity recorded by the DELPHI experiment. No
convincing evidence for a signal is found, and results are presented in the form of 
mass-dependent upper bounds on coupling factors (in units of model-independent 
reference cross-sections) for all processes, allowing interpretation of the data
in a large class of models.

\end{shortabs}
\vfill
\begin{center}
\DpSubmit \ \\ 
\DpComment \ \\
\DpEMail \ \\
\end{center}
\vfill
\clearpage
\headsep 10.0pt
\addtolength{\textheight}{10mm}
\addtolength{\footskip}{-5mm}
\begingroup
%
\newcommand{\DpName}[2]{\hbox{#1$^{\ref{#2}}$},\hfill}
\newcommand{\DpNameTwo}[3]{\hbox{#1$^{\ref{#2},\ref{#3}}$},\hfill}
\newcommand{\DpNameThree}[4]{\hbox{#1$^{\ref{#2},\ref{#3},\ref{#4}}$},\hfill}
\newskip\Bigfill \Bigfill = 0pt plus 1000fill
\newcommand{\DpNameLast}[2]{\hbox{#1$^{\ref{#2}}$}\hspace{\Bigfill}}
%
\footnotesize
\noindent
\DpName{J.Abdallah}{LPNHE}
\DpName{P.Abreu}{LIP}
\DpName{W.Adam}{VIENNA}
\DpName{P.Adzic}{DEMOKRITOS}
\DpName{T.Albrecht}{KARLSRUHE}
\DpName{T.Alderweireld}{AIM}
\DpName{R.Alemany-Fernandez}{CERN}
\DpName{T.Allmendinger}{KARLSRUHE}
\DpName{P.P.Allport}{LIVERPOOL}
\DpName{U.Amaldi}{MILANO2}
% ...
\normalsize
\endgroup
\titlefoot{Department of Physics and Astronomy, Iowa State
     University, Ames IA 50011-3160, USA
    \label{AMES}}
\titlefoot{Physics Department, Universiteit Antwerpen,
     Universiteitsplein 1, B-2610 Antwerpen, Belgium \\
     \indent~~and IIHE, ULB-VUB,
     Pleinlaan 2, B-1050 Brussels, Belgium \\
     \indent~~and Facult\'e des Sciences,
     Univ. de l'Etat Mons, Av. Maistriau 19, B-7000 Mons, Belgium
    \label{AIM}}
\titlefoot{Physics Laboratory, University of Athens, Solonos Str.
     104, GR-10680 Athens, Greece
    \label{ATHENS}}
\titlefoot{Department of Physics, University of Bergen,
     All\'egaten 55, NO-5007 Bergen, Norway
    \label{BERGEN}}
\titlefoot{Dipartimento di Fisica, Universit\`a di Bologna and INFN,
     Via Irnerio 46, IT-40126 Bologna, Italy
    \label{BOLOGNA}}
\titlefoot{Centro Brasileiro de Pesquisas F\'{\i}sicas, rua Xavier Sigaud 150,
     BR-22290 Rio de Janeiro, Brazil \\
     \indent~~and Depto. de F\'{\i}sica, Pont. Univ. Cat\'olica,
     C.P. 38071 BR-22453 Rio de Janeiro, Brazil \\
     \indent~~and Inst. de F\'{\i}sica, Univ. Estadual do Rio de Janeiro,
     rua S\~{a}o Francisco Xavier 524, Rio de Janeiro, Brazil
    \label{BRASIL}}
\titlefoot{Coll\`ege de France, Lab. de Physique Corpusculaire, IN2P3-CNRS,
     FR-75231 Paris Cedex 05, France
    \label{CDF}}
\titlefoot{CERN, CH-1211 Geneva 23, Switzerland
    \label{CERN}}
\titlefoot{Institut de Recherches Subatomiques, IN2P3 - CNRS/ULP - BP20,
     FR-67037 Strasbourg Cedex, France
    \label{CRN}}
\titlefoot{Now at DESY-Zeuthen, Platanenallee 6, D-15735 Zeuthen, Germany
    \label{DESY}}
\titlefoot{Institute of Nuclear Physics, N.C.S.R. Demokritos,
     P.O. Box 60228, GR-15310 Athens, Greece
    \label{DEMOKRITOS}}
\titlefoot{FZU, Inst. of Phys. of the C.A.S. High Energy Physics Division,
     Na Slovance 2, CZ-180 40, Praha 8, Czech Republic
    \label{FZU}}
\titlefoot{Dipartimento di Fisica, Universit\`a di Genova and INFN,
     Via Dodecaneso 33, IT-16146 Genova, Italy
    \label{GENOVA}}
\titlefoot{Institut des Sciences Nucl\'eaires, IN2P3-CNRS, Universit\'e
     de Grenoble 1, FR-38026 Grenoble Cedex, France
    \label{GRENOBLE}}
\titlefoot{Helsinki Institute of Physics, P.O. Box 64,
     FIN-00014 University of Helsinki, Finland
    \label{HELSINKI}}
\titlefoot{Joint Institute for Nuclear Research, Dubna, Head Post
     Office, P.O. Box 79, RU-101 000 Moscow, Russian Federation
    \label{JINR}}
\titlefoot{Institut f\"ur Experimentelle Kernphysik,
     Universit\"at Karlsruhe, Postfach 6980, DE-76128 Karlsruhe,
     Germany
    \label{KARLSRUHE}}
\titlefoot{Institute of Nuclear Physics PAN,Ul. Radzikowskiego 152,
     PL-31142 Krakow, Poland
    \label{KRAKOW1}}
\titlefoot{Faculty of Physics and Nuclear Techniques, University of Mining
     and Metallurgy, PL-30055 Krakow, Poland
    \label{KRAKOW2}}
\titlefoot{Universit\'e de Paris-Sud, Lab. de l'Acc\'el\'erateur
     Lin\'eaire, IN2P3-CNRS, B\^{a}t. 200, FR-91405 Orsay Cedex, France
    \label{LAL}}
\titlefoot{School of Physics and Chemistry, University of Lancaster,
     Lancaster LA1 4YB, UK
    \label{LANCASTER}}
\titlefoot{LIP, IST, FCUL - Av. Elias Garcia, 14-$1^{o}$,
     PT-1000 Lisboa Codex, Portugal
    \label{LIP}}
\titlefoot{Department of Physics, University of Liverpool, P.O.
     Box 147, Liverpool L69 3BX, UK
    \label{LIVERPOOL}}
\titlefoot{Dept. of Physics and Astronomy, Kelvin Building,
     University of Glasgow, Glasgow G12 8QQ
    \label{GLASGOW}}
\titlefoot{LPNHE, IN2P3-CNRS, Univ.~Paris VI et VII, Tour 33 (RdC),
     4 place Jussieu, FR-75252 Paris Cedex 05, France
    \label{LPNHE}}
\titlefoot{Department of Physics, University of Lund,
     S\"olvegatan 14, SE-223 63 Lund, Sweden
    \label{LUND}}
\titlefoot{Universit\'e Claude Bernard de Lyon, IPNL, IN2P3-CNRS,
     FR-69622 Villeurbanne Cedex, France
    \label{LYON}}
\titlefoot{Dipartimento di Fisica, Universit\`a di Milano and INFN-MILANO,
     Via Celoria 16, IT-20133 Milan, Italy
    \label{MILANO}}
\titlefoot{Dipartimento di Fisica, Univ. di Milano-Bicocca and
     INFN-MILANO, Piazza della Scienza 2, IT-20126 Milan, Italy
    \label{MILANO2}}
\titlefoot{IPNP of MFF, Charles Univ., Areal MFF,
     V Holesovickach 2, CZ-180 00, Praha 8, Czech Republic
    \label{NC}}
\titlefoot{NIKHEF, Postbus 41882, NL-1009 DB
     Amsterdam, The Netherlands
    \label{NIKHEF}}
\titlefoot{National Technical University, Physics Department,
     Zografou Campus, GR-15773 Athens, Greece
    \label{NTU-ATHENS}}
\titlefoot{Physics Department, University of Oslo, Blindern,
     NO-0316 Oslo, Norway
    \label{OSLO}}
\titlefoot{Dpto. Fisica, Univ. Oviedo, Avda. Calvo Sotelo
     s/n, ES-33007 Oviedo, Spain
    \label{OVIEDO}}
\titlefoot{Department of Physics, University of Oxford,
     Keble Road, Oxford OX1 3RH, UK
    \label{OXFORD}}
\titlefoot{Dipartimento di Fisica, Universit\`a di Padova and
     INFN, Via Marzolo 8, IT-35131 Padua, Italy
    \label{PADOVA}}
\titlefoot{Rutherford Appleton Laboratory, Chilton, Didcot
     OX11 OQX, UK
    \label{RAL}}
\titlefoot{Dipartimento di Fisica, Universit\`a di Roma II and
     INFN, Tor Vergata, IT-00173 Rome, Italy
    \label{ROMA2}}
\titlefoot{Dipartimento di Fisica, Universit\`a di Roma III and
     INFN, Via della Vasca Navale 84, IT-00146 Rome, Italy
    \label{ROMA3}}
\titlefoot{DAPNIA/Service de Physique des Particules,
     CEA-Saclay, FR-91191 Gif-sur-Yvette Cedex, France
    \label{SACLAY}}
\titlefoot{Instituto de Fisica de Cantabria (CSIC-UC), Avda.
     los Castros s/n, ES-39006 Santander, Spain
    \label{SANTANDER}}
\titlefoot{Inst. for High Energy Physics, Serpukov
     P.O. Box 35, Protvino, (Moscow Region), Russian Federation
    \label{SERPUKHOV}}
\titlefoot{J. Stefan Institute, Jamova 39, SI-1000 Ljubljana, Slovenia
     and Laboratory for Astroparticle Physics,\\
     \indent~~Nova Gorica Polytechnic, Kostanjeviska 16a, SI-5000 Nova Gorica, Slovenia, \\
     \indent~~and Department of Physics, University of Ljubljana,
     SI-1000 Ljubljana, Slovenia
    \label{SLOVENIJA}}
\titlefoot{Fysikum, Stockholm University,
     Box 6730, SE-113 85 Stockholm, Sweden
    \label{STOCKHOLM}}
\titlefoot{Dipartimento di Fisica Sperimentale, Universit\`a di
     Torino and INFN, Via P. Giuria 1, IT-10125 Turin, Italy
    \label{TORINO}}
\titlefoot{INFN,Sezione di Torino, and Dipartimento di Fisica Teorica,
     Universit\`a di Torino, Via P. Giuria 1,\\
     \indent~~IT-10125 Turin, Italy
    \label{TORINOTH}}
\titlefoot{Dipartimento di Fisica, Universit\`a di Trieste and
     INFN, Via A. Valerio 2, IT-34127 Trieste, Italy \\
     \indent~~and Istituto di Fisica, Universit\`a di Udine,
     IT-33100 Udine, Italy
    \label{TU}}
\titlefoot{Univ. Federal do Rio de Janeiro, C.P. 68528
     Cidade Univ., Ilha do Fund\~ao
     BR-21945-970 Rio de Janeiro, Brazil
    \label{UFRJ}}
\titlefoot{Department of Radiation Sciences, University of
     Uppsala, P.O. Box 535, SE-751 21 Uppsala, Sweden
    \label{UPPSALA}}
\titlefoot{IFIC, Valencia-CSIC, and D.F.A.M.N., U. de Valencia,
     Avda. Dr. Moliner 50, ES-46100 Burjassot (Valencia), Spain
    \label{VALENCIA}}
\titlefoot{Institut f\"ur Hochenergiephysik, \"Osterr. Akad.
     d. Wissensch., Nikolsdorfergasse 18, AT-1050 Vienna, Austria
    \label{VIENNA}}
\titlefoot{Inst. Nuclear Studies and University of Warsaw, Ul.
     Hoza 69, PL-00681 Warsaw, Poland
    \label{WARSZAWA}}
\titlefoot{Fachbereich Physik, University of Wuppertal, Postfach
     100 127, DE-42097 Wuppertal, Germany
    \label{WUPPERTAL}}
\addtolength{\textheight}{-10mm}
\addtolength{\footskip}{5mm}
\clearpage
\headsep 30.0pt
\end{titlepage}
%
\pagenumbering{arabic} 
\setcounter{footnote}{0} %
\large
\section{Introduction}

As is well known, the Standard Model of electroweak interactions
describes the available data with considerable accuracy, only lacking
evidence for the Higgs boson as confirmation of its scalar sector \cite{E-W}. 

A number of extensions to the scalar sector of the Standard Model 
allow the current level of agreement between prediction and
measurement to be preserved. Beyond the simplest one-doublet scalar sector of the
Standard Model, any model with arbitrary numbers of Higgs doublets and
singlets will satisfy the above conditions, in particular concerning
the relation between the electroweak gauge boson masses and the
SU(2)$\times$U(1) mixing angle. To satisfy the constraint given by the
apparent weakness of flavour-changing neutral currents, it is
generally imposed in addition that every fermion couples to at most
one Higgs doublet \cite{FCNC}.

Within this framework, the simplest extensions of the Standard Model are
the so-called Two-Higgs Doublet Models (2HDM), of which various types
exist, depending on the choice of the scalar couplings to
fermions. The first type assumes that one doublet only couples to
fermions while the other one couples to gauge bosons. At LEP2, the resulting
final states include decays of the lightest Higgs boson into photon pairs, which are
studied in \cite{DELphobic}. The second and most studied type assumes
that one doublet couples to the up-type fermions (neutrinos and the
u, c, t~quarks) while the other one couples to down-type fermions
(charged leptons and the d, s and b~quarks). Depending on the mixing of
the two doublets, the dominant decays of the lightest Higgs boson will be either
c~quarks and/or gluons (these final states are searched for in
\cite{DELblind}), or b~quarks and $\tau$~leptons. This last case is
the focus of this work.

There is a third possible choice of couplings, in which
one Higgs doublet couples to leptons only, while the other couples to
quarks. In this case, the dominant Higgs boson decay modes
may be leptonic, leading, when Higgs bosons are produced in pairs or radiated 
off primary $\tau$~leptons, to the striking four-$\tau$ final state.

This paper presents searches for final states occurring in the scenarios decribed above,
when Higgs bosons are produced through the Yukawa process, in pairs, or in association with Z
bosons. The first section of this work introduces our conventions, describes the data sets 
and some aspects common to all analyses. Section~\ref{allLEP1} describes searches for the Yukawa process in LEP1 data; the four-b, 
four-$\tau$, and \bb\tautau\ final states are addressed. The searches for 
final states with at least four b~quarks or $\tau$~leptons at LEP2 are described in Section~\ref{allLEP2}. 
In all final states, the Higgs boson mass domain is explored from threshold to the kinematic limit.
Our results are summarized in Section~\ref{results}, and include a reinterpretation of the DELPHI Standard 
Model Higgs boson search \cite{DELhiggs}, constraining the hZ process, when h decays into b~quark or 
$\tau$~lepton pairs. Section~\ref{conclusions} concludes the paper.

Neutral Higgs bosons beyond the Standard Model have also been searched for by the other LEP Collaborations \cite{otherhiggs}.
The present paper considers additional final states (i.e. the four-$\tau$ final state, in Higgs boson pair production and in the Yukawa process),
and revisits more usual final states by extending the searched mass range.

\subsection{Signals considered in this paper}
\label{higgsprod}

\begin{figure}[p]
\begin{center}
\caption{Higgs boson production processes at LEP.}
\begin{tabular}{lll}
\epsfig{file=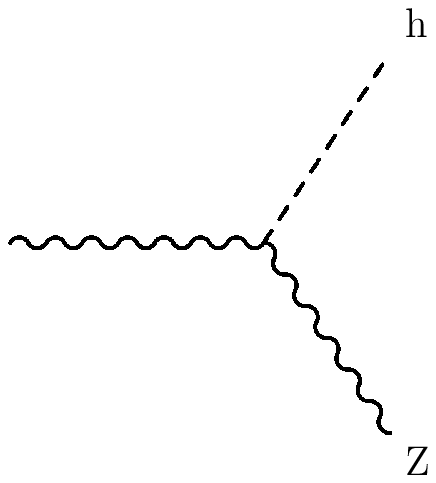,width=.25\textwidth} &
\epsfig{file=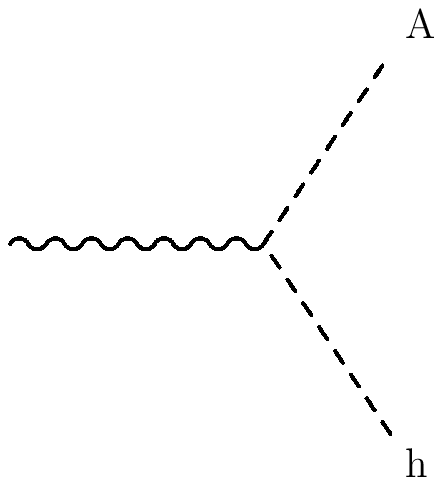,width=.25\textwidth} &
\epsfig{file=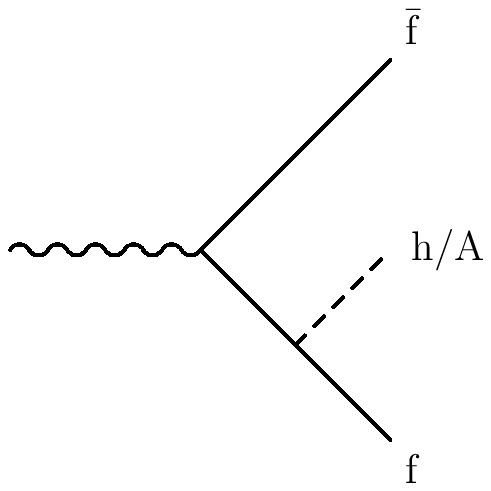,width=.25\textwidth} \\
Bjorken process & Pair production & Yukawa process \\
\end{tabular}
\label{diag1}
\end{center}
\end{figure}

The extension of the Standard Model Higgs sector by at least one
doublet significantly enriches its phenomenology. The Higgs boson
spectrum consists of a number of CP-even Higgs bosons (denoted h), CP-odd Higgs
bosons (A) and pairs of charged scalars $\mathrm{H^{\pm}}$. Neutral Higgs boson
production mechanisms at LEP are the Bjorken process (\epem \ra hZ),
pair production (\epem \ra hA) and Yukawa radiation off heavy fermions
(\epem \ra \ff h and \epem \ra \ff A).
The cross-sections of the first two, gauge-mediated processes
are (up to kinematic factors) bounded by the Standard Model hZ
cross-section; mixing of Higgs doublets induces partial or total
suppression with respect to this reference. The third,
fermion-mediated process can be significantly enhanced compared to the
Standard Model \ff h cross-section, which is too
low to be observed at LEP. Diagrams of these processes are displayed
in Figure \ref{diag1}.

\begin{figure}[p]
\begin{center}
\caption{Non-fermionic Higgs boson decay modes.}
\begin{tabular}{lll}
\epsfig{file=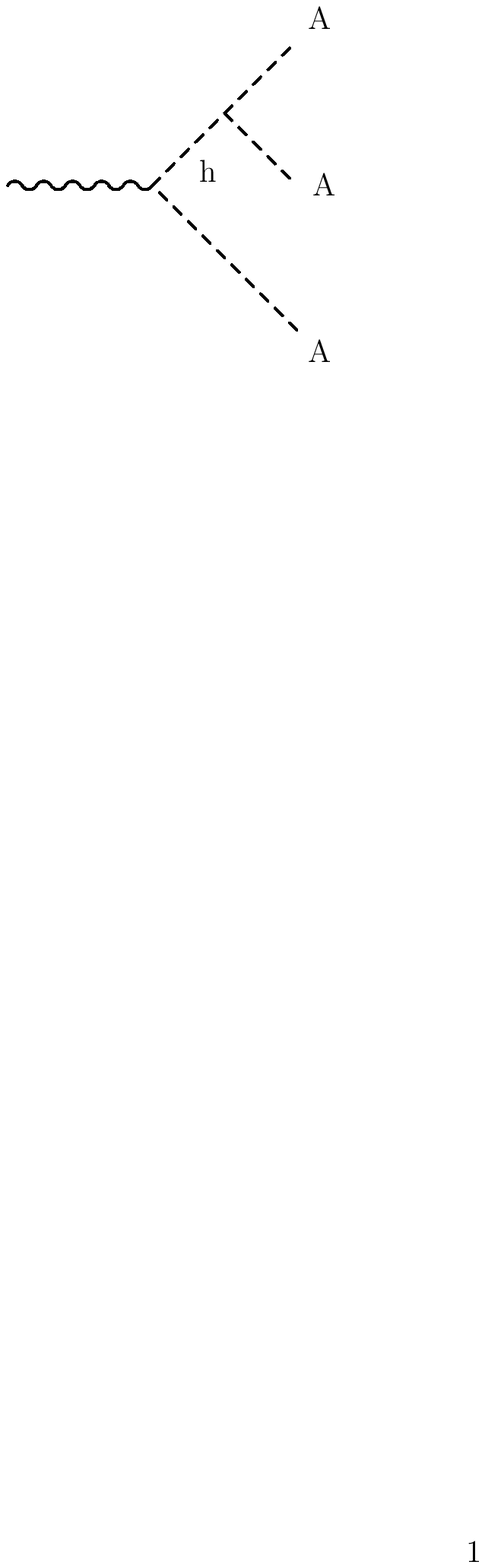,width=.25\textwidth} &
\epsfig{file=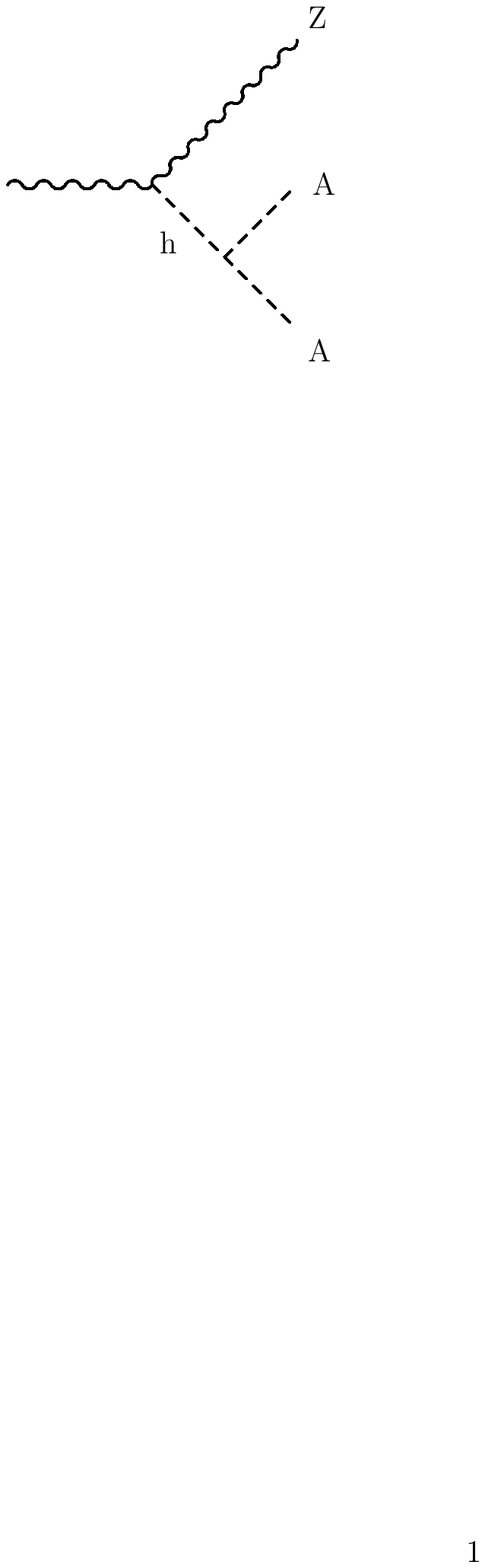,width=.25\textwidth} &
\epsfig{file=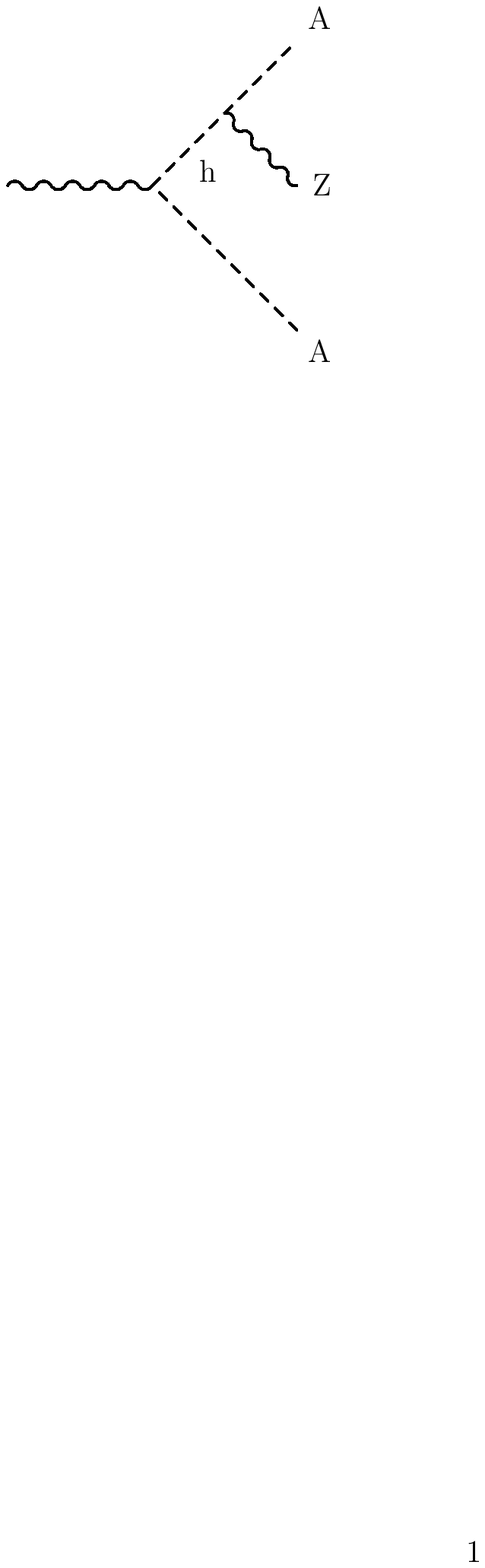,width=.25\textwidth} \\
\end{tabular}
\begin{tabular}{ll}
\epsfig{file=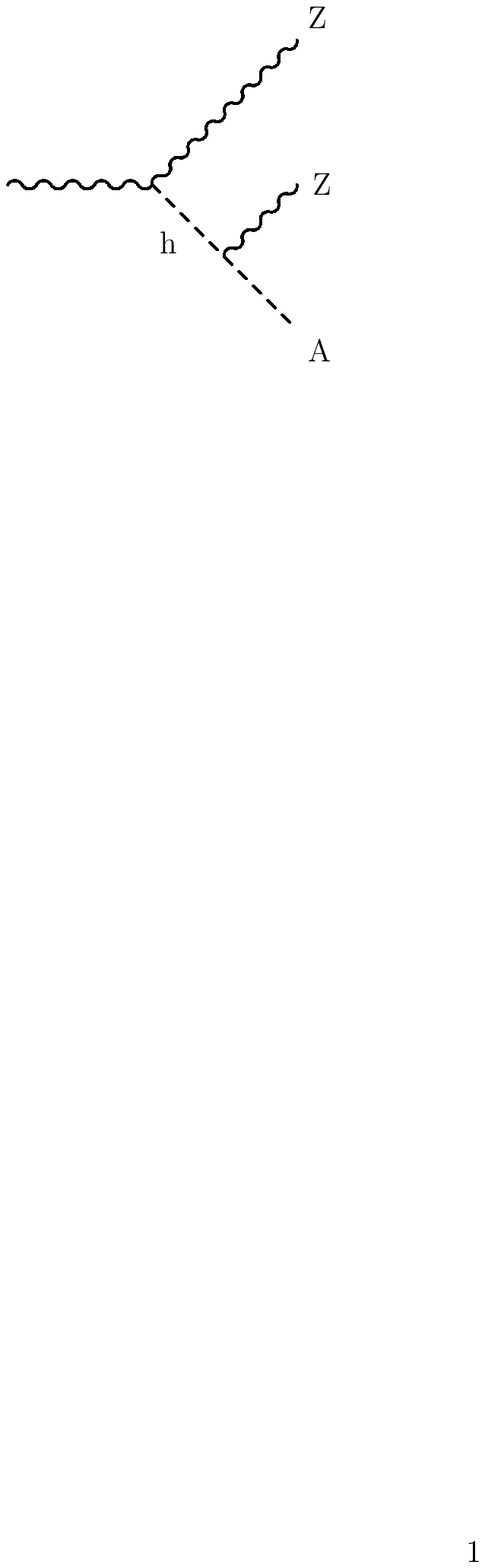,width=.25\textwidth} &
\epsfig{file=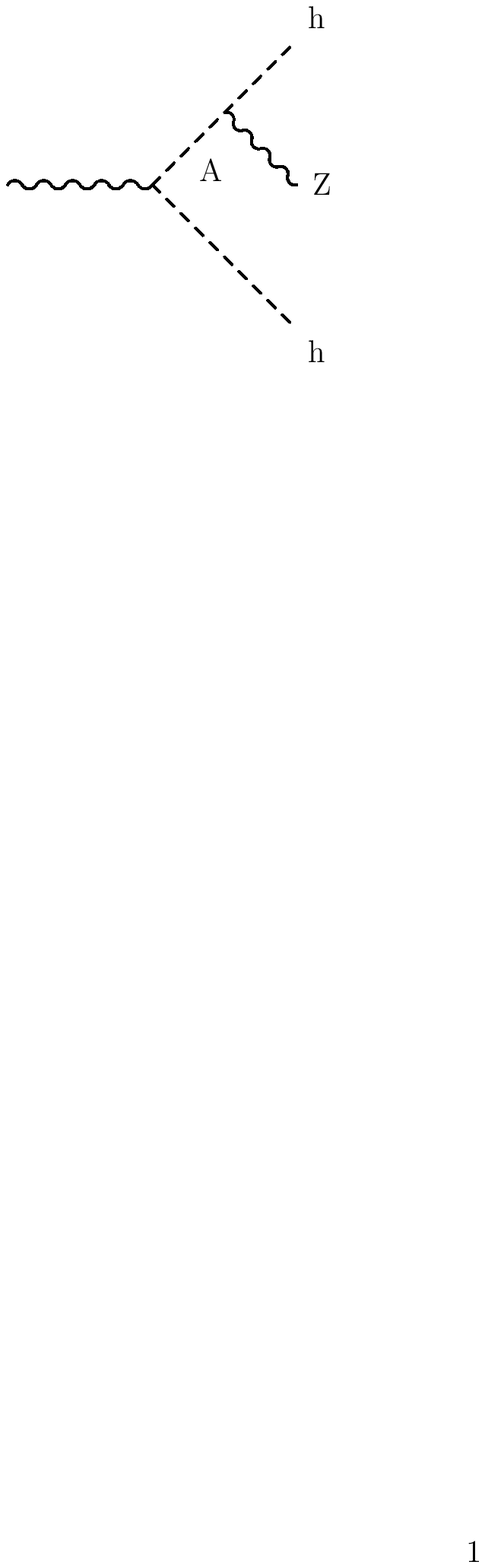,width=.25\textwidth} \\
\end{tabular}
\label{diag2}
\end{center}
\end{figure}

Depending on their mass hierarchy, there are a number of
production and decay chains involving Higgs bosons (see also Figure \ref{diag2}):
\begin{enumerate}
 \item {\epem \ra hA \ra (AA)A and \epem \ra hZ \ra (AA)Z when \mh\ $>$ 2 \mA;}
 \item {\epem \ra hA \ra (AZ)A and \epem \ra hZ \ra (AZ)Z when \mh\ $>$ \mZ\ + \mA;}
 \item {\epem \ra hA \ra h(hZ) when \mA\ $>$ \mZ\ + \mh.}
\end{enumerate}

\noindent Among these, only processes 1 and 3 are explicitly studied
here. Note however that the h(hZ) and the (AZ)A processes involve
exactly the same vertices, which means that all distributions are expected to be similar
if \mh\ and \mA\ are exchanged; as a consequence, our
h(hZ) results will be directly translated to the (AZ)A case with
swapped h and A masses. On the other hand, the (AZ)Z process is of
very small relevance to LEP, since given the available centre-of-mass energies and the presence
of two Z bosons in the final states, the open mass domain for h and A
is very small.  

We limit our analysis to decays of the lighter Higgs boson into b~quarks 
or $\tau$~leptons, and only the dominant hadronic Z decays are
considered. We take the threshold of Higgs boson decays to b~quarks to
be 12~\GeVcc, which slightly exceeds twice the mass of the lightest
B mesons. If, due to strong interaction corrections, this threshold appears to be
higher, it is enough to truncate our results at the relevant Higgs
boson mass values.

Further details of the phenomenology (explicit expressions for
production rates and branching fractions) are model dependent (see for example
\cite{HHG} for descriptions). It is however important to note that
extensions of the Higgs sector beyond two doublets do not increase
the list of available final states. We therefore choose the universal
approach to extract, for each process and as a function of the Higgs
boson masses, upper bounds on the production cross-section times the branching fraction into
the considered final state. These
bounds will be expressed in terms of reference cross-sections, defined below for the three
primary processes.

Any final state initiated by \epem \ra hZ is
conveniently expressed in terms of the Standard Model hZ cross-section
(we use the computation from \cite{gkwsig})
and suppression factors arising from mixing of the Higgs doublets
and branching fractions (hereafter denoted R and BR, respectively). 
Given what is said above, we have:

\begin{eqnarray*}
\mathrm{\sigma_{hZ \mra \mbb Z}} &=& \mathrm{\sigma_{hZ}^{SM} \times R_{hZ} \times BR(h\mra \mbb)} \\
               &\equiv& \mathrm{\sigma_{hZ}^{SM} \times C^{2}_{Z(h\mra bb)}};\\
\mathrm{\sigma_{hZ \mra \mtautau Z}} &=& \mathrm{\sigma_{hZ}^{SM} \times R_{hZ} \times BR(h\mra \mtautau)}\\
               &\equiv& \mathrm{\sigma_{hZ}^{SM} \times C^{2}_{Z(h\mra \tau\tau)}};\\
\mathrm{\sigma_{(AA)Z \mra 4b+jets}} &=& \mathrm{\sigma_{hZ}^{SM} \times BR(Z \mra hadrons )
                              \times R_{hZ} \times BR(h\mra AA) 
                              \times BR^{\,2}(A\mra \mbb)} \\
                     &\equiv& \mathrm{\sigma_{hZ}^{SM} \times BR(Z \mra hadrons)
                              \times C^{2}_{Z(AA\mra 4b)}}.
\end{eqnarray*}

\noindent In the particular case of the 2HDM of type II,
characterized by two mixing angles $\alpha$, $\beta$ and the two Higgs doublets
coupling to the up- and down-type fermions respectively, we would have
$\mathrm{R_{hZ}} = \sin^{2}(\alpha-\beta)$, $\Gamma(\mathrm{h \mra \mbb,\mtautau}) \propto
|\sin\alpha / \cos\beta|^{2}$, and $\mathrm \Gamma(A \mra \mbb) \propto
\tan^{2}\beta$. The factorization of the cross-section into a reference
cross-section and a term C$^{2}$ containing all details about the
Higgs sector is general. Our results will be expressed in terms of
$\mathrm C^{2}_{Z(h\mra bb)}$, $\mathrm C^{2}_{Z(h\mra \tau\tau)}$, and $\mathrm C^{2}_{Z(AA\mra 4b)}$\footnote{To 
keep the notation compact, we drop the distinction between particle and anti-particle
in the expressions of the C$^2$ factors.}.

The reference cross-section for \epem \ra hA is
obtained by computing this process in the absence of any mixing in the
Higgs sector (using {\tt HZHA}~\cite{HZHA}), and depends only on electroweak constants 
and the h and A Higgs boson masses. It is thus well-suited to express our results 
in a general way. The processes that interest us are:

\begin{eqnarray*}
\mathrm{\sigma_{hA \mra 4f}} &=& \mathrm{\sigma_{hA}^{ref} \times R_{hA}
                                     \times BR(h\mra \mff)
                                     \times BR(A\mra \mff)} \\
                            &\equiv& \mathrm{\sigma_{hA}^{ref} \times
                                     C^{2}_{hA\mra 4f}} ; \\
\mathrm{\sigma_{(AA)A \mra 6b}} &=& \mathrm{\sigma_{hA}^{ref} \times R_{hA}
                          \times BR(h\mra AA)
                          \times BR^{\,3}(A\mra \mbb)} \\
                 &\equiv& \mathrm{\sigma_{hA}^{ref} \times
                          C^{2}_{hA\mra 6b}} ;  \\
\mathrm{\sigma_{h(hZ) \mra 4b+jets}} &=& \mathrm{\sigma_{hA}^{ref} \times R_{hA}
                               \times BR(A\mra hZ)
                               \times BR^{\,2}(h\mra \mbb)
                               \times BR(Z\mra hadrons)} \\
                      &\equiv& \mathrm{\sigma_{hA}^{ref}  \times BR(Z\mra hadrons) \times
                               C^{2}_{Z(hh\mra 4b)}};
\end{eqnarray*}

\noindent where f stands for b or $\tau$. In the 2HDM, we would have $\mathrm R_{hA}=\cos^{2}(\alpha-\beta)$. 
Our upper bounds will be set on 
$\mathrm C^{2}_{hA\mra 4b}$, $\mathrm C^{2}_{hA\mra 4\tau}$, $\mathrm C^{2}_{hA\mra 6b}$, 
and $\mathrm C^{2}_{Z(hh\mra 4b)}$. 

Reference cross-sections for the Yukawa process are obtained in a
similar way. The Standard Model
\epem \ra \ff h (f=b,$\tau$) cross-section is used for h
production. Computing this cross-section with a suitable
(pseudo-scalar) \ff A vertex gives the reference for A
production (both cross-sections are taken from \cite{JKMK}). We obtain:

\begin{eqnarray*}
\mathrm{\sigma_{\mbb h \mra 4b}} &=& \mathrm{\sigma^{SM}_{\mbb h}
                               \times R_{\mbb h}
                               \times BR(h\mra \mbb)}\\
                      &\equiv& \mathrm{\sigma^{SM}_{\mbb h} \times
                               C^{2}_{bb(h\mra bb)}};\\
\mathrm{\sigma_{\mbb h\mra \mbb \mtautau}} &=& \mathrm{\sigma^{SM}_{\mbb h}
                                    \times R_{\mbb h}
                                    \times BR(h\mra \mtautau)}\\
                      &\equiv& \mathrm{\sigma^{SM}_{\mbb h} \times
                                    C^{2}_{bb(h\mra \tau\tau)}} ;\\
\mathrm{\sigma_{\mtautau h\mra 4\tau}} &=& \mathrm{\sigma^{SM}_{\mtautau h}
                                    \times R_{\mtautau h}
                                    \times BR(h\mra \mtautau)}\\
                      &\equiv& \mathrm{\sigma^{SM}_{\mtautau h} \times
                                    C^{2}_{\tau\tau(h\mra \tau\tau)}} ; \\
\end{eqnarray*}

\noindent and similar expressions for Yukawa production of A bosons. Again
$\mathrm C^{2}_{bb(h\mra bb)}$, $\mathrm C^{2}_{bb(h\mra \tau\tau)}$, $\mathrm C^{2}_{\tau\tau(h\mra \tau\tau)}$ and
the similar expressions for A contain all terms specific to the Higgs sector under
consideration. In 2HDM(II), the vertex enhancement factors
$\mathrm R_{\mbb h}$ and $\mathrm R_{\mbb A}$ are $|\sin\alpha/\cos\beta|^2$
and $\tan^{2}\beta$, respectively. Note that since the Z couples much more strongly to b quarks than to $\tau$ leptons,
the \bb(h,A \ra~\tautau) process always has larger cross-section than the mirror \tautau(h,A \ra~\bb) process. This last
process is not considered.

For the hZ and hA initiated processes, the C$^2$ factors are always
products of rotation matrix elements and branching ratios, and
therefore always satisfy C$^2 < 1$. The Yukawa processes may have
C$^2 > 1$ as well, as illustrated by the 2HDM(II) example above.

Our results may be interpreted in a large number of models and situations. Results on the decay
h \ra AA can be applied to H \ra hh as well, provided this last channel is open. In the case of
CP violation in the Higgs sector, pair production of the two lightest Higgs bosons h$_1$ and h$_2$ is different
from the CP-conserving \epem \ra hA only by an additional form factor that can be absorbed in $\mathrm{R_{hA}}$. 
Similarly, CP-violating Yukawa production of the lightest Higgs boson, \epem \ra \ff h$_1$, can always be 
written as a weighted sum of the CP-conserving \ff h and \ff A cross-sections \cite{pilaf}, and can be bounded from below: 

\begin{eqnarray*}
\mathrm{\sigma_{\mff h_1}} &=& \mathrm{\frac{R^{S}_{\mff h_1}}{R_{\mff h}} \times \sigma_{\mff h} + \frac{R^{P}_{\mff h_1}}{R_{\mff A}} \times \sigma_{\mff A}} \\
              &>& \mathrm{\left( \frac{R^{S}_{\mff h_1}}{R_{\mff h}} + \frac{R^{P}_{\mff h_1}}{R_{\mff A}} \right) \times min(\sigma_{\mff h},\sigma_{\mff A})} \\
         &\equiv& \mathrm{R_{\mff h_1} \times min(\sigma_{\mff h},\sigma_{\mff A})},
\end{eqnarray*}

\noindent where $\mathrm R^{S}_{\mff h_1}$ and $\mathrm R^{P}_{\mff h_1}$ are scalar and pseudoscalar effective couplings of the lightest Higgs boson
to the primary fermion, and $\mathrm R_{\mff h}$ and $\mathrm R_{\mff A}$ are defined above; therefore, comparing a CP-violating model 
prediction for \epem \ra \ff h$_1$ (summarized in $\mathrm R_{\mff h_1}$, and taking branching fractions into account) to our 
weakest exclusion among the corresponding \epem \ra \ff h and \ff A processes always yields a conservative answer.

On the contrary, our results on \epem \ra hZ do assume standard quantum numbers for the Higgs
boson, as a non-standard Higgs boson parity would imply different polarization of the associated Z particle, and
hence different polar angle distributions for the final bosons. The signal selection efficiency is thus affected,
and our results in this domain should be used with care.

The results also apply to the production of non-Higgs scalar particles. The cross-sections and the analyses presented here 
however assume that the produced scalars have negligible width (less than 1 GeV).

\subsection{Data samples and simulation}
\label{dataset}

The data used in this analysis amount to 79.4~pb$^{-1}$ collected by DELPHI at LEP1, in 1994 and
1995, and 611.2~pb$^{-1}$ collected at the highest LEP2 energies in the
years 1998 to 2000. The subsamples and corresponding centre-of-mass energies are listed in
Table~\ref{ene}. 

A detailed description of the DELPHI detector layout and performance can be found 
in \cite{delperf}. The data analysed in this paper were taken in
optimal conditions up to the last period of the year 2000, when
DELPHI was affected by the failure of one of the twelve sectors of its
main tracking device, the Time Projection Chamber (TPC). The
tracking algorithm was adapted, and tracks crossing the flawed region were 
recovered with the silicon Vertex Detector,
the Inner Detector, and the Outer Detector. This modification was fully incorporated in the
physics events simulation \cite{DELhiggs}. 

Large Monte Carlo samples of background and signal events have been produced 
using the {\tt PYTHIA}\cite{PYTJET}, {\tt KK2f}\cite{KK2F}, 
{\tt EXCALIBUR}\cite{EXCALIBUR}, {\tt WPHACT}\cite{WPHACT} and 
{\tt HZHA} event generators. The size of the two-quark (QCD) and
four-fermion Standard Model background samples represent about 50 times
the luminosity collected at LEP2, and two to five times the luminosity
collected at LEP1.

Yukawa events were simulated on the Z resonance with a generator based on \cite{JKMK}.
The h and A bosons were radiated off primary $\tau$~leptons and b~quarks, and decayed into 
$\tau$~lepton or b~quark pairs. The signal samples contain 10000 events each,
with Higgs boson mass values ranging from threshold up to 50~\GeVcc.

The available indirect Higgs boson decay channels were simulated for
the LEP2 analyses. (AA)A \ra 6b events were simulated with \mA\ between 12 and 50 \GeVcc\ and \mh\ between
30 and 170 \GeVcc; (AA)Z \ra (4b)\qq\ events were simulated with \mA\ between 12 and 50 \GeVcc\ and \mh\
between 30 and 105 \GeVcc; h(hZ) \ra \bb(\bb\qq) events were simulated with \mh\ between 12 and 30 \GeVcc\
and \mA\ from 110 to 170 \GeVcc. The direct decay processes hA \ra 4$\tau$ and hA \ra 4b were simulated
over the whole kinematically allowed mass range. 

The LEP2 background events were simulated at all centre-of-mass energies listed in Table~\ref{ene}.
The LEP2 signal events were generated at $\sqrt{s}=$200 GeV, in mass steps of 5~\GeVcc\ close to the 
decay thresholds, and 10~\GeVcc\ elsewhere. Dedicated samples for systematic uncertainty evaluation were 
generated at all LEP2 centre-of-mass energies, for a reduced number of mass points. All LEP2 signal
samples contain 2000 events.

All generated events used {\tt PYTHIA} for decay and hadronization and 
were processed through the detailed DELPHI simulation program~\cite{delsim}.

\begin{table}
\caption{Centre-of-mass energies and corresponding luminosities used in the analysis.
The first and second number for the year 2000 correspond to the luminosity recorded before and 
after the failure of one TPC sector, respectively.} 
\begin{center}
\begin{tabular}{c|c|cccc|c}
year & $1998$ & \multicolumn{4}{c|}{$1999$} & 2000 \\ 
\hline
$\sqrt{s}$ (GeV) & $189$ & $192$ & $196$ & $200$ & $202$ & 202 to 208 \\
${\cal{L}}$ (pb$^{-1}$) & $158.0$ & $25.9$ & $76.9$ & $84.2$ & $41.1$ & $164.1 + 61.0$  \\
\end{tabular}
\end{center}
\label{ene}
\end{table}

\subsection{Methods common to all analyses}
\label{methods}

Unless stated otherwise, charged particles are selected if their
momentum is greater than 100 MeV/c, and if their measured distance to
the interaction point is less than 4 cm in the transverse plane,
and less than 4 cm/$\sin\theta$ along the beam direction, where $\theta$ is the particle polar angle. 
Neutral particles are defined as calorimetric clusters not associated to tracks, and are
selected if their measured energy is larger than 200 MeV in the
electromagnetic calorimeter, or larger than 300 MeV in the hadron calorimeter.

The analyses described below select $\tau$ particles, and the
selection criteria rely partly on the identification of their leptonic
decay products. Muons are identified in the muon chambers,
where signals coincide with the extrapolation of tracks measured in
the central detectors. Muons are also characterized by energy deposits
in the hadron calorimeter, compatible with minimum-ionizing
particles. Electrons are identified mainly by energy loss
measurements in the TPC, shower profile variables in the
electromagnetic calorimeter, and by comparing the measured track
momentum and associated calorimeter energy. In the analyses searching for $\tau$'s
at LEP1, the DELPHI standard identification tag is used for both lepton flavours, 
with performances given in \cite{delperf}. In the four-$\tau$ search at LEP2, the lepton 
selections are very similar to those developed for the analysis of fully leptonic W pair decays \cite{wwpaper}.

The method used to select b~quark jets is described in detail in
\cite{btag-summ}. Variables that discriminate between fragmented
b~quarks (leading to long lived B hadrons) and ordinary jets are
combined into a single variable, hereafter denoted $x_b$ for events
and $x_{bi}$ for the jet of $i$-th largest b-likeness (in four-jet
events, $x_{b1}$ is the highest jet b-tagging value, and $x_{b4}$ is
the lowest). Contributions to this variable are the compatibility of
tracks with the primary vertex, based on their measured impact
parameter; the transverse momentum of identified leptons with respect to the jet
axis; and the rapidity, effective mass, and fraction of the jet
momentum, of particles assigned to a possible reconstructed secondary
vertex.

All search results presented in this work are interpreted using a
modified frequentist technique based on the extended
likelihood ratio \cite{alrmc}. For a given experiment, the test
statistic $Q$ is defined as the likelihood ratio of the signal+background hypothesis ($s+b$) to
the background hypothesis ($b$), computed from the number of observed and expected events
in both hypotheses. Individual events
may also carry a signal-to-background ratio based on a measured
discriminating variable, such as the reconstructed mass (this possibility is used in
the LEP2 four-$\tau$ search). Probability
density functions (PDFs) for $Q$ in the $b$ and $s+b$ hypotheses are
built using Monte Carlo sampling of the (Poisson-distributed)
background and signal expectations, and of the optional discriminating
variable distributions. The confidence levels $\mathrm{CL_b}$ and
$\mathrm{CL_{s+b}}$ are defined as the integrals of the $b$ and $s+b$
PDFs for $Q$ between $-\infty$ and the actually observed value
$Q_{obs}$. The confidence level in the signal hypothesis,
$\mathrm{CL_s}$, is conservatively approximated by the ratio
$\mathrm{CL_{s+b}/CL_b}$. 1-$\mathrm{CL_s}$ measures the confidence
with which the signal hypothesis can be rejected, and will be larger
than 0.95 for an exclusion confidence of 95\%.

 
\section{LEP1 data analysis}
\label{allLEP1}

This section describes the search for the Yukawa process in LEP1 Data. The four-b, \bb\tautau, 
and four-$\tau$ final states are analysed.

\subsection{The four-b final state}
\label{4bLEP1}

This section describes a search for neutral Higgs boson production
in the four-b channel. The analysis is focused on the Yukawa process, and subsequently applied
to Higgs boson pair production.

Let us first discuss the issue of the background estimation. An irreducible background contribution
originates from events with two primary b quarks and a gluon splitting into a second b quark pair, i.e. Z \ra \bb(g \ra \bb).
This gluon splitting happens with a probability $\mathrm g_{bb}$.
The most recent theoretical estimate is $\mathrm{g_{bb}^{th}}=1.75 \pm 0.40 \times 10^{-3}$ \cite{Seymour}. 
In the simulation we use $\mathrm{g_{bb}=1.5 \times 10^{-3}}$, the default value  in \cite{PYTJET}, somewhat below
the theoretically preferred value. This quantity has also been measured by the LEP and SLD Collaborations, with an 
average result of $\mathrm{g_{bb}^{exp}}=2.74 \pm 0.42 \times 10^{-3}$ \cite{worldgbb}. 

The available measurements are however not insensitive to four-jet events with light Higgs boson decays to b quark 
pairs which, if present, would contaminate the selected samples and lead to an overestimation of the measured 
$\mathrm g_{bb}$ value. This possibility was not taken into account in \cite{worldgbb}. The efficiency 
of these analyses on Higgs boson events has not been estimated, and therefore the $\mathrm g_{bb}$ 
measurements potentially contain a contribution from Higgs boson events.

Our strategy is therefore to keep the value of $\mathrm{g_{bb}=1.5 \times 10^{-3}}$ in the simulation. The possible
presence of an excess in the data can then be interpreted in two alternative ways: either by attributing the
excess to gluon splitting events and estimate the additional contribution to $\mathrm g_{bb}$ 
(this is not the focus of this paper, and will be done only indicatively in the following), or by 
attributing the excess to the signal and obtain conservative limits on Higgs boson production. Considering 
the large uncertainties on the various estimates of $\mathrm g_{bb}$, we do not use this channel for signal 
discovery.

The analysis itself is described in the following. For Higgs boson masses of about half of the Z mass we expect 
a four-jet topology, whereas close to threshold only three jets may be reconstructed. Taking this 
into account we develop two parallel selection procedures, corresponding to event reconstructions in three 
and four jets respectively.

At first, the events are required to contain at least six charged
particles. At this preselection stage we force the reconstruction of three jets, 
and the 2 \ra 3 jet transition point $y_{23}$ of the Durham algorithm
\cite{durham} should be greater than 0.01. For all reconstructed jets,
the b-tagging values $x_{bi}$ are computed as described in Section~\ref{methods}, and
ordered from higher to lower b-likeness. The b-tagging variable of
the most b-tagged jet, $x_{b1}$, is required to be greater than 0. 

The preselection eliminates all backgrounds but hadronic Z decays.
Non-b hadronic events are significantly reduced as well and represent about
10\% of the remaining sample. After this step, all events are reconstructed as four-jet events.

\begin{figure}[p]
\begin{center}
\epsfig{file=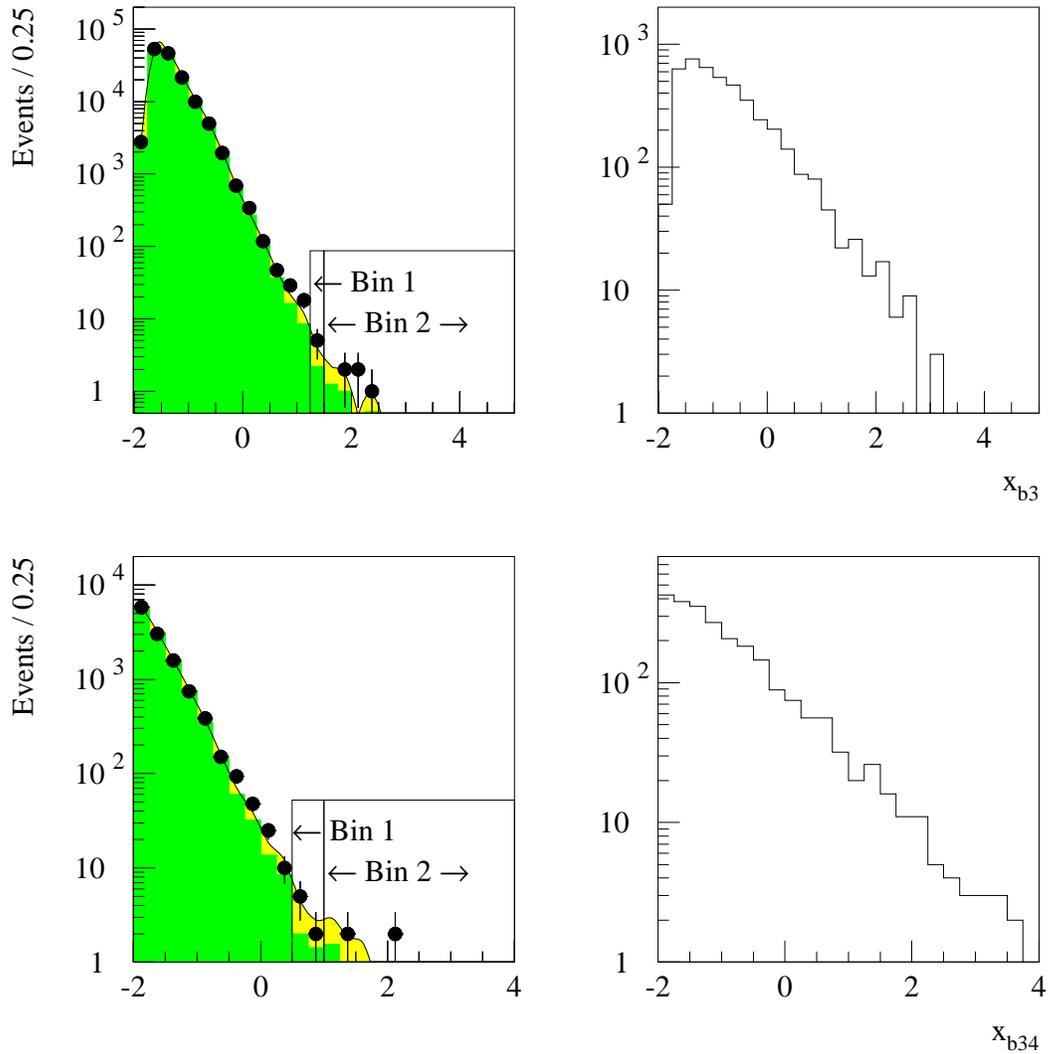,width=.9\textwidth}
\caption{Comparison between data and simulation for the distributions of the final b-tagging
variables of the Yukawa four-b analyses, as defined in the text (left). The points are the data. 
On top of the dark histogram, representing the Standard Model \qq\ background with
$\mathrm{g_{bb}=1.5\,10^{-3}}$, a fit to the data suggests a larger
gluon splitting value (see text). Distributions expected
for a \bb(h~\ra\bb) signal are shown on the right, with arbitrary normalization.}
\label{fg-datap-lot}
\end{center}
\end{figure}

The remaining b-tagging discriminating power is contained in the least
tagged jets. The final selection relies on \xbiii\ in the three-jet
topology, and on the sum $x_{b34}=x_{b3}+x_{b4}$ in the four-jet
topology. The distributions of these variables are shown in Figure~\ref{fg-datap-lot}. 
In both the three-jet and four-jet analyses, two channels are defined for the
final analysis (denoted Bin 1 and Bin 2, see again Figure~\ref{fg-datap-lot}). They are chosen to have a similar
expected background, and a signal efficiency of at least 1\% to 2\% in Bin 2.

Numerical comparisons between the data and the simulation
are shown in Table \ref{tab4bL1}. The 1\% difference seen at the preselection 
level is explained by residual imperfections of the b-tagging efficiency simulation \cite{btag-summ}.
At the end of the analysis, an excess of data is observed in all channels. One explanation could be
the possible underestimation of the gluon splitting probability.

\begin{table}
\caption{Number of observed and expected background events in the Yukawa four-b analyses,
at various steps of the selection; $\mathrm{g_{bb}}=1.5 \times 10^{-3}$.}
\begin{center}
\begin{tabular}{lll|r|r}
Cut   &       &                       & Total background & Data (94-95)\\ \hline
preselection   &       &                       & 141128 $\pm$ 207 & 142527      \\
three-jet topology:&       & $       \xbiii > -2 $ & 140705 $\pm$ 206 & 142042      \\
               & Bin 1 & $   1.5>\xbiii>1.25 $ &    2.2 $\pm$ 0.9 & 5           \\
               & Bin 2 & $       \xbiii> 1.5 $ &    3.2 $\pm$ 1.1 & 5           \\
four-jet topology:&       & $        \xbiv > -2 $ &  11421 $\pm$ 17  &  11848      \\
               & Bin 1 & $1.0\,\, >\xbiv>0.5 $ &    3.4 $\pm$ 1.1 & 7           \\ 
               & Bin 2 & $       \xbiv > 1.0 $ &    3.5 $\pm$ 1.0 & 4           \\ 
\end{tabular}
\label{tab4bL1}
\end{center}
\end{table}

Efficiencies for h and A production in the Yukawa process are shown in Table~\ref{bbbb_Y}. For the
interpretation of results, the three-jet or four-jet analysis is chosen at each mass point as a
function of the expected exclusion performance.

The selection developed for this search is directly applied to
pair production of neutral Higgs bosons, with efficiencies given
in Table~\ref{ha1-eff}. The efficiencies are evaluated for both
three-jet and four-jet analyses, and found to be almost always better in
the second case. The four-jet analysis is retained for the interpretation 
of the results, described in Section~\ref{res:direct}. 

The systematic uncertainty related to the residual differences between
b-tagging efficiency in data and in the simulation is estimated using
Ref.~\cite{btag-summ}, where it is shown that the difference is limited
to $\pm$10\% for high purity b jet selection. This uncertainty is assumed,
and added in quadrature to the statistical uncertainty from the
limited size of the simulation samples. Considering the
conservative assumptions on data and background described above, no
further systematic uncertainty is assumed. 

A fit to the \bb\ and \bb g \ra 4b components of the data is performed as a cross-check. 
An independent sample of four-b events with gluon splitting is introduced,
and its normalization is adjusted so that its addition to the standard simulation 
(with $\mathrm{g_{bb}=1.5 \times 10^{-3}}$) reproduces the observation.
In the three-jet analysis, the additional contribution is found to be $(3.0 \pm 0.7) \times 10^{-3}$, 
bringing the total gluon splitting value to $(4.5\pm 0.7) \times 10^{-3}$. In the four-jet analysis,
$(3.2\pm 0.7) \times 10^{-3}$ is found, leading to a total of $(4.7\pm 0.7) \times 10^{-3}$.
The result is displayed in Figure~\ref{fg-datap-lot} as well. This 
estimation of $\mathrm{g_{bb}}$ is purely indicative.

\subsection{The \bb\tautau\ final state}
\label{2b2tLEP1}

In the \bb\tautau\ final state of the Yukawa process (i.e., \bb(h \ra \tautau)), the Higgs boson 
decay products often have high momentum, and appear as a collimated slim jet. We therefore reconstruct 
three jets in this final state, of which one is expected to contain a pair of $\tau$~leptons of low decay
multiplicity. The two other jets, initiated by b~quarks, are expected to have higher multiplicity.

As in the previous analysis, event reconstruction is forced into three jets using the
Durham algorithm. The b-tagging algorithm is then
applied to evaluate the b-likeness at both the event and jet levels. The jets are
ordered according to their b-tagging value; the two jets with highest value are assumed
to be b-jets.

At the preselection level we require the total charged multiplicity in the event
to be at least 10. As before, the Durham parameter $y_{23}$ is required to be greater
than $0.01$. The event b-tagging variable $x_b$ must be greater than
0. The cosine of the angle between the two b-jets should satisfy
$\cos{\alpha_{12}} < 0.9$. The preselection eliminates
almost all non-hadronic background components, leaving mostly
Z \ra \bb\ events. 

Furthermore, $x_{b1}$ is required to be greater than 0 and $x_{b2}$ to be greater than -1.
The jet with lowest b-tagging value is supposed to correspond to
the $\tau$ pair. Events with gluon radiation may fake the signal, but
gluon jets usually have high multiplicity, whereas we expect the
$\tau$ pair to be narrow and have low multiplicity. For the remaining cuts, only charged 
particles of momentum greater than 1~\GeVc\ are taken into account. 

\begin{figure}
\begin{center}
\epsfig{figure=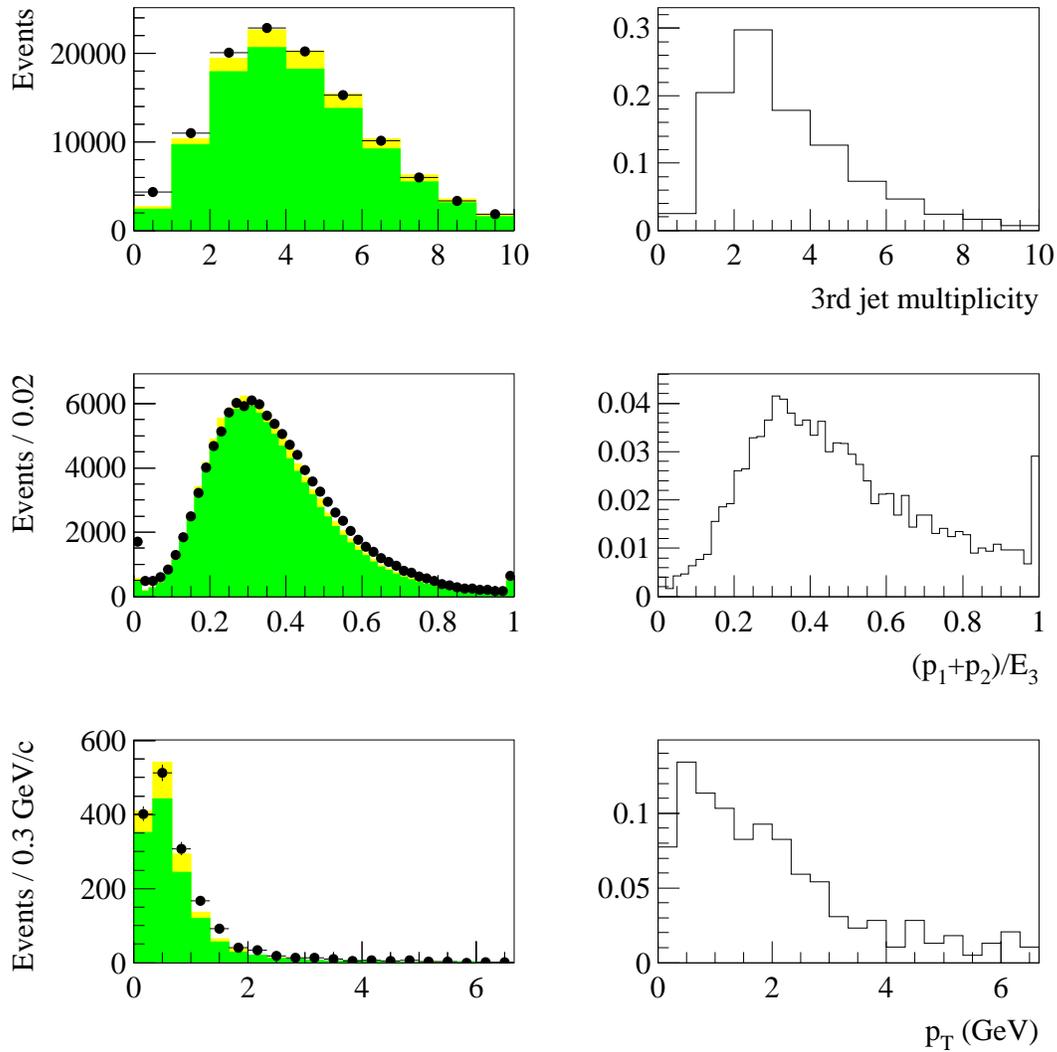,width=0.9\textwidth}
\end{center}
\caption{Comparison between data and simulation for the distributions of some variables used in the
\bb\tautau\ analysis, at the preselection level. On the left, the points are the data, the dark histograms 
represent the Standard Model \qq\ background, and the light histograms 
represent the \qq\lplm\ (l=e, $\mu$, $\tau$) contribution. The histograms on the right show 
distributions for a \bb(h~\ra\tautau) signal, with arbitrary normalization.}
\label{btdatamc}
\end{figure}

The jet of lowest b-likeness is required to
have a charged multiplicity of 1, 2 or 3, and a total multiplicity of at least 2. 
Its broadness, defined as the cosine of the largest angle between two 
particles in the jet, $\vert\cos{\theta}\vert$, should be larger
than~$0.64$. The sum of momentum fractions of the two particles with the
highest momentum in this jet, denoted $(p_{1}+p_{2})/E_{3}$ (where $E_{3}$ is the energy of the jet of 
lowest b-likeness), should be greater than~$0.5$. Furthermore, we require at least one leptonic $\tau$ decay, 
by demanding an identified lepton (muon or electron), with $\mathrm{p_{T}}>$1~\GeVc\ (where $\mathrm{p_{T}}$ 
is defined as the transverse momentum of the lepton with respect to the jet axis). Figure~\ref{btdatamc} 
illustrates three of the selection variables described above.

Seven events are selected in the data, whereas 10.6$\pm$2.3 events are expected from background
processes. 
Along the whole selection procedure, hadronic Z decays are the dominant background contribution; 
less than 5\% arise from four-fermion processes. Numerical comparisons between data and simulation 
are shown in Table~\ref{tabbtL1}.

\begin{table}
\caption{Number of observed and expected background events in the Yukawa
\bb\tautau\ analysis, at various steps of the selection.}
\begin{center}
 \begin{tabular}{l|rr}
Cut & Total background & Data (94-95) \\ 
\hline
preselection            & 120015 $\pm$  285 & 116485 \\
$x_{b1}, x_{b2}$        &  38385 $\pm$  161 &  36195 \\
$3^{rd}$ jet multiplicity      &  10015 $\pm$   83 &   9808 \\
$3^{rd}$ jet broadness  &   2143 $\pm$   38 &   2033 \\
lepton ID               &  461.9 $\pm$ 17.9 &    430 \\
lepton $\mathrm{p_{T}}$ &   10.6 $\pm$  2.3 &      7 \\
 \end{tabular}
 \label{tabbtL1}
\end{center}
\end{table}

The selection efficiencies for \bb(h \ra \tautau) and \bb(A \ra \tautau) are given in
Table \ref{bbtt_Y}. The small difference in rejection
between data and expected background, evaluated at the preselection
level and for each selection variable, leads to a systematic uncertainty of 3.0\% on
the background expectation, and to 3.8\% on the signal
efficiency. These values are added in quadrature to the statistical
errors given in Tables \ref{tabbtL1} and \ref{bbtt_Y}.

\subsection{The four-$\tau$ final state}
\label{4tLEP1}

In this section we describe a search for Higgs boson production in the
four-$\tau$ channel, via the Yukawa process. This final state
can be dominant in models where Higgs doublets couple preferentially
to leptons. Since the one-prong $\tau$ decay is largely
dominant, a first analysis, sensitive to events with four charged
particles seen in the detector, is described below. Nevertheless, when four $\tau$'s
are present, the probability that one of them decays into three
charged particles is significant. To account for these events,
a complementary analysis is developed and is described in the second 
part of this section. These four-prong and six-prong decays represent
respectively 53.1\% and 37.8\% of all events with four $\tau$~leptons.

Due to the nature of the final states considered here (i.e., low
multiplicity and low visible energy) the acceptance criteria for
reconstructed particles are tightened compared to the description given in
Section \ref{methods}. Charged particles are now selected if their momentum
is larger than 400 \MeVc, their angle with respect to the beam axis is
larger than $20^{\circ}$, they are seen in the TPC and finally,
their impact parameter along the beam axis is less than 3 cm.

\subsubsection{The four-prong selection}

The following series of preselection cuts are used to reject events from
beam-gas interactions and from $\gamma\gamma$ collisions. 
Only events with exactly four reconstructed charged particles are considered.
The total electric charge of the particles must be 0. The sum of the
impact parameters with respect to the beam-spot must be less than
300~$\mathrm \mu m$ in the transverse plane. The pair of oppositely charged particles of lowest
invariant mass, denoted $m_\pm$ in the remaining of this section,
must be separated from at least one of the remaining charged particles by more than $90^{\circ}$. 
The invariant mass $m_\pm$ must be larger than 200 \MeVcc. The missing momentum along the
beam axis  must be less than $35\%\sqrt{s}$. Finally, either the
missing transverse momentum must be larger than $5\%\sqrt{s}$, or
the visible mass must be greater than 25~\GeVcc.

At this stage, the main background consists of
Z \ra \tautau\ events, where the $\tau$'s have decayed into one prong and three prongs, respectively. This
background is reduced by requiring the lowest triplet invariant mass to be
greater than 2~\GeVcc. The remaining \tautau\ events have both $\tau$'s decayed into three prongs, 
when one charged particle is missed in each
hemisphere. To reject them, the visible mass recoiling against
$m_\pm$ is required to be larger than 2~\GeVcc.

Remaining backgrounds come from low-multiplicity hadronic Z
decays and four-fermion events. These background components are reduced
by requiring the pair of charged particles recoiling against
$m_\pm$ to have a mass larger than 10\% of the total visible
mass. Furthermore, the neutral multiplicity must
not exceed six. Four-fermion events not containing $\tau$~leptons are
rejected by requiring the visible mass to be less than 60~\GeVcc. 
One of the particles in $m_\pm$ should be
identified as an electron or muon, and the other one should not be identified as
a lepton of the same flavour; the remaining two charged particles
should not both be identified as electrons or muons. Finally, the cut on the
invariant mass $m_\pm$ is tightened to $m_\pm >$1~\GeVcc.

\begin{figure}
\begin{center}
\epsfig{figure=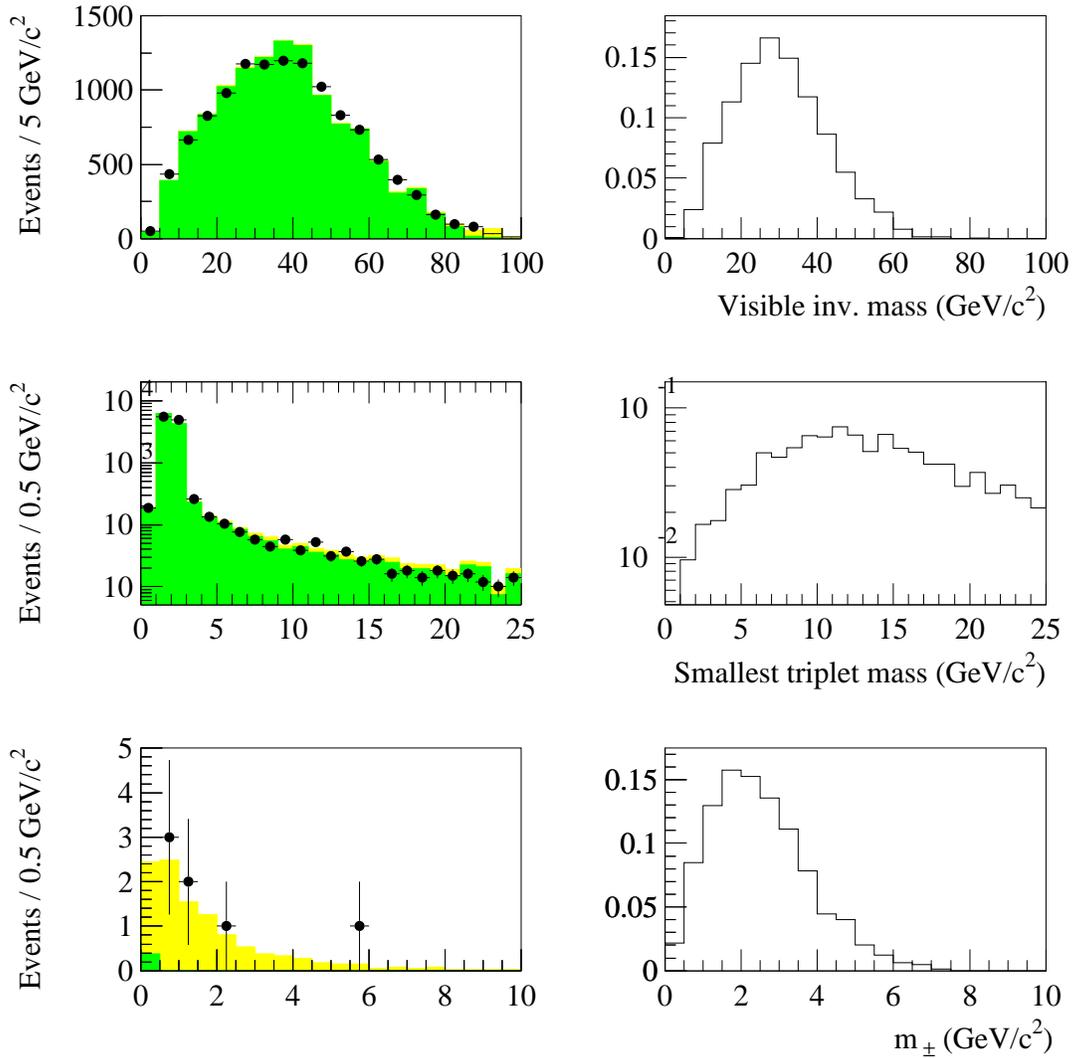,width=0.9\textwidth}
\end{center}
\caption{Comparison between data and simulation for the distributions of some variables used in the
four-$\tau$ four-prong analysis. The visible mass and the lowest triplet mass are
shown at the preselection level. The $m_\pm$ mass is shown just before the final 
selection cut. On the left, the points are the data, the dark histograms represent the
Standard Model \qq\ and \tautau\ backgrounds, and the light histograms
represent the various four-fermion contributions. The histograms on the right show
distributions for a \tautau(h~\ra\tautau) signal, with arbitrary normalization.}
\label{datamc4p}
\end{figure}

Distributions of the visible mass and of the lowest triplet mass are
displayed in Figure~\ref{datamc4p} at the preselection level.
The distribution of the $m_\pm$ invariant mass is also shown, just before the last cut is applied, with
seven observed events and 10.8$\pm$1.0 expected events.

After all selection cuts are applied, four events are observed in the data, while
4.1$\pm$0.5 are expected from background, all of which
are genuine four-fermion events; the contribution from four-lepton
events with at least one $\tau$ pair amounts to 3.8$\pm$0.5 events and
the remaining originates from four-lepton events with electrons and
muons only.

Comparisons between data and simulated background samples are shown in
Table~\ref{tab4pr}. Signal efficiencies vary from 3\% to 6\%,
going from low to high signal mass (see Table~\ref{tttt_Y4p} for details). These efficiencies 
correspond to 5.7\% and 11.8\% of the true four-prong decays of the signal.

\subsubsection{The six-prong selection}

Exactly six reconstructed charged particles are required in this
search. The remaining preselection criteria against beam-gas and
$\gamma\gamma$ events are applied as above.

Since one of the $\tau$~leptons is expected to decay in the three-prong
mode, the lowest triplet invariant mass should not exceed
$\mathrm m_{\tau}$; the cut is applied at 1.8~\GeVcc. Moreover, this triplet
is required to have momentum larger than 3~\GeVc. It is then treated
as a pseudo-particle, and the six-prong topology becomes a
pseudo-four-prong one.

To reject low multiplicity hadronic Z decays and $\tau$
pair decays into six prongs, the system recoiling against the
triplet of lowest mass should have a mass greater than 4~\GeVcc, and
the total multiplicity must be less than 13. The visible mass is
required to be less than 60~\GeVcc. The pair of oppositely charged particles of
lowest invariant mass must pass the cut $m_\pm >$1~\GeVcc\ (here, the pair may contain the
pseudo-particle made by the triplet of lowest invariant mass).

\begin{figure}
\begin{center}
\epsfig{figure=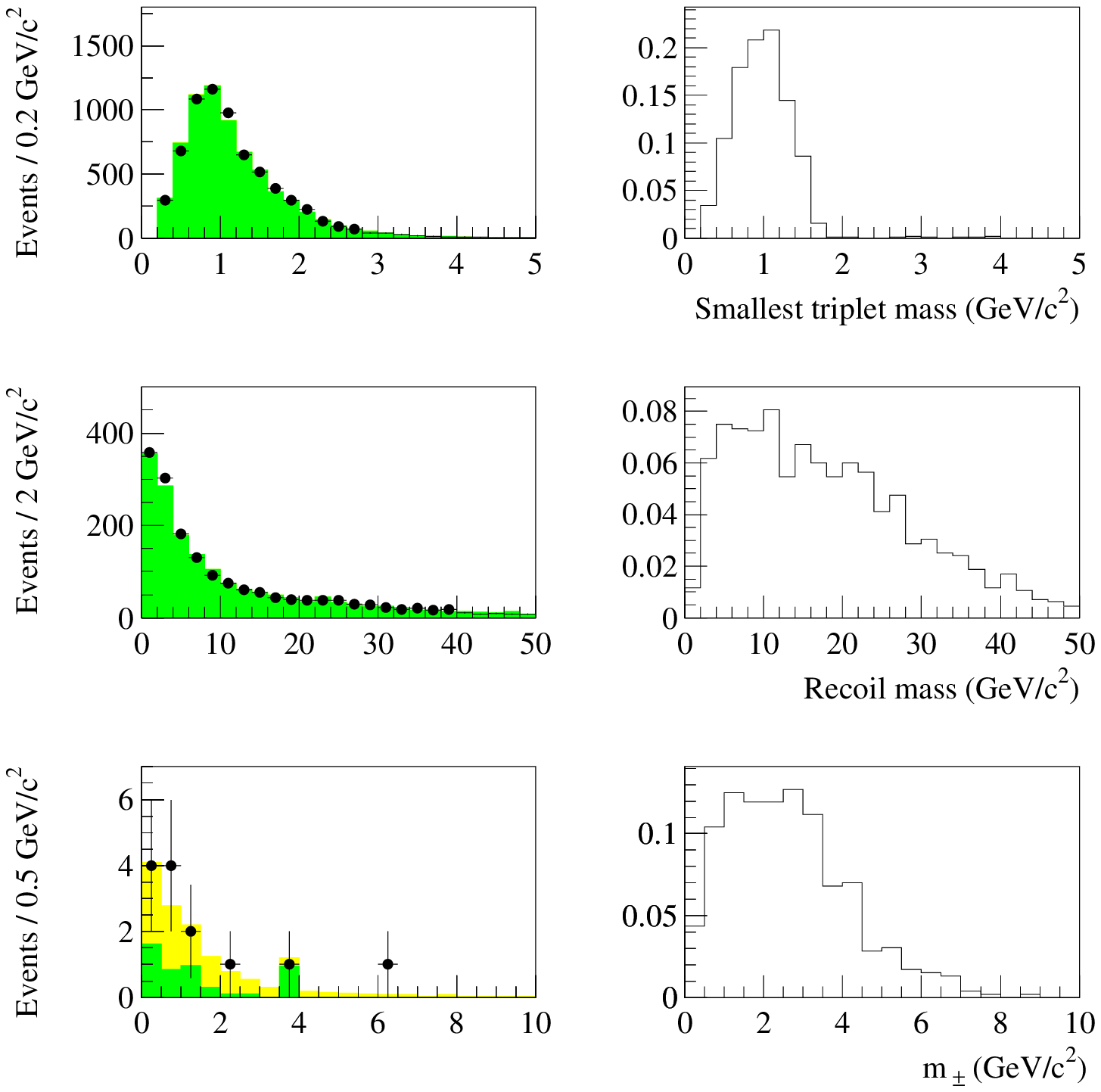,width=0.9\textwidth}
\end{center}
\caption{Comparison between data and simulation for the distributions of some variables used in the
four-$\tau$ six-prong analysis. The lowest triplet mass and the mass of the system 
recoiling against it are shown at the preselection level. The $m_\pm$ mass is 
shown just before the final selection cut. On the left, the points are the data, the dark histograms represent the
Standard Model \qq\ and \tautau\ backgrounds, and the light histograms
represent the four-fermion contributions. The histograms on the right show 
distributions for a \tautau(h~\ra\tautau) signal, with arbitrary normalization.}
\label{datamc6p}
\end{figure}

Distributions of the minimal triplet mass, and of the mass of the
three charged particles recoiling against it, are displayed at the
preselection level in Figure~\ref{datamc6p}.
The distribution of the invariant mass $m_\pm$ is also shown, just before the final cut is applied.
At this level, 13 events are observed and 14.2$\pm$2.9 are expected.

After all cuts, four events are observed, while 6.0$\pm$1.5
are expected from the simulation. Of these, 3.4$\pm$1.4 are hadronic Z
decays, 1.9$\pm$0.2 are four-lepton events with at least one
$\tau$ pair. The remaining contribution comes from four-fermion events
with two quarks and two leptons. 

The cut-by-cut evolution of the data and simulated background samples is shown in Table~\ref{tab6pr}. Signal efficiencies
vary from 2.5\% at low mass, to 5.6\% at high mass, corresponding to
6.3\% to 14.9\% of the true six-prong decays of the signal. Details can be found in
Table~\ref{tttt_Y6p}.

\paragraph{}
Systematic uncertainties on the expected backgrounds and on the signal efficiencies
are estimated as in Section~\ref{2b2tLEP1}. Each selection cut described above is applied in turn at the 
preselection level, and the difference in rejection between the data and the simulation is 
attributed to the imperfect modelling of the corresponding distribution.
The resulting uncertainties amount to 8\% on backgrounds and 5\% on signals in the four-prong analysis,
and to 3.5\% on backgrounds, and 3\% on signals in the six-prong analysis.

\begin{table}[htbp]

\caption{Four-$\tau$ final state at LEP1. Number of observed and expected
  background events, at various stages of the four-prong analysis.}
\begin{center}
\vspace{1ex}
\begin{tabular}{l|rrrrr}
Cut & \multicolumn{1}{c}{\tautau}
    & \multicolumn{1}{c}{\qq}
    & \multicolumn{1}{c}{4f}
    & \multicolumn{1}{c}{Total}
    & \multicolumn{1}{c}{Data (94-95)} \\
\hline
preselection    & 10586.0 &  1148.1 &   177.3 & 11911.4$\pm$168.5 & 11876 \\
anti-\tautau    &     8.7 &   444.3 &   152.1 &   605.1$\pm$ 17.7 &   574 \\
anti-\qq        &     3.3 &    20.8 &   121.7 &   145.8$\pm$  6.7 &   137 \\
final selection &         &         &     4.1 &     4.1$\pm$  0.5 &     4 \\
\end{tabular}
\end{center}
\label{tab4pr}

\caption{Four-$\tau$ final state at LEP1. Number of observed and expected
  background events, at various stages of the six-prong analysis.}
\begin{center}
\vspace{1ex}
\begin{tabular}{l|rrrrr}
Cut & \multicolumn{1}{c}{\tautau}
    & \multicolumn{1}{c}{\qq}
    & \multicolumn{1}{c}{4f}
    & \multicolumn{1}{c}{Total}
    & \multicolumn{1}{c}{Data (94-95)} \\
\hline
preselection               & 935.1 & 5744.4 & 80.5 & 6760.0$\pm$84.8 & 6733 \\
anti-\tautau, \qq           &   4.3 &   52.3 &  5.2 &   61.8$\pm$11.7 &   58 \\
final selection            &       &    3.4 &  2.6 &    6.0$\pm$ 1.5 &    4 \\
\end{tabular}
\end{center}
\label{tab6pr}

\end{table}

 
\section{LEP2 data analysis}
\label{allLEP2}

The searches for final states with at least four b quarks or with exactly four 
$\tau$ leptons in LEP2 data are described in what follows.

\subsection{Final states with b quarks}
\label{4bLEP2}

This section describes a search for cascade decays of neutral
Higgs bosons. The considered decay chains are hA \ra (AA)A, hZ \ra (AA)Z,
hA \ra (AZ)A and hA \ra h(hZ). The lightest Higgs boson is assumed to decay into b quark
pairs. The final state will contain six quarks, of which at least four are b~quarks.
The analysis developed here is also applied to the direct decay hA \ra 4b.

Events with cascade decays a priori lead to a six-jet final state. However,
when the mass of the lighter Higgs boson approaches 2\mb, the decay jets may not be resolved.
This then leads to a three-jet topology in the (AA)A channel,
or to a four-jet topology in the (AA)Z or h(hZ) channels.

Due to the large range of masses and topologies that are searched for,
different signals often differ more among themselves
than from the background. Instead of analysing each topology 
individually, we have designed a polyvalent method exploiting only the
presence of at least four b~quarks.

The preselection used in this analysis has been developed for 
Standard Model Higgs boson searches in hadronic events \cite{DELhiggs}, and is 
briefly outlined here. Multiplicity and energy flow cuts eliminate
radiative and $\gamma\gamma$ events, and significantly reduce the QCD
background. Selected events are then forced into a four-jet configuration using the Durham 
algorithm, and the mass of each jet is required to exceed 1.5\,\GeVcc.

The rest of the analysis does not rely on event shapes, and uses only b-tagging information. 
Variables with large discriminating power are the secondary vertex multiplicity \nvg\label{nvg}, 
the b-likeness variables \xbi\label{xbi} and \xbii\label{xbii}, and the
b-likeness sum $x_{b34} = x_{b3}+x_{b4}$\label{xbiv}. Considering the total number of
secondary vertex hypotheses $N_v$, which includes secondary vertices
failing the fit-quality selection (see \cite{btag-summ}),
achieves supplementary discrimination. A combined variable, denoted B in the following, 
is defined as the sum of the logical values of the following conditions (each satisfied
condition increases the value of B by 1 unit): 
$$ 
\mathrm{B}=(\nvg>2)+(\nva>5)+(\xbi>2)+(\xbii>0)+(\xbtf>-2).
$$

\begin{figure}
\begin{center}
\epsfig{figure=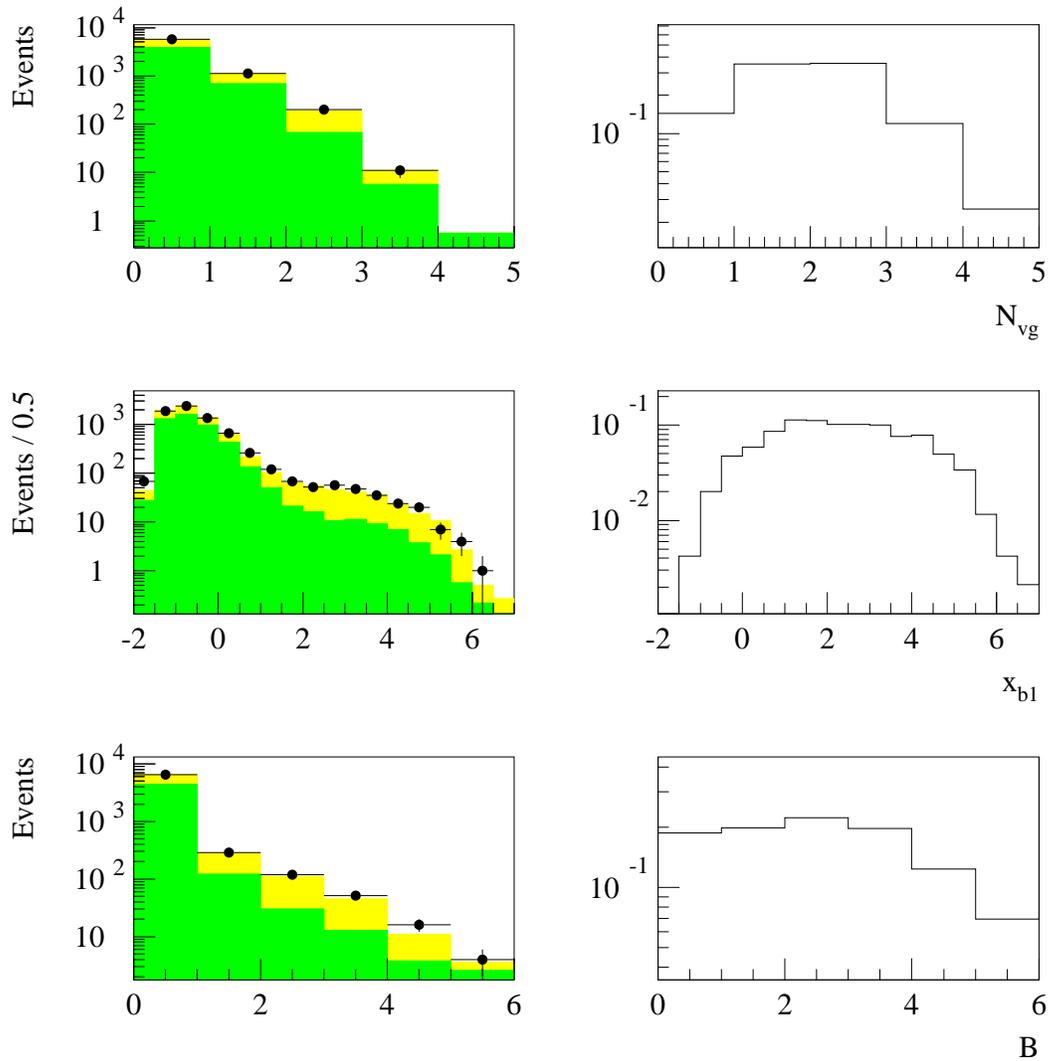,width=0.9\textwidth}
\end{center}
\caption{Comparison between data and simulation for the distributions of some variables used in the
         LEP2 four-b search, at the preselection level. On the left, the points are the data, the dark 
         histograms show the Standard Model four-fermion background, and the light 
         histograms represent the two-fermion \qq\ contribution. The histograms on
         the right show distributions for a (AA)Z~\ra~4b+jets signal, with arbitrary normalization.} 
\label{fg-bcod}
\end{figure}

For the final selection, B is required to be greater than 3.
A preselection-level data to simulation comparison of the distributions of some analysis
variables is shown in Figure~\ref{fg-bcod}. Numerical comparisons between the data and the simulation
are shown in Tables~\ref{comp0} and~\ref{comp4}.

\begin{table}

\caption{Final states with b~quarks. Comparison between data and simulation at
the preselection level. The data sets 2000a and 2000b correspond to data taken
before and after the failure of TPC sector 6, respectively.}

\begin{center}
\begin{tabular}{l|rrrr}
Data set          & 4f &   \qq    & Total &  Data  \\ \hline
$189$ GeV &  $1144.1$ &  $739.6$ &  $1883.7 \pm 28.3 $ & $1896$ \\
$192$ GeV &   $198.3$ &  $105.6$ &   $303.9 \pm  4.2 $ &  $319$ \\
$196$ GeV &   $595.1$ &  $298.2$ &   $893.3 \pm 14.3 $ &  $919$ \\
$200$ GeV &   $655.2$ &  $312.5$ &   $967.7 \pm 14.5 $ &  $949$ \\
$202$ GeV &   $318.2$ &  $144.2$ &   $462.4 \pm  6.9 $ &  $465$ \\
 2000a   &  $1295.1$ &  $563.1$  &  $1858.2 \pm 27.9 $ & $1826$ \\
 2000b   &   $447.5$ &  $192.0$  &   $639.5 \pm  9.6 $ &  $632$ \\ \hline
all energies & $4653.6$ & $2355.2$& $7008.8 \pm 46.4 $ & $7006$ \\
\end{tabular}
\label{comp0}
\end{center}

\caption{Comparison between data and simulation for events satisfying B $>3$ (final selection).
The data sets 2000a and 2000b correspond to data taken before and after the failure of TPC sector 6, respectively.}
\begin{center}
\begin{tabular}{l|rrrr}
Data set          & 4f & \qq      & Total & Data  \\ \hline
$189$ GeV  & $  1.4$  & $  1.6$ & $  3.0 \pm 0.7 $  & $  2$ \\
$192$ GeV  & $  0.2$  & $  0.5$ & $  0.7 \pm 0.3 $  & $  2$ \\
$196$ GeV  & $  1.1$  & $  1.0$ & $  2.1 \pm 0.4 $  & $  2$ \\
$200$ GeV  & $  1.0$  & $  1.0$ & $  2.0 \pm 0.3 $  & $  2$ \\
$202$ GeV  & $  0.3$  & $  0.5$ & $  0.8 \pm 0.2 $  & $  1$ \\
 2000a    & $  2.1$  & $  1.6$ & $  3.7 \pm 0.6 $  & $ 10$ \\
 2000b    & $  0.6$  & $  0.6$ & $  1.2 \pm 0.2 $  & $  1$ \\ \hline 
all energies & $6.8$  & $  6.9$ & $ 13.7 \pm 1.8 $  & $ 20$ \\
\end{tabular}
\label{comp4}
\end{center}

\caption{Breakdown of the excess observed in 2000a, and the corresponding expected background. 
Three centre-of-mass energy windows are used, namely $\sqrt{s}<205.5$, $205.5<\sqrt{s}<207.1$, and $\sqrt{s}>207.1$.}
\begin{center}
\begin{tabular}{l|rr}
Energy window (GeV) & Exp. bg & Data \\ \hline
$\sqrt{s}<205.5$       & 1.6 $\pm$ 0.3 & 5 \\
$205.5<\sqrt{s}<207.1$ & 1.9 $\pm$ 0.4 & 4 \\
$\sqrt{s}>207.1$       & 0.2 $\pm$ 0.1 & 1 \\
\end{tabular}
\label{comp5}
\end{center}

\end{table}

The excess observed in the data of 2000a after the last cut (see Table~\ref{comp4})
has been verified to be unrelated to any spurious event reconstruction problem. Its 
possible meaning will be discussed in Section~\ref{res:direct}. The breakdown of this sample in 
centre-of-mass energy windows, as shown in Table~\ref{comp5}, does not indicate a high mass signal 
appearing at the highest centre-of-mass energy. The data taken in 1998, 1999, and 2000b, agree with 
the Standard Model background expectation.

Since the signal samples were generated at only one centre-of-mass
energy (namely $\sqrt{s}=$200\,GeV), a procedure is designed to
estimate the efficiencies at the other energies. To do so, the
four-momenta of the primary bosons are rescaled to correspond to
the desired centre-of-mass energy, and all particles coming from
the primary pair are boosted accordingly. Rescaled events are analysed
using the analysis chain described above. The validity of this
procedure was verified using a few dedicated signal samples simulated at the extreme
centre-of-mass energies corresponding to the analysed data set, i.e. 189 and 208 GeV. 
The method proves to have a precision of $\pm$2\%.

The signal efficiencies for the simulated mass points are given in
Tables~\ref{aaa-eff},\,\ref{aaz-eff}, and \,\ref{hhz-eff}. The
efficiency for any arbitrary mass point is obtained by
linear interpolation between the three closest simulated points.
The analysis described above is also directly applied to the hA \ra 4b
channel, with resulting efficiencies given in Table~\ref{ha2-eff}.

In addition to the uncertainties already quoted, a systematic error is included
accounting for residual imperfections in the b-tagging description in the simulation.
An uncertainty of $\pm$5\% is assumed \cite{btag-summ}. 

Uncertainties on the gluon splitting probability have much smaller impact
(as in Section~\ref{4bLEP2}, we use $\mathrm g_{bb} = 1.5 \times 10^{-3}$). Compared to 
the LEP1 four-b analysis, the present selection needs to preserve high signal efficiency.
The background rejection is thus much weaker, and the fraction of events predicted to contain gluon splitting
into \bb\ after the last cut is only 2\%. Assuming 50\% uncertainty on this fraction contributes
an uncertainty of 1\% on the background estimate.

\subsection{The four-$\tau$ final state}
\label{4tLEP2}

This final state consists of four narrow jets of low multiplicity coming from the $\tau$ decays. 
When the h or A boson mass decreases, the decay products are often observed as
a single jet, due to the low angle between the decay $\tau$~leptons. Three independent analysis streams 
are developed to provide sensitivity  to the whole (\mh,\mA) mass plane: a four-jet, a three-jet 
and a two-jet stream, respectively adapted to the case where both bosons are heavy, one boson is light, 
or both h and A are light.

Some criteria are common to all analyses. A charged-particle
multiplicity between 4 and 8 is required, to reject lepton pairs and
hadronic events. Algorithms used in the lepton identification are the
same as those used in the selection of fully-leptonic W
pairs~\cite{wwpaper}. The four-lepton background is rejected by
requiring that the momentum of the most energetic identified muon or electron, if present,
is less than $0.25 \sqrt{s}$. If a second muon or electron is identified, it should have momentum less 
than $0.15 \sqrt{s}$. In the following, jets are defined as clusters of particles (of which at least one is charged) 
contained in a cone with a $15^{\circ}$ opening angle. The analysis streams are now described in turn.

\subsubsection{The four-jet stream}
\label{fourjetstream}

The four-jet analysis is derived from that of the four-$\tau$ final state
applied in the search for doubly charged Higgs bosons (Section~3.1
of Ref.~\cite{hpphmm}), but discarding all mass cuts. Events are 
clustered into jets, and each jet is required to be separated from the others by at least $15^{\circ}$. 
Only events with four reconstructed jets are accepted and every jet is considered as a $\tau$ candidate.

To improve the reconstruction of the $\tau$ energy, the $\tau$ momenta are rescaled, imposing energy
and momentum conservation while preserving the measured directions. If any rescaled 
jet momentum is negative, the event is rejected.

The two-photon background is reduced by the following requirements: the momenta of the jets have to be larger than 
$0.01 \sqrt{s}$, the visible energy outside a cone of 
$25^\circ$ around the beam-axis is required to be greater than $0.15 \sqrt{s}$, 
and the total energy of neutral particles should be less than $0.35 \sqrt{s}$. 

After all cuts only one event is observed in the data, while 
1.9 events are expected from background processes. Efficiencies around
$40-50\%$ are obtained for h and A masses higher than $\sim$50~\GeVcc.

The rescaled $\tau$ momenta are used to
reconstruct the Higgs boson masses after the jets are paired according to their charges and the dijet masses.
The charge of a jet is defined as the sum of the charges of the jet particles if this sum is found to be $\pm$1, 
and as the charge of the most energetic charged particle of the jet otherwise. The pairing is chosen so as to 
minimise the 
difference between the two reconstructed dijet masses. After pairing, the sum of the dijet masses is used as a
discriminating variable in the confidence level computations (Section~\ref{methods}).

\subsubsection{The three-jet stream}

Events enter this stream if three jets are found after clustering is performed as in the 
four-jet stream. Each jet is considered as a $\tau$~candidate, and should again be separated from the 
others by at least $15^{\circ}$. To reject the two-photon background, the same criteria as described 
in Section~\ref{fourjetstream} are used.

Additional cuts are applied to reduce the remaining $Z \gamma^\star$ background. The absolute value of the 
cosine of the missing momentum 
polar angle should be less than 0.9. All jets should have polar angle between 
$20^{\circ}$ and $160^{\circ}$. For signal events, the three reconstructed jets are expected to be in the 
same plane. Therefore, the sum of the three angles between the jets, $\alpha_{123}$, 
is required to be greater than $357^{\circ}$. Finally, the lowest
jet-jet angle, $\alpha_1$, is required to be greater than $25^{\circ}$.

Six events are selected in the data, while 6.5 events are expected from the
background. The efficiency for $\mathrm{m_A}$=4~\GeVcc\ and $\mathrm{m_h}$ greater than 60
\GeVcc\ is about $40\%$. 

The final discriminating variable for the confidence level computations is the 
highest reconstructed Higgs boson mass, since the other one is expected to be 
low. This mass is calculated by rescaling the momentum of the jets, imposing
energy and momentum conservation while keeping the jet directions fixed. The pairing is then chosen as
follows. If only one jet has an electric charge equal to 0, the 
mass is given by the opposite jet pair. In other cases, the mass
is given by the two jets, if they exist, containing only one charged particle; or by the two jets 
with opposite charges, if the third one has an electric charge greater than 1 in absolute value. 
If none of these configurations is present, the mass is given by the two jets of opposite charges 
and with nearest rescaled $\tau$~momenta.

\subsubsection{The two-jet stream}

If an event is not classified in the two previous streams, it is a candidate for
the two-jet analysis. Only events with either four or six charged 
particles, and with total electric charge zero, are accepted in this 
stream. 

Every neutral particle energy is added to the 
momentum of the nearest charged particle, if it is distant by less than $15^{\circ}$. Neutral 
particles making angles larger than $15^{\circ}$ with all charged particles are not recombined.

A charged multiplicity of six signals that one of the $\tau$~leptons has decayed into three prongs. 
To ensure this is the case, the lowest triplet invariant mass should not exceed 1.4~\GeVcc\
and its momentum should be greater than 5~\GeVc.

At this stage, events are grouped into four $\tau$ candidates, coming from either the
four charged particles, or the three charged particles plus the opposite triplet of lowest mass.
The two-photon background is reduced by requiring all $\tau$~candidate momenta to
be larger than $0.005 \sqrt{s}$, and the visible energy outside  
a cone of  $25^\circ$ around the beam-axis is required to be between 
$0.15 \sqrt{s}$ and $0.8 \sqrt{s}$. In addition, events with the third lowest
angle between $\tau$~candidates, $\alpha_3$, less than $70^{\circ}$ are rejected. 
Finally, the polar angles of all $\tau$~candidates must lie between $25^\circ$ and $155^\circ$, while
at least one must have a polar angle between $50^\circ$ and $130^\circ$.

Six events are selected in the data, in agreement with the 9.5 events expected from the background 
processes. The efficiency for $\mathrm{m_A}$=4~\GeVcc\ and $\mathrm{m_h}$=4~\GeVcc\ is 37\%. The 
mass estimation often fails in this topology, and it is not possible to 
reconstruct either the h mass or the A mass. The second lowest angle between $\tau$~candidates is
chosen as final discriminating variable in the confidence level computations.

\begin{figure} [p]
\begin{center}
\epsfig{figure=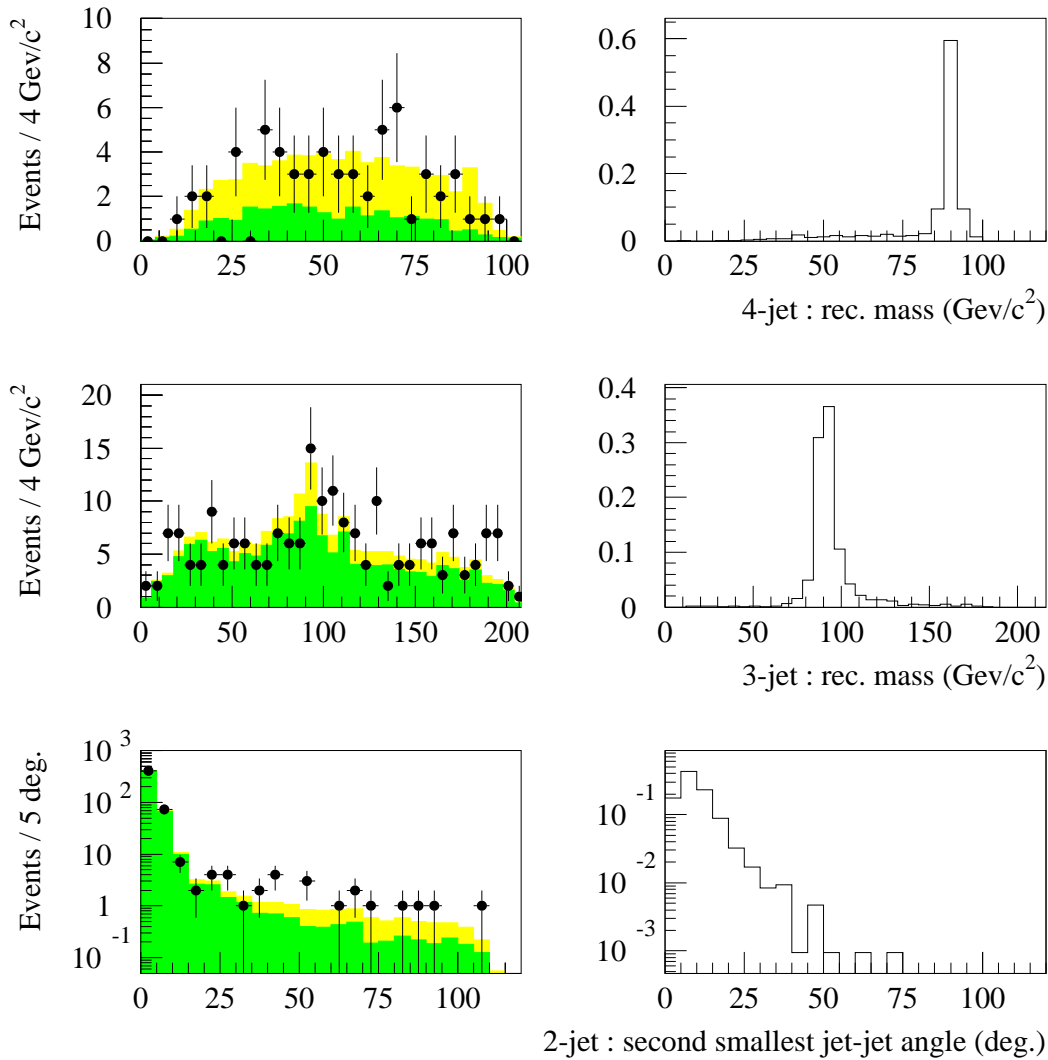,width=0.9\textwidth}
\end{center}
\caption{Comparison between data and simulation for the mass distributions
         used in the statistical interpretation of the four-$\tau$ analyses. On the left, the points are the data,
	 the light histograms represent 
	 the four-lepton contributions, and the dark histograms represent the remaining
         two- and four-fermion processes. The four-jet and three-jet discriminants are shown at their
         respective preselection level; the two-jet discriminant is shown after the 
         $\gamma\gamma$ rejection. The histograms on the right show distributions 
         for three example signals: (\mh,\mA) $=$ (90,90), (90,4), and (4,4)~\GeVcc\ for the
         four-jet, three-jet and two-jet analysis respectively. Normalization is arbitrary.}
\label{hA4t}
\end{figure}

\paragraph{}
Good agreement between the data and the expected background 
is observed for each analysis, as illustrated in Figure \ref{hA4t}. Combining all streams, 
13 events are selected in the data, whereas 18.0$\pm$1.2 events are
expected from the Standard Model background processes. Details are shown in
Tables \ref{tab:cutevol4jet}, \ref{tab:cutevol3jet} and
\ref{tab:cutevol2jet}. The efficiencies of the four-$\tau$ analysis streams are shown 
in Table~\ref{eff4tau} for representative simulated mass points. 

All results contain statistical and systematic
uncertainties added in quadrature. Systematic uncertainties are
estimated by varying the simulated charged particle momenta, jet-jet angles and particle
identification variables in a range given by the residual differences
between their distributions in data and simulation. Because of the large amount of missing energy 
in this final state, the efficiencies are expected to vary slowly with $\sqrt{s}$. Using a few 
dedicated signal samples simulated at different centre-of-mass energies corresponding 
to the analysed data set, this is verified to be true up to $\pm$1.5\%. Taking this into account, the 
total systematic uncertainty amounts to about $\pm$3\% for signal efficiencies, and to $\pm$10-13\% 
for the background; these last numbers are dominated by the finite Monte Carlo statistics.

\begin{table}[htbp]

\caption{Four-$\tau$ final state. Number of observed and expected background events, at various stages of the four-jet
analysis stream, for the total 189-208 GeV sample.}
\begin{center}
\vspace{1ex}
\begin{tabular}{l|rrrr}
Cut    & \multicolumn{1}{c}{four-lepton}
    & \multicolumn{1}{c}{others}
    & \multicolumn{1}{c}{Total}
    & \multicolumn{1}{c}{Data} \\
\hline
four-jet preselection     &  44.0 & 23.4 &  67.4       &   59   \\
anti $\gamma\gamma$       &  28.9 &  2.1 &  31.0       &   26   \\
anti four-lepton             &   1.7 &  0.2 & 1.9$\pm$0.2 &    1   \\
\end{tabular}
\end{center}
\label{tab:cutevol4jet}


\caption{Four-$\tau$ final state. Number of observed and expected background events, at various stages of the three-jet
analysis stream, for the total 189-208 GeV sample.}
\begin{center}
\vspace{1ex}
\begin{tabular}{l|rrrr}
Cut    & \multicolumn{1}{c}{four-lepton}
    & \multicolumn{1}{c}{others}
    & \multicolumn{1}{c}{Total}
    & \multicolumn{1}{c}{Data} \\
\hline
three-jet preselection      &  39.2 & 153.4  & 192.6  &  199    \\
anti four-lepton               &   9.6 &  90.8  & 100.4  &   98    \\
anti $\gamma\gamma$         &   5.9 &  12.5  &  18.4  &   22    \\
$\alpha_{123}$, $\alpha_1$ cuts
                            &   2.7 &   3.9 & 6.6$\pm$0.7 &    6 \\
\end{tabular}
\end{center}
\label{tab:cutevol3jet}


\caption{Four-$\tau$ final state. Number of observed and expected background events, at various stages of the two-jet
analysis stream, for the total 189-208 GeV sample.} 
\begin{center}
\vspace{1ex}
\begin{tabular}{l|rrrr}
Cut     & \multicolumn{1}{c}{four-lepton}
    & \multicolumn{1}{c}{others} 
    & \multicolumn{1}{c}{Total}
    & \multicolumn{1}{c}{Data} \\
\hline
$\tau$ selection          &  31.3 & 1299.9  &  1331.2     & 1358  \\
anti $\gamma\gamma$       &  14.0 &  502.4  &   516.4     &  517  \\
$\alpha_3$                &   3.7 &   11.8  &    15.5     &   13  \\
Jet angular cuts          &   1.6 &    7.9  & 9.5$\pm$1.0 &    6  \\
\end{tabular}
\end{center}
\label{tab:cutevol2jet}

\end{table}


 
\section{Results}
\label{results}

The results from the analyses described above are summarised in this section.
The Yukawa process, hA and hZ production 
followed by direct Higgs boson decays into fermions, and cascade decays are discussed in turn.
The excess found in the LEP2 b-tagging analysis is discussed. Since no obvious signal
is found, the observations are interpreted in terms of excluded cross-sections, using the 
conventions described in Section~\ref{higgsprod}.
For all final states, the tables given in Appendix B provide
explicit numerical upper bounds on the corresponding C or C$^2$ factors.
All the limits presented in the following are at the 95\% confidence level (CL).

\subsection{Search for the Yukawa process at LEP1}
\label{res:yukawa}

Results of the Yukawa production analyses of Section~\ref{allLEP1} are
presented in the form of mass-dependent upper bounds on the C$^{2}$
factors defined in the introduction. Reference cross-sections for
Yukawa production of h and A are obtained using \cite{JKMK}. In all
cases, the mass range between production threshold and 50~\GeVcc\ is
considered, and the C$^{2}$ values excluded at exactly 95\% CL are
determined. Since these values are very large, the numbers given in
Table \ref{cpl_Yuk} and the corresponding
figures refer to C rather than to C$^{2}$. The former corresponds to
the matrix-element enhancement factor, when 100\% branching fraction 
into the relevant final state is assumed.

\begin{figure} [p]
\begin{center}
\epsfig{figure=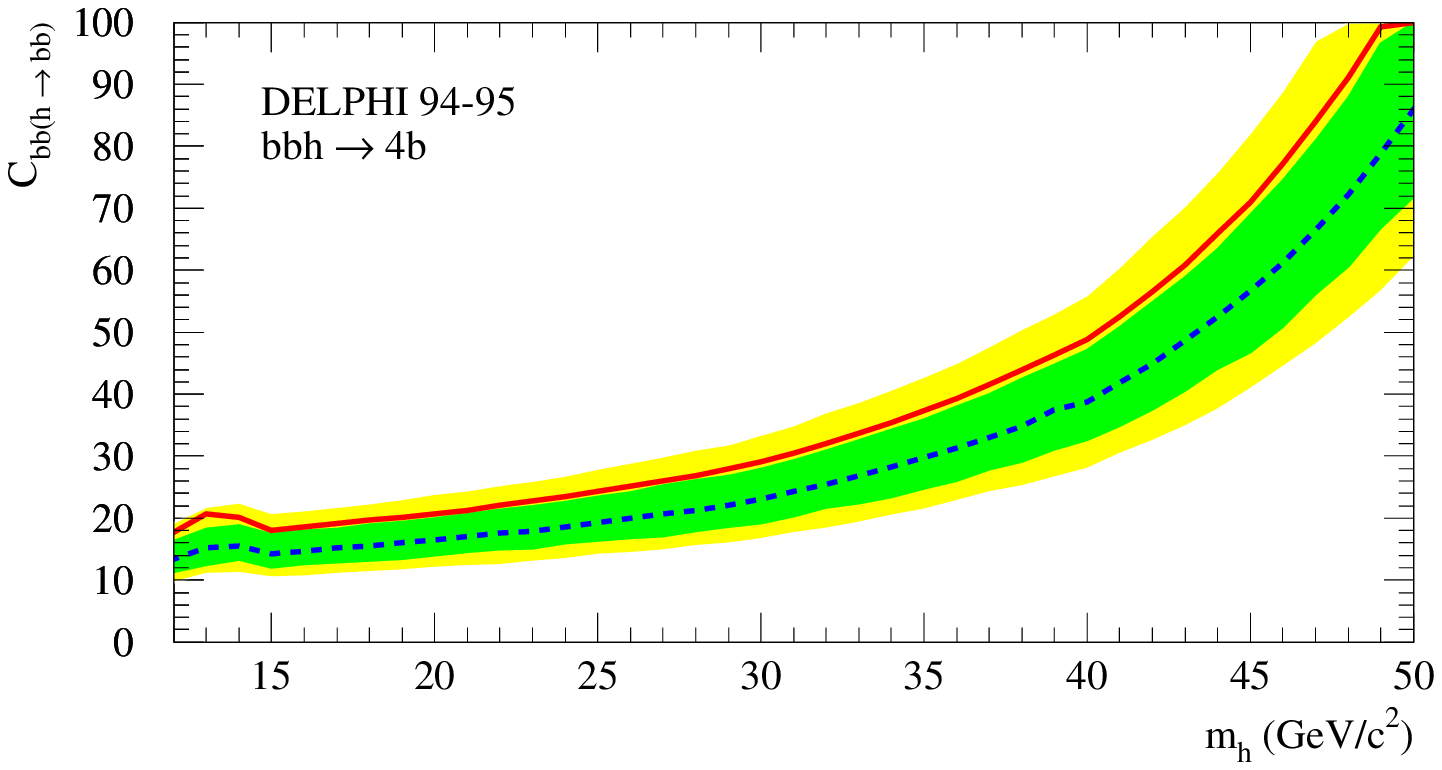, width=.9\textwidth}
\epsfig{figure=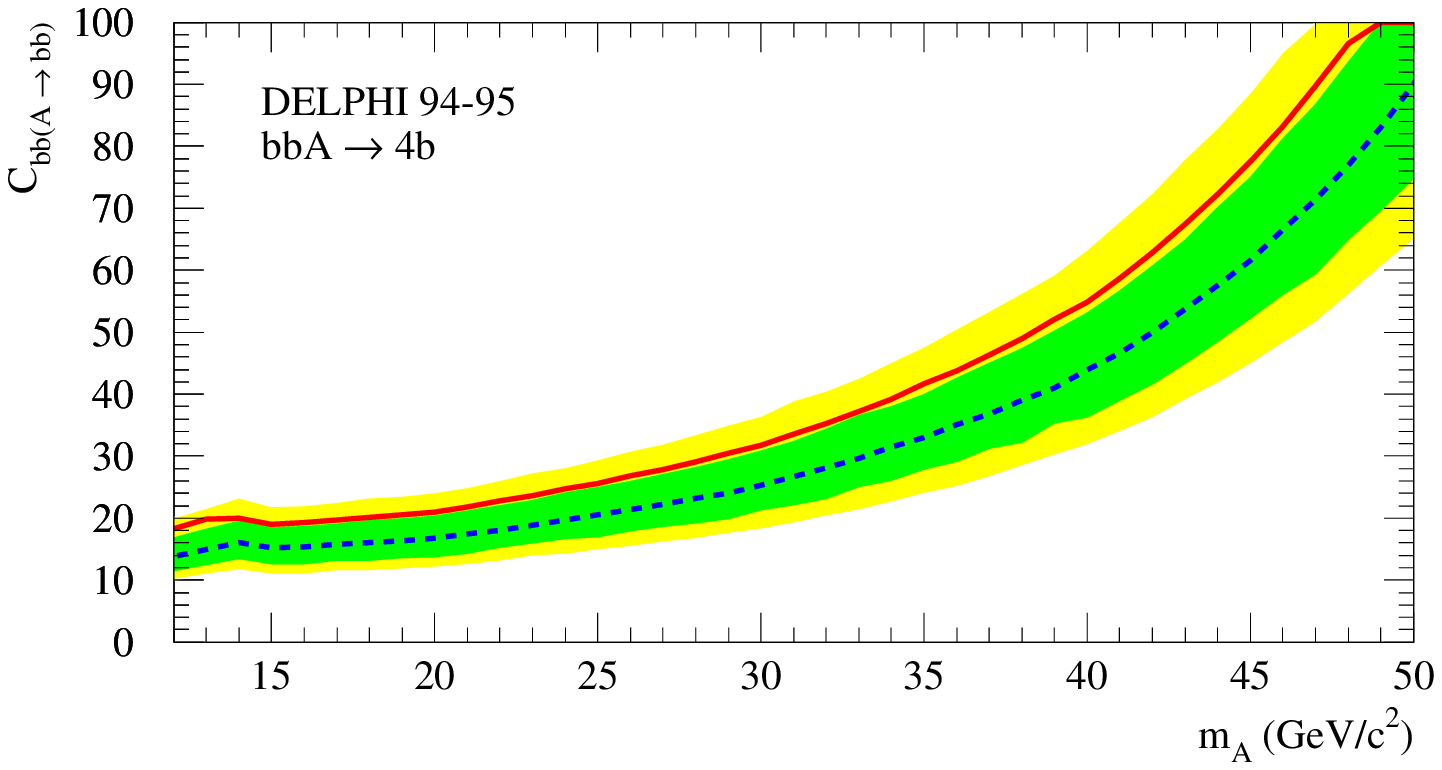, width=.9\textwidth}
\end{center}
\caption{Upper limits on $\mathrm C_{bb(h\mra bb)}$ (top) and $\mathrm C_{bb(A\mra bb)}$ (bottom),
         defined in Section~\ref{higgsprod}.
         The dashed line shows the average expectation for background experiments, 
         and the full line shows the observation. The bands correspond to the 
         68.3\% and 95.0\% confidence intervals for background-only experiments.
         The excess observed in the data translates into an exclusion slightly
         weaker than expected. The discrepancy is about 1.2 standard deviations in the mass range
         $\mathrm m_{h,A} > 15$ GeV, where the four-jet analysis is used. For lower masses the three-jet analysis
         is used, with a discrepancy just below 2 standard deviations.}
\label{yuk4b}
\end{figure}

\begin{figure} [p]
\begin{center}
\epsfig{figure=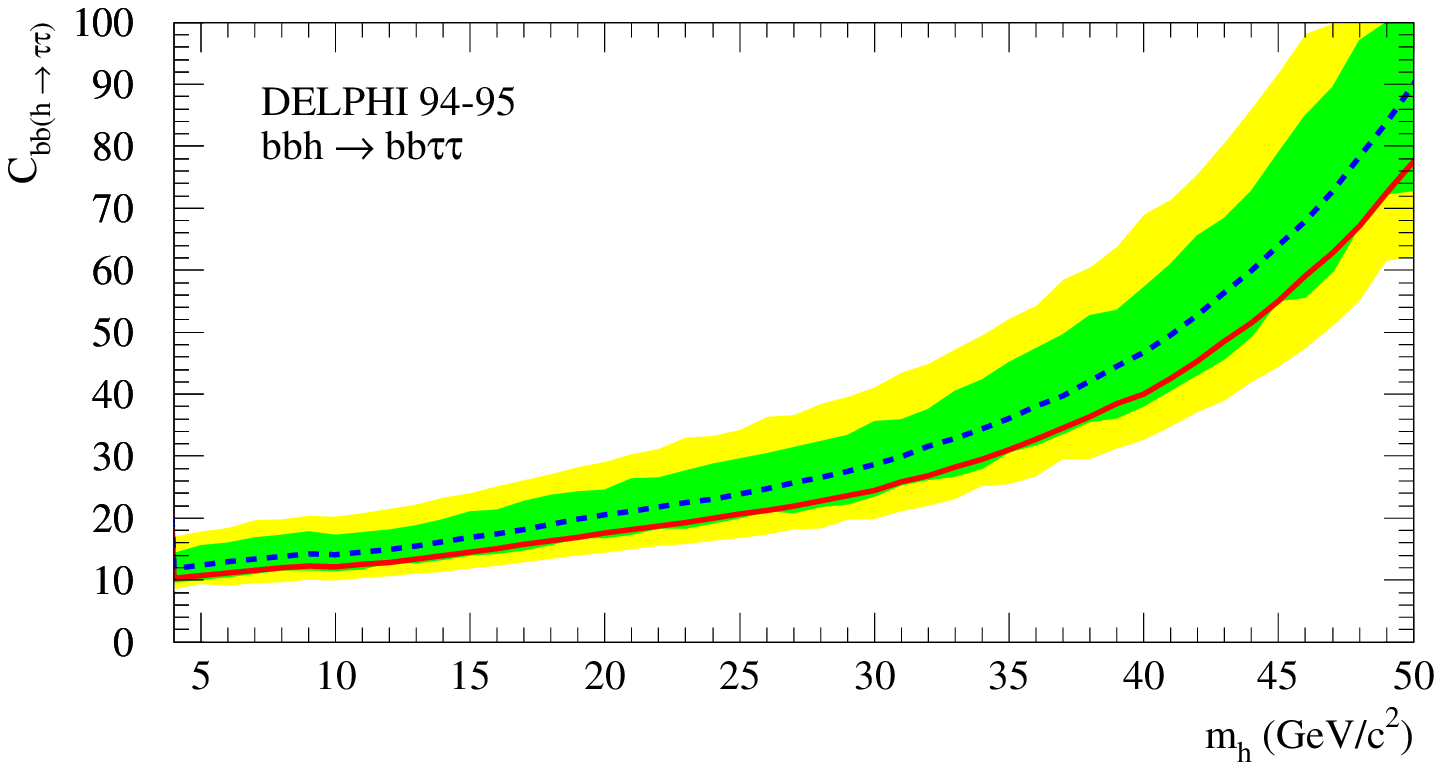, width=.9\textwidth}
\epsfig{figure=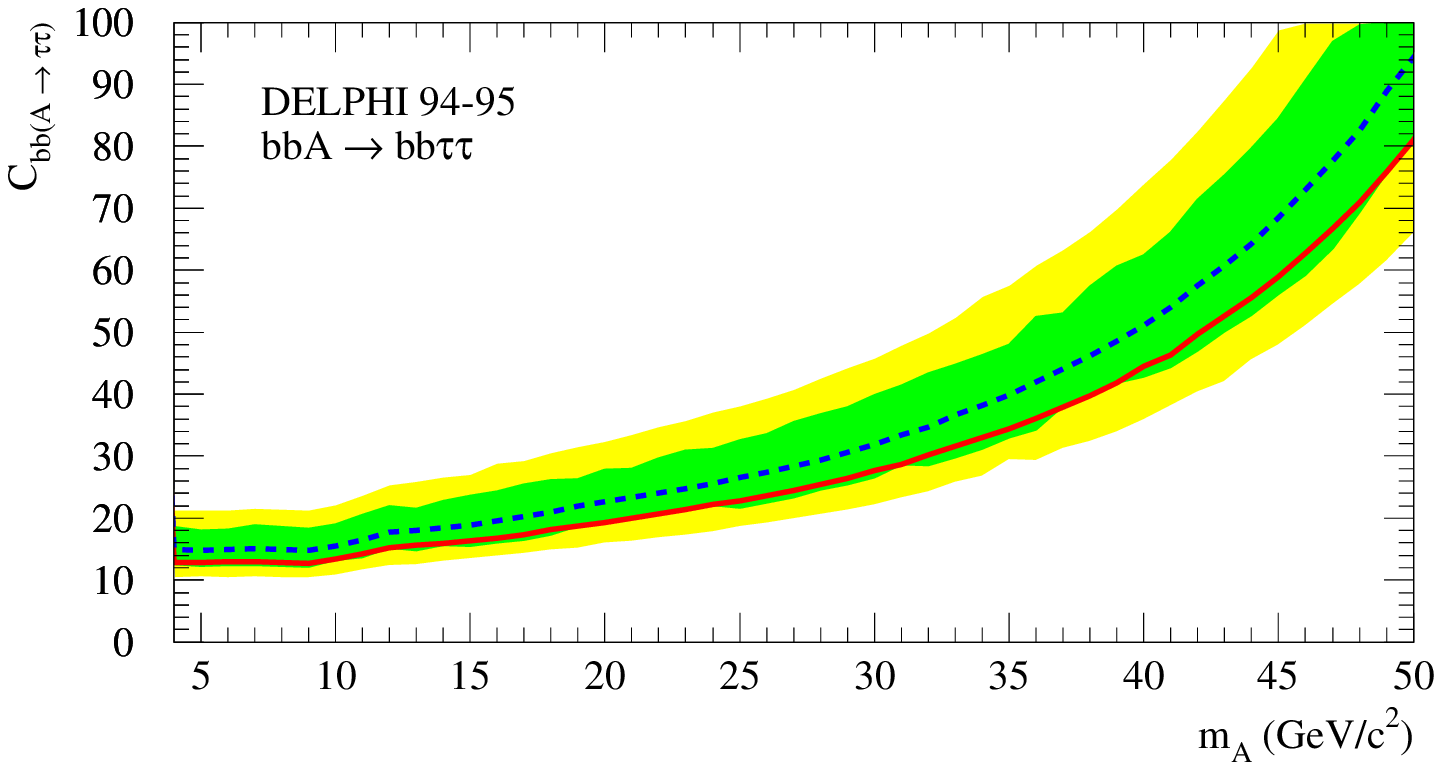, width=.9\textwidth}
\end{center}
\caption{Upper limits on $\mathrm C_{bb(h\mra \tau\tau)}$ (top) and $\mathrm C_{bb(A\mra \tau\tau)}$ 
         (bottom), defined in Section~\ref{higgsprod}. 
         The dashed line shows the average expectation for background experiments, 
         and the full line shows the observation.The bands correspond to the 
         68.3\% and 95.0\% confidence intervals for background-only experiments.}
\label{yukbt}
\end{figure}

\begin{figure} [p]
\begin{center}
\epsfig{figure=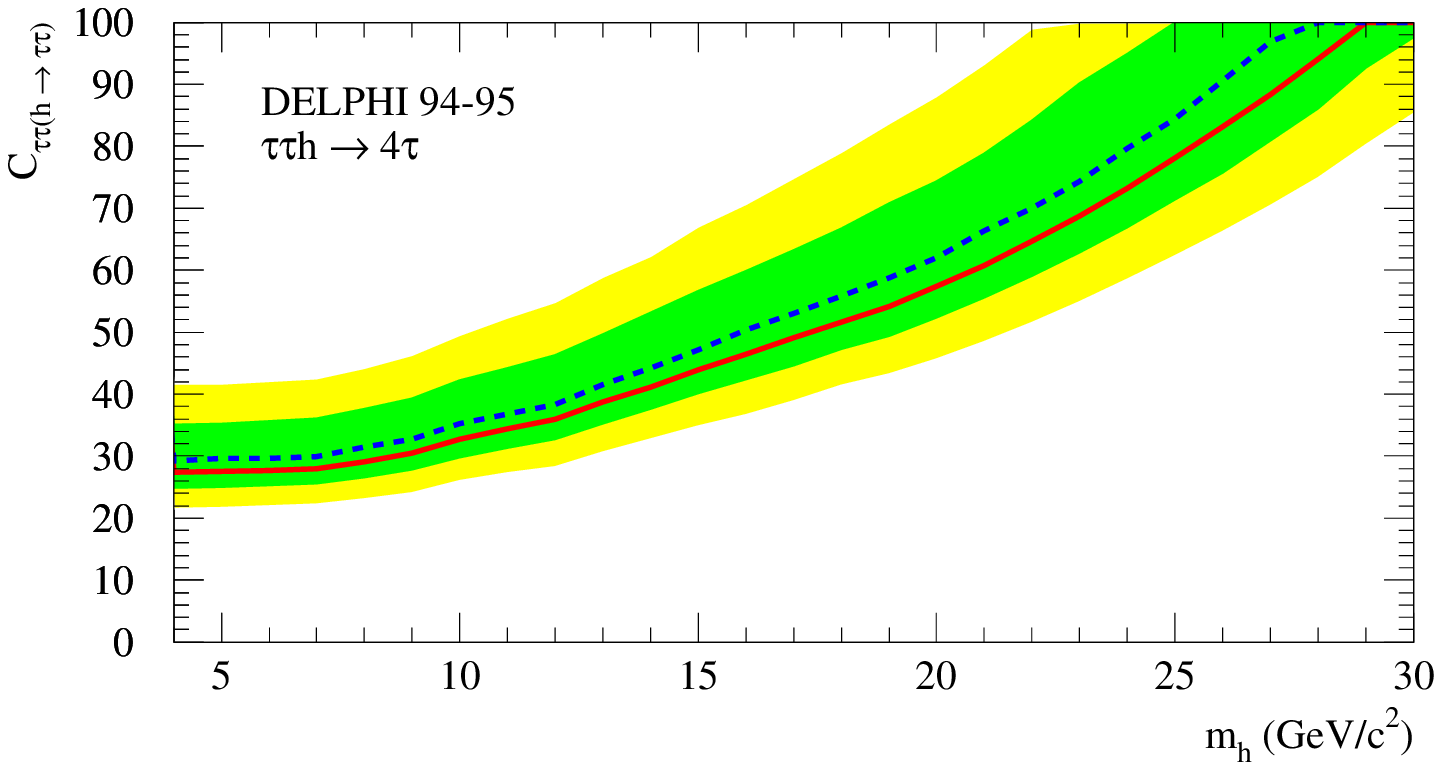, width=.9\textwidth}
\epsfig{figure=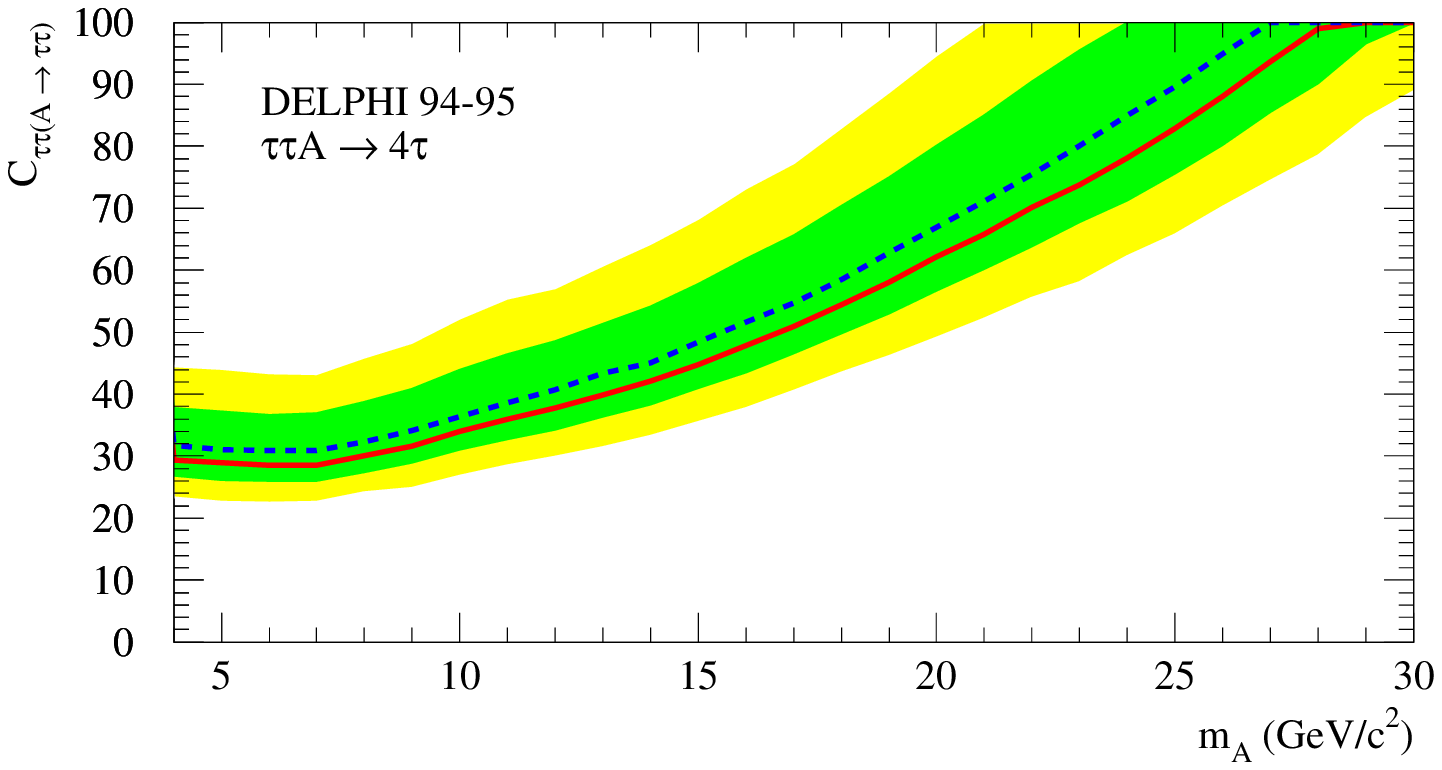, width=.9\textwidth}
\end{center}
\caption{Upper limits on $\mathrm C_{\tau\tau(h\mra \tau\tau)}$ (top) and 
         $\mathrm C_{\tau\tau(A\mra \tau\tau)}$ (bottom), defined in Section~\ref{higgsprod}. 
         The dashed line shows the average expectation for background experiments, 
         and the full line shows the observation.The bands correspond to the 
         68.3\% and 95.0\% confidence intervals for background-only experiments.}
\label{yuk4t}
\end{figure}

The four-b Yukawa results on $\mathrm C_{bb(h\mra bb)}$ and $\mathrm C_{bb(A\mra bb)}$, shown
in Figure~\ref{yuk4b}, are obtained by combining Bin 1 and Bin 2 as
independent channels, either in the three-jet analysis or in the four-jet
analysis, keeping the analysis with the best expected exclusion sensitivity
at each mass point. The \bb\tautau\ channel leads to the upper bounds on 
$\mathrm C_{bb(h\mra \tau\tau)}$ and $\mathrm C_{bb(A\mra \tau\tau)}$ displayed in
Figure~\ref{yukbt}. Results on the four-$\tau$ channel are shown in
Figure~\ref{yuk4t}. Upper bounds are placed on $\mathrm C_{\tau\tau(h\mra \tau\tau)}$ and 
$\mathrm C_{\tau\tau(A\mra \tau\tau)}$ by combining the independent four-prong and six-prong analyses.

The slight deficit in the \bb\tautau\ channel translates into an exclusion
slightly stronger than expected. On the contrary, the excess in the four-b
channel induces an exclusion which is slightly weaker (at $1\sigma$) than expected
from the simulation. The four-$\tau$ channel result is in agreement with
the background hypothesis.

In the four-b analysis, the inclusion of Bin 1 improves the sensitivity on
$\mathrm C_{bb(h\mra bb)}$ by 10\%, compared to using Bin 2 alone. In the four-jet
analysis, Bin 2 excludes signals larger than 7 events, which could be
compared to our previous result \cite{tmp-7-120}, where the limit was
set at 50.4 events. The improvement in sensitivity on $\mathrm C_{bb(h\mra
bb)}$ and $\mathrm C_{bb(A\mra bb)}$ is nearly threefold over the whole mass
range. The three-jet analysis has better expected performance than the
four-jet analysis in the very low mass region (below \mh,\mA $\sim$ 15~\GeVcc).

As the figures indicate, the four-b and the \bb\tautau\ channels
have similar intrinsic sensitivity (the expected exclusions are
similar). This is not the case for the four-$\tau$ channel. Although the
signal to background ratio in this channel is better than that in the
four-b and \bb\tautau\ channels (as can be seen from Tables~\ref{tab4pr},
\ref{tab6pr}, \ref{tttt_Y4p} and \ref{tttt_Y6p}), the much weaker coupling of the Z 
boson to the primary $\tau$~leptons induces weaker sensitivity on
$\mathrm C_{\tau\tau(h\mra\tau\tau)}$ and $\mathrm C_{\tau\tau(A\mra\tau\tau)}$. 

Numerical values for the observed exclusions are given in Table~\ref{cpl_Yuk}.

\subsection{hA and hZ production: direct decays}
\label{res:direct}

Higgs boson production in the hA \ra 4b and hA \ra 4$\tau$ channels
is assessed using the results of the analyses described in Sections~\ref{4bLEP1}, \ref{4bLEP2} 
and \ref{4tLEP2}, as well as those of the searches for the hA \ra 4b process in the
MSSM framework at all LEP2 energies, as described in Ref.~\cite{DELhiggs}. 
Exclusion limits are also given for the hZ process 
when the Higgs boson decays into b~quark pairs or $\tau$~lepton pairs, 
using the results of the searches for the hZ process applied to all LEP2 
data samples, as described in~\cite{DELhiggs}.

The C$^2$ factor for each process is defined in the
introduction. The Higgs boson mass domain is then scanned, and at each
point the C$^2$ value excluded at exactly 95\% CL is determined.

Event rates for the hA process are computed with the {\tt HZHA}
generator~\cite{HZHA}, and using interpolation of the signal efficiencies
(Tables~\ref{ha2-eff} and \ref{eff4tau}, Appendix \ref{efficiencies}). Rates for the hZ production 
process are determined as described in~\cite{DELhiggs}. The combination of 
data at different centre-of-mass  energies is done assuming the expected
evolution of the hA and hZ production cross-sections with energy.

\subsubsection{The four-b search}

Figure~\ref{contours} shows the results of the search for hA \ra 4b.
The LEP1 data analysis presented in Section~\ref{4bLEP1} is combined with the
LEP2 analyses of Section~\ref{4bLEP2} and of Ref.~\cite{DELhiggs}. 
As these last two analyses are not independent, only the analysis
with the best expected exclusion power is kept at each mass point
and at each centre-of-mass energy. While the analysis
presented in this paper has good performance over the whole mass plane, the
MSSM analysis~\cite{DELhiggs} has optimal sensitivity when \mh $\sim$ \mA\ and
provides better results in this region. 

A strong sensitivity is obtained both at high mass from LEP2 data, and
in the lower mass region where the LEP1 data contribute
significantly. In the case of no suppression (i.e. full strength
production, and 100$\%$ branching into  four b~quarks, i.e. $\mathrm C^{2}_{hA \mra 4b}$=1), the
search excludes a region roughly given by \mh,\mA $>$ 12~\GeVcc,
\mh,\mA $<$ 130~\GeVcc\ when the opposite mass is small,
and \mh $+$\mA $<$ 180~\GeVcc\ when the h and A masses are
similar. When the suppression factor is less than 5\%, the
excluded region is obtained essentially from LEP1 data. 

The consistency of the numerical excess found in the data of 2000a, with the 
data recorded in 1998, 1999, and 2000b, is estimated in the following way.
The excess is attributed to a signal, and used to normalize its cross-section. It is 
then possible to confront the signal hypothesis with the data surviving the selections in the 
complementary data sets.
Given 6.3 signal events in 2000a, the number of signal events expected in the other data sets 
depends on the nature of the signal and its mass. The primary signal process is taken to 
be \epem \ra hA, since its cross-section rises more slowly with energy than the hZ 
cross-section. The conclusions made for hA are then a fortiori valid for hZ.
Three mass hypotheses are considered, namely (\mh,\mA)=(90,90), (95,95), and (100,100)~\GeVcc.
The corresponding rates are summarized in Table~\ref{excess}, correctly taking into account 
the kinematic thresholds. In each case the confidence levels in the background and signal hypotheses 
are given.

In all cases, a signal corresponding to the observed excess in 2000a would produce a visible 
signal in the other data sets. Since the observations are background-like, and have confidence 
levels in the signals of 19\% at best, we conclude that the excess of 2000a is not confirmed by 
the remaining data.

\begin{figure} [p]
\begin{center}
\epsfig{figure=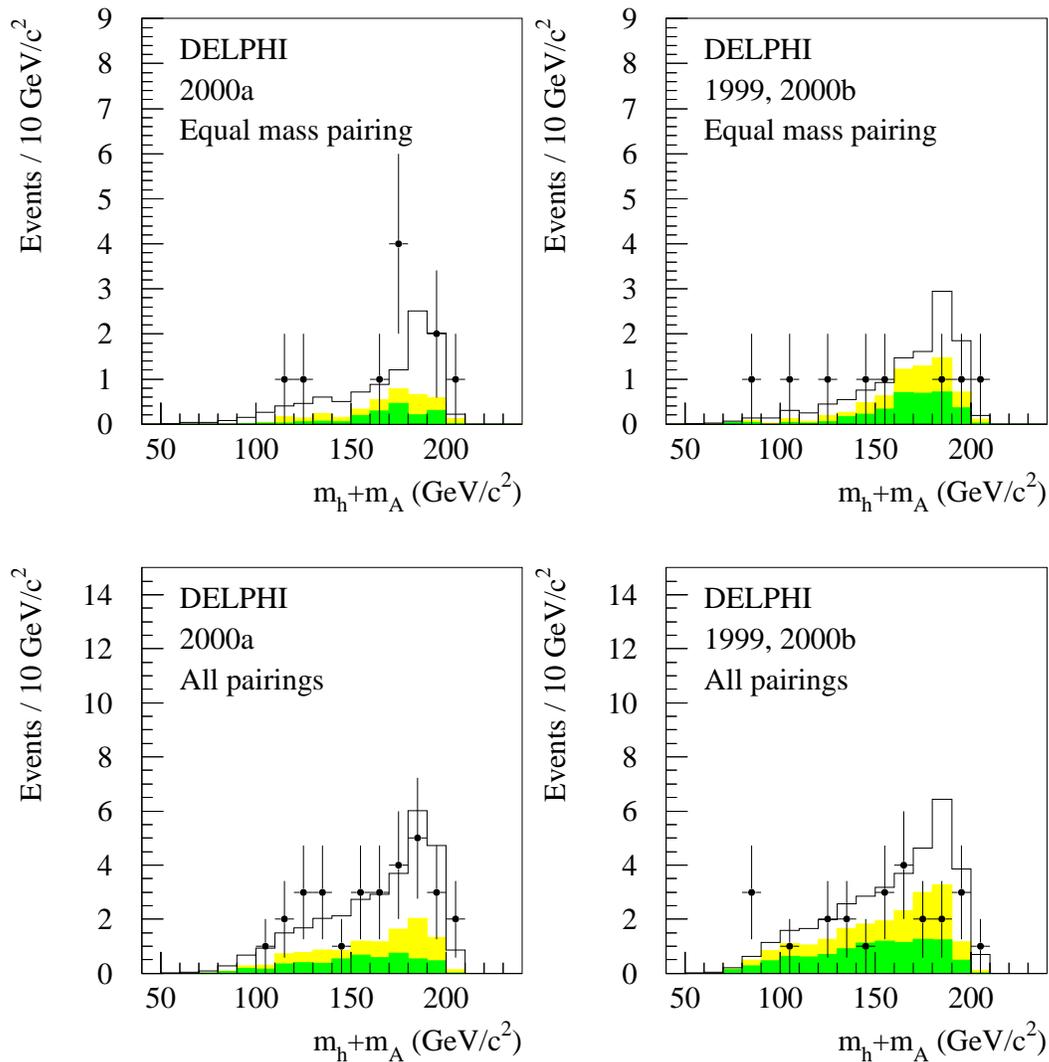, width=.9\textwidth}
\end{center}
\caption{Distributions of \mh $+$\mA\ for the data taken in 2000a (left)
and for the complementary data set (right). For both datasets, the mass distributions are given with the jet 
pairing chosen to minimize the dijet mass difference (above), and including all pairings (below).
The points are the data; the light and dark histograms represent the Standard Model four-fermion and 
\qq\ backgrounds, respectively. An hA signal (\mh $=$\mA $=$ 95 \GeVcc) is superimposed; it is 
normalized to the excess observed in 2000a, and to 
the corresponding expectation for the complementary dataset. The 1998 data are below the signal threshold and 
discarded.}
\label{excess1}
\end{figure}

As a further illustration, the expected and observed mass distributions are shown in 
Figure~\ref{excess1}, for the 2000a data set, and the 
complementary 1999 and 2000b sets. Shown is the 
distribution obtained when choosing the jet pairing so that the dijet mass difference is 
minimized; an example signal with (\mh,\mA)=(95,95)~\GeVcc\ is superimposed, normalized as above 
(a lower mass signal is strongly disfavoured according to the results of Table~\ref{excess}). 
The mass distribution when all pairings enter (i.e., each event contributes three times) is also shown.

\begin{table}
\caption{Numerical study of the excess observed in period 2000a. In this data set,
10 events are observed while 3.7$\pm$0.6 are expected (Table~\ref{comp4}). The entire excess 
is attributed to a signal, and predictions are made for the complementary data sets, for three 
mass hypotheses of an example hA signal. For every hypothesis, the observation and expected 
background correspond to the complementary data taken above threshold (see Table~\ref{comp4}), and the corresponding 
confidence levels in the background and signal hypotheses are given.}
\begin{center}
\begin{tabular}{rr|rrrrr}
       Mass (\GeVcc)      & Rate (2000a) & Rate (compl.) & Bkg. & Data & $\mathrm{CL_b}$ & $\mathrm{CL_s}$ \\ \hline
  (\mh,\mA)=(90,90) &           6.3 &  $\sim 10$    & 10.0 &   10 &            46\% &            10\% \\
                  (95,95) &           6.3 &  $\sim  5$    &  6.8 &    8 &            63\% &            19\% \\
                (100,100) &           6.3 &  $\sim  3$    &  2.0 &    2 &            41\% &             8\% \\
\end{tabular}
\label{excess}
\end{center}
\end{table}

\begin{figure}[htbp]
\begin{center}
\epsfig{figure=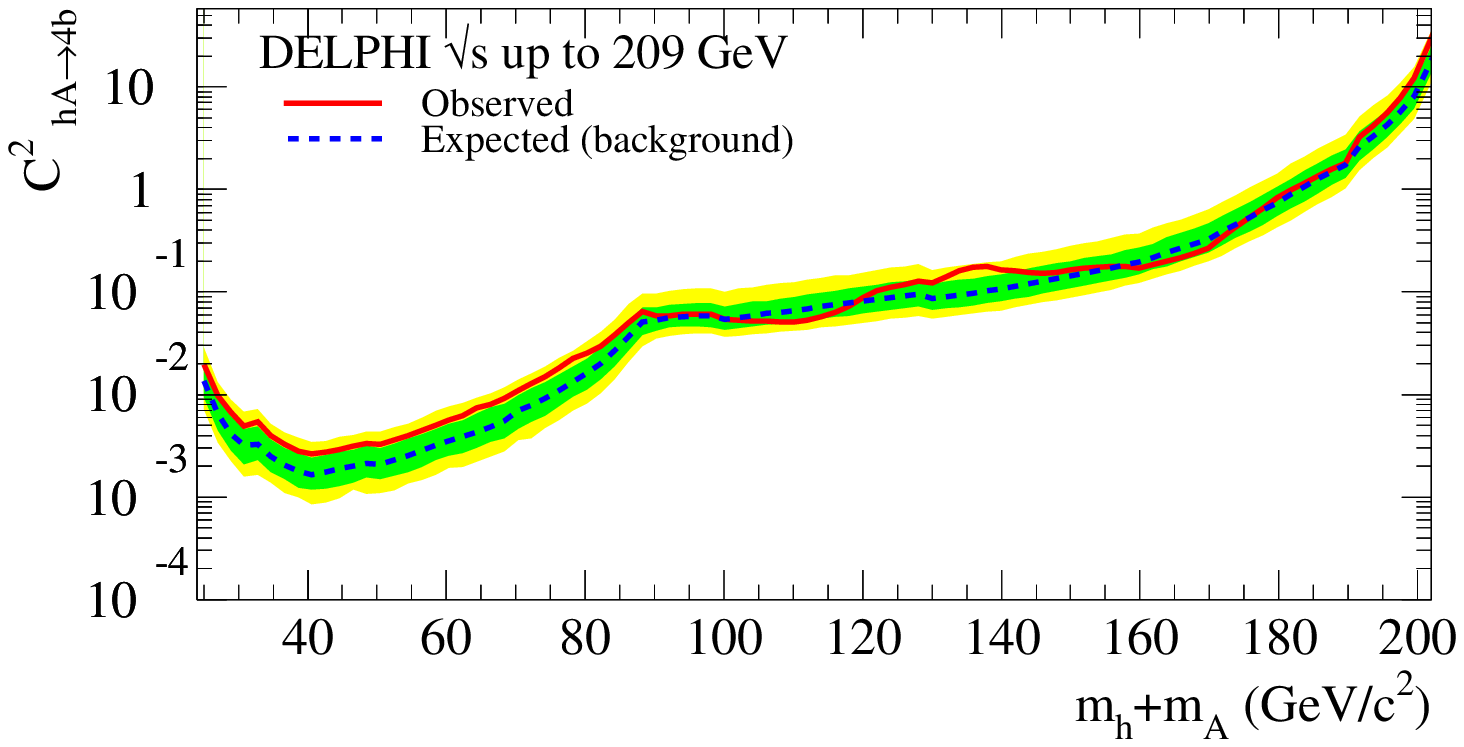,width=.9\textwidth}
\epsfig{figure=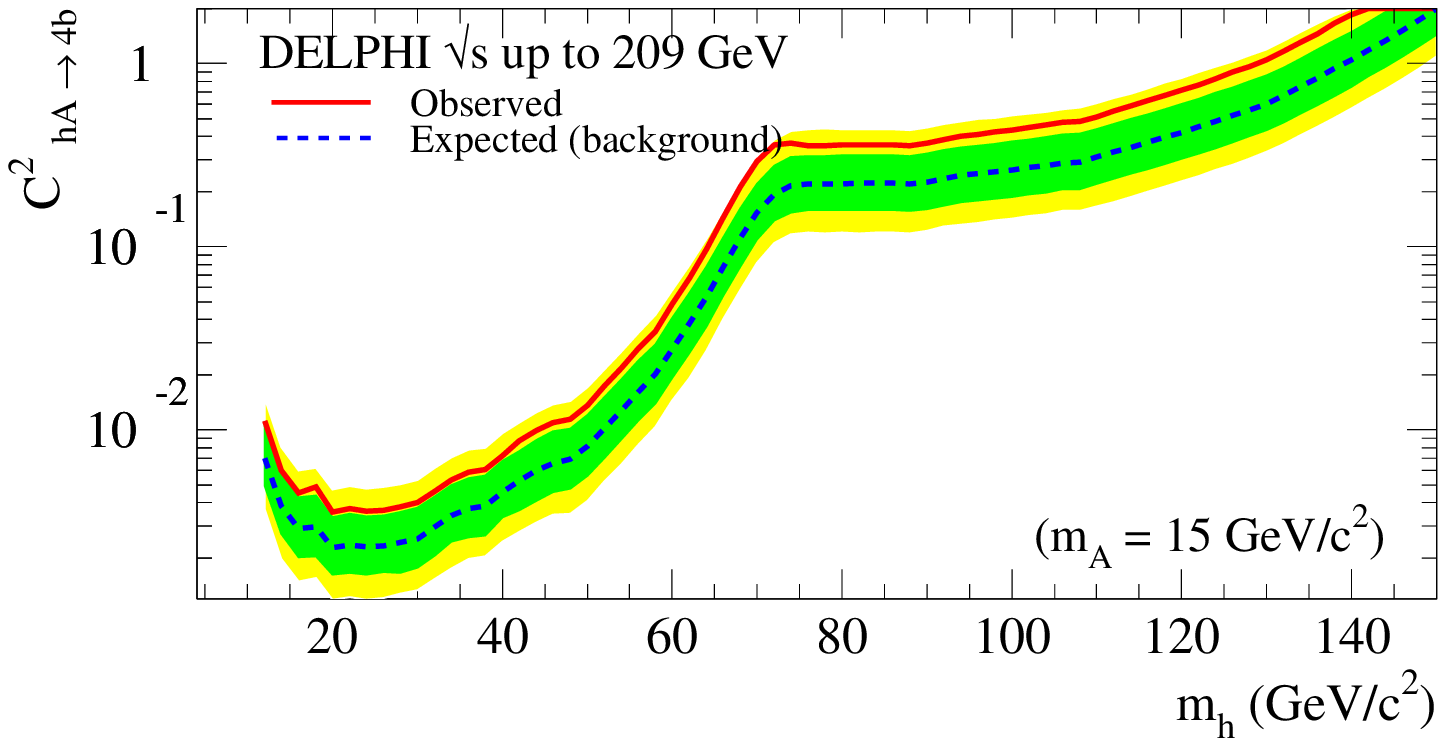,width=.9\textwidth}
\caption[]{
95\% CL upper bounds on the reduction factor $\mathrm C^2_{hA \mra 4b}$, as
defined in Section~\ref{higgsprod}. Results are presented for h and A bosons with 
equal masses (top) and with one mass fixed to 15~\GeVcc\ (bottom).
The limits observed in the data (full curve) are shown together with 
the expected median limits in background process experiments 
(dashed curve). The bands correspond to the 
68.3\% and 95.0\% confidence intervals for background-only experiments.
}
\label{slice4b}
\end{center}
\end{figure}

The upper limit on $\mathrm C^{2}_{hA \mra 4b}$ as a function of \mh $+$\mA\ is shown in 
Figure~\ref{slice4b} for equal h and A masses as well as for large mass differences. In these figures, the 
observed result is compared to the expected limits, allowing a comparison 
of the data with the SM background predictions. The agreement is well within 
2 standard deviations over the whole range of mass hypotheses in the case
of equal h and A masses: there, the results are given by the LEP1 
analysis of Section~\ref{4bLEP1} up to about 90~\GeVcc\ in \mh $+$\mA,
with limits on $\mathrm C^{2}_{hA \mra 4b}$ between $\sim$ 0.1\% and 10\%,
and by the LEP2 MSSM analysis of~\cite{DELhiggs} at higher masses,
with limits on $\mathrm C^{2}_{hA \mra 4b}$ around 10\% up to 160~\GeVcc.
For full strength production and decay, a mass limit on \mh\ and 
\mA\ of 90.9~\GeVcc\ is reached.
In the case of large mass differences between the h and A bosons,
the results are given by the LEP1 and the LEP2 analyses presented in this paper.
As a result of the excess observed in the data of 2000a, there is a disagreement
between the data and the SM background prediction in the upper limit on $\mathrm C^{2}_{hA \mra 4b}$. 
When \mA\ is fixed at 15~\GeVcc (Figure~\ref{slice4b}) the disagreement
amounts to 1.6 standard deviations for any \mh\ above $\sim$70~\GeVcc. This also translates
into a mass limit of 127.8~\GeVcc on \mh $+$\mA, whereas 138.0~\GeVcc\ is expected on 
average from background experiments.

Numerical values for the observed exclusions are given in Table~\ref{cpl_hA4b}.

\subsubsection{The four-$\tau$ search}

\begin{figure} [p]
\begin{center}
\epsfig{figure=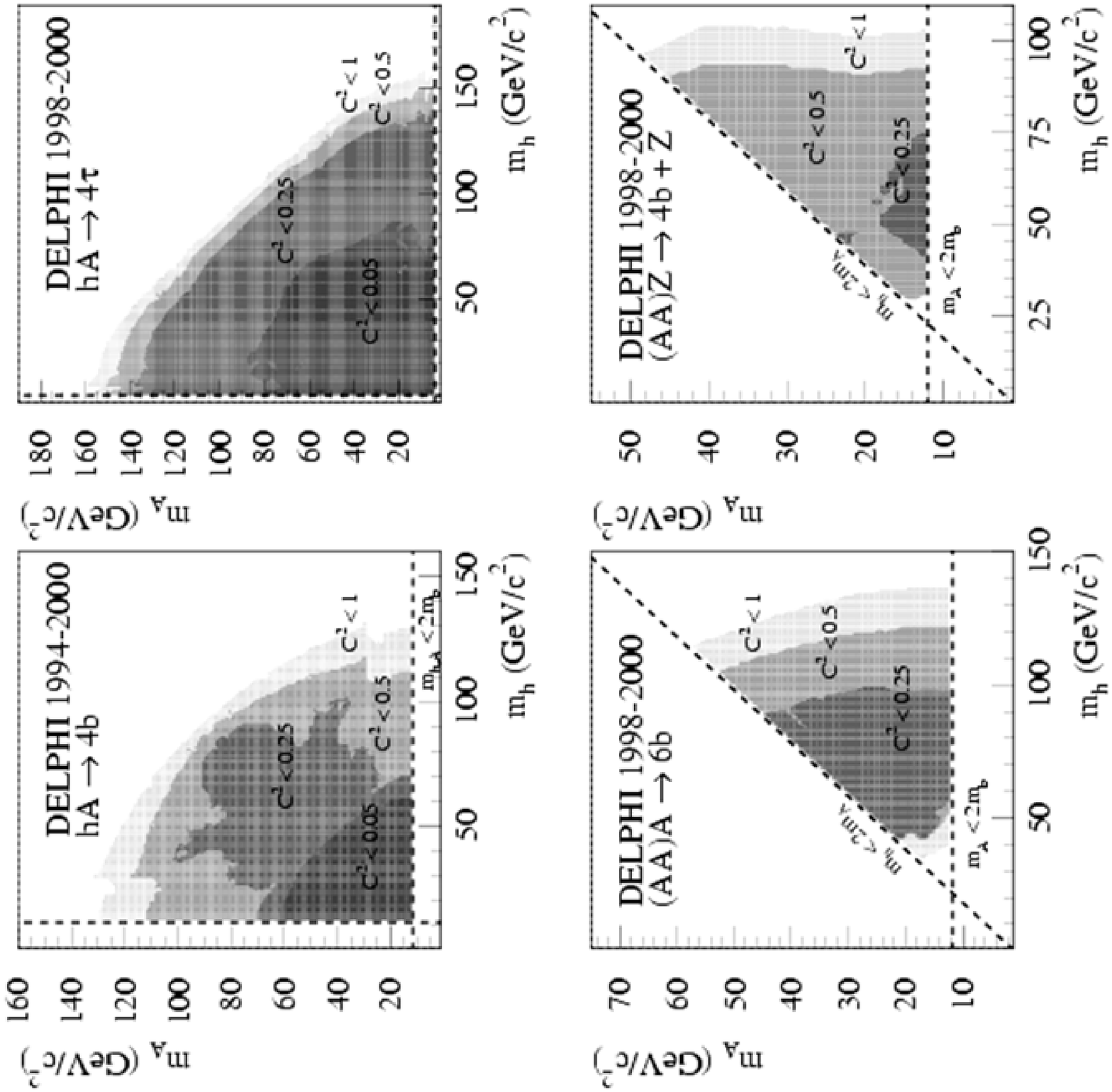, width=.9\textwidth,angle=-90}
\end{center}
\caption{Excluded couplings in the (\mh,\mA) plane. 
Upper left: hA \ra 4b ($\mathrm C^{2}\equiv C^{2}_{hA\mra 4b}$); 
upper right: hA \ra 4$\tau$ ($\mathrm C^{2}\equiv C^{2}_{hA\mra 4\tau}$); 
lower left: (AA)A \ra 6b ($\mathrm C^{2}\equiv C^{2}_{hA\mra 6b}$);
lower right: (AA)Z \ra 4b+jets ($\mathrm C^{2}\equiv C^{2}_{Z(AA\mra 4b)}$). The $\mathrm C^{2}$
parameters are defined in Section~\ref{higgsprod}.
The three outer embedded regions correspond to excluded $\mathrm C^{2}$ values of 1, 0.5, and 0.25
respectively; for the hA \ra 4b final state (which includes LEP1 results)
and the hA \ra 4$\tau$ final state, the innermost region corresponds to 
excluded couplings smaller than 0.05.}
\label{contours}
\end{figure}

\begin{figure}[htbp]
\begin{center}
\epsfig{figure=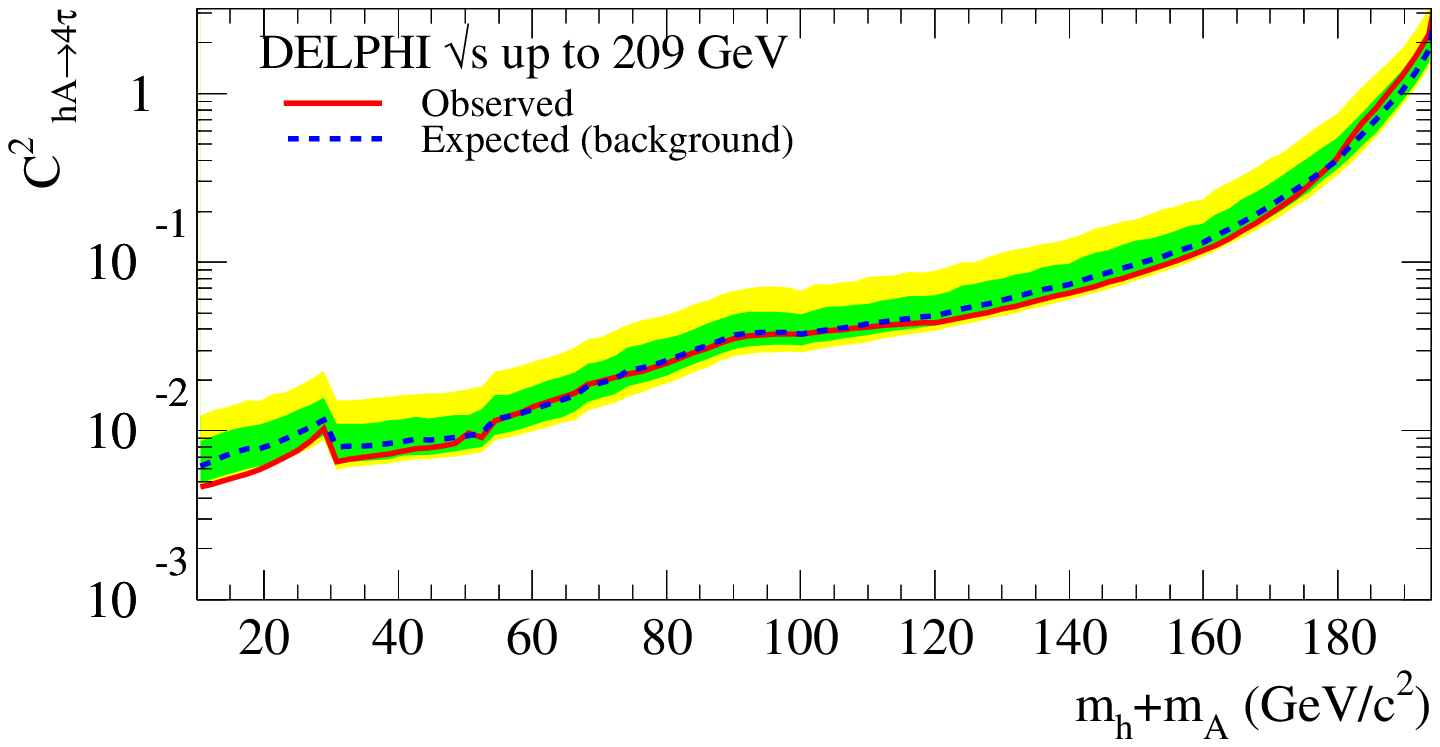,width=.9\textwidth}
\epsfig{figure=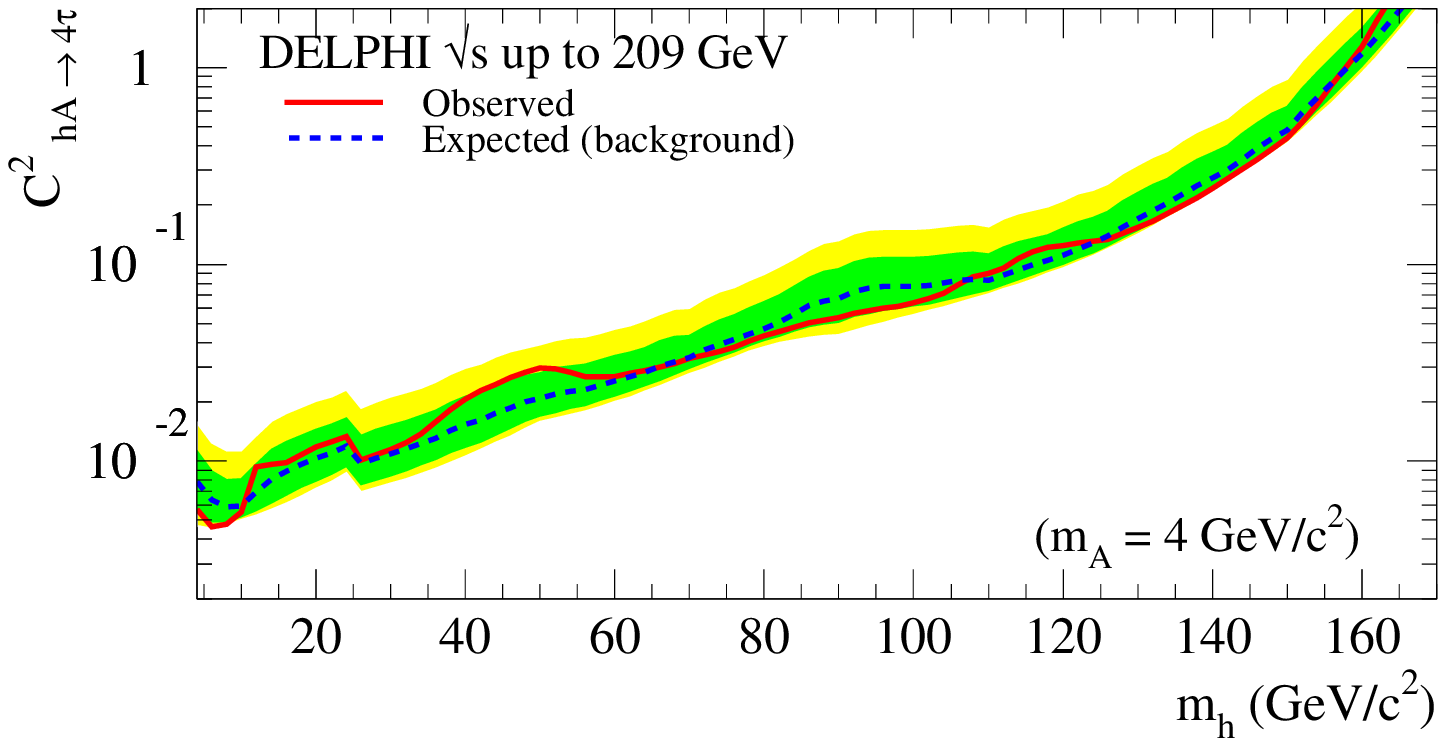,width=.9\textwidth}
\caption[]{
95\% CL upper bounds on the reduction factor $\mathrm C^2_{hA \mra 4\tau}$, as
defined in the text.
Results are presented in the four-$\tau$ channel for h and A bosons with 
equal masses (top) and with one mass fixed to 4~\GeVcc\ (bottom).
The limits observed in the data (full curve) are shown together with 
the expected median limits in background process experiments 
(dashed curve). The bands correspond to the 
68.3\% and 95.0\% confidence intervals for background-only experiments.
}
\label{slice4t}
\end{center}
\end{figure}

The results of the hA \ra 4$\tau$ analysis are shown in the 
(\mh,\mA) plane in Figure~\ref{contours} and as
a function of \mh $+$\mA\ for mass-degenerate h and A bosons
in Figure~\ref{slice4t}. In the case of no suppression,
this very sensitive search allows a large range of masses to be excluded, 
from the \tautau\ threshold up to around 10~\GeVcc\ below the
kinematical limit. For equal h and A masses, this translates into 
a mass limit of 93.6~\GeVcc. Limits on $\mathrm C^{2}_{hA \mra 4\tau}$ are very strong,
e.g. below 10\% up to 140~\GeVcc\ in \mh $+$\mA\ for equal masses,
allowing large portions of the mass plane to be excluded even up to
$\mathrm C^{2}_{hA \mra 4\tau}\sim$0.25, as shown in Figure~\ref{contours}.
Finally, Figure~\ref{slice4t} also shows the results when one
Higgs boson mass is fixed at 4~\GeVcc. In this case, full strength
production is excluded up to \mh,\mA $=$ 158.1~\GeVcc.

Numerical values for the observed exclusions are given in Table~\ref{cpl_hA4t}.

\subsubsection{hZ with h \ra \bb\ and h \ra \tautau}

\begin{figure}[htbp]
\begin{center}
\epsfig{figure=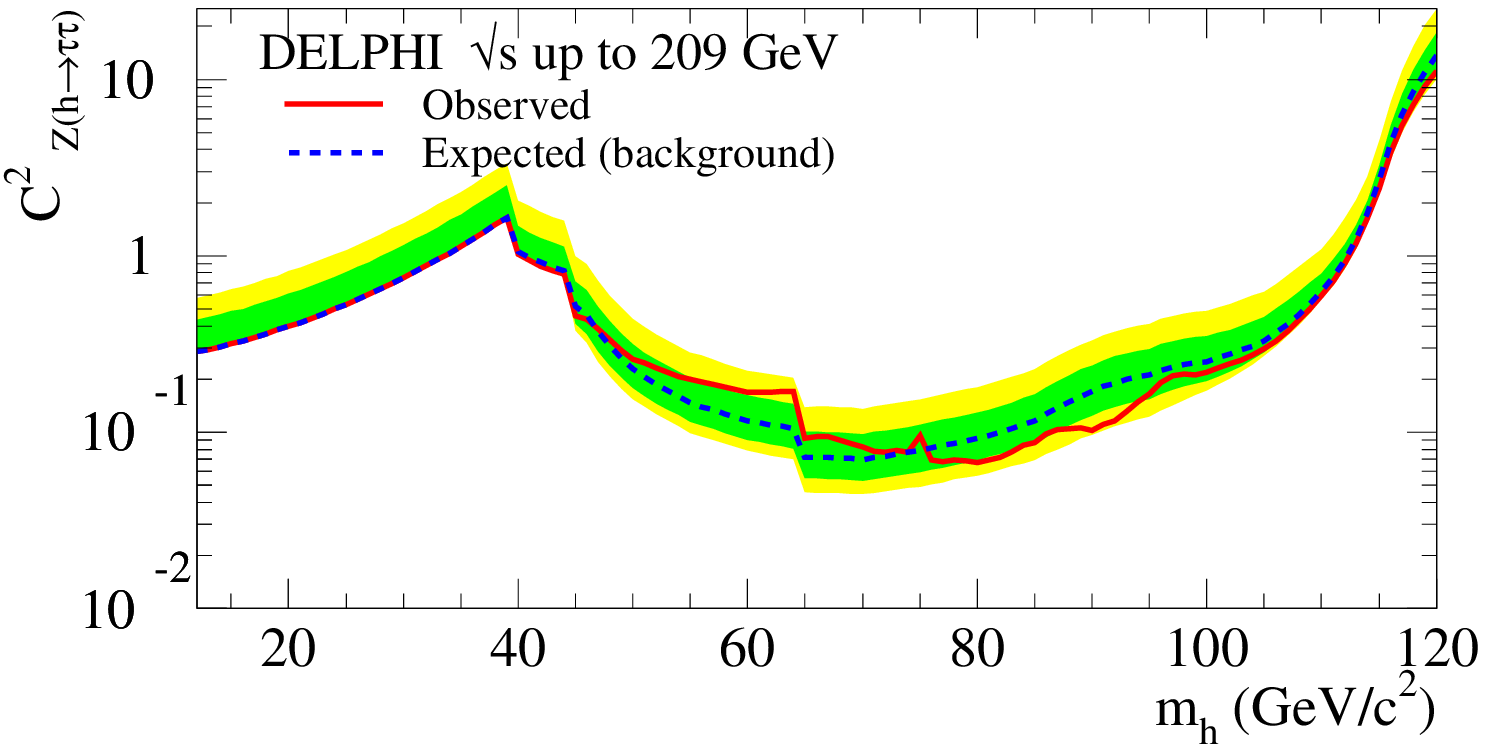,width=.9\textwidth}
\epsfig{figure=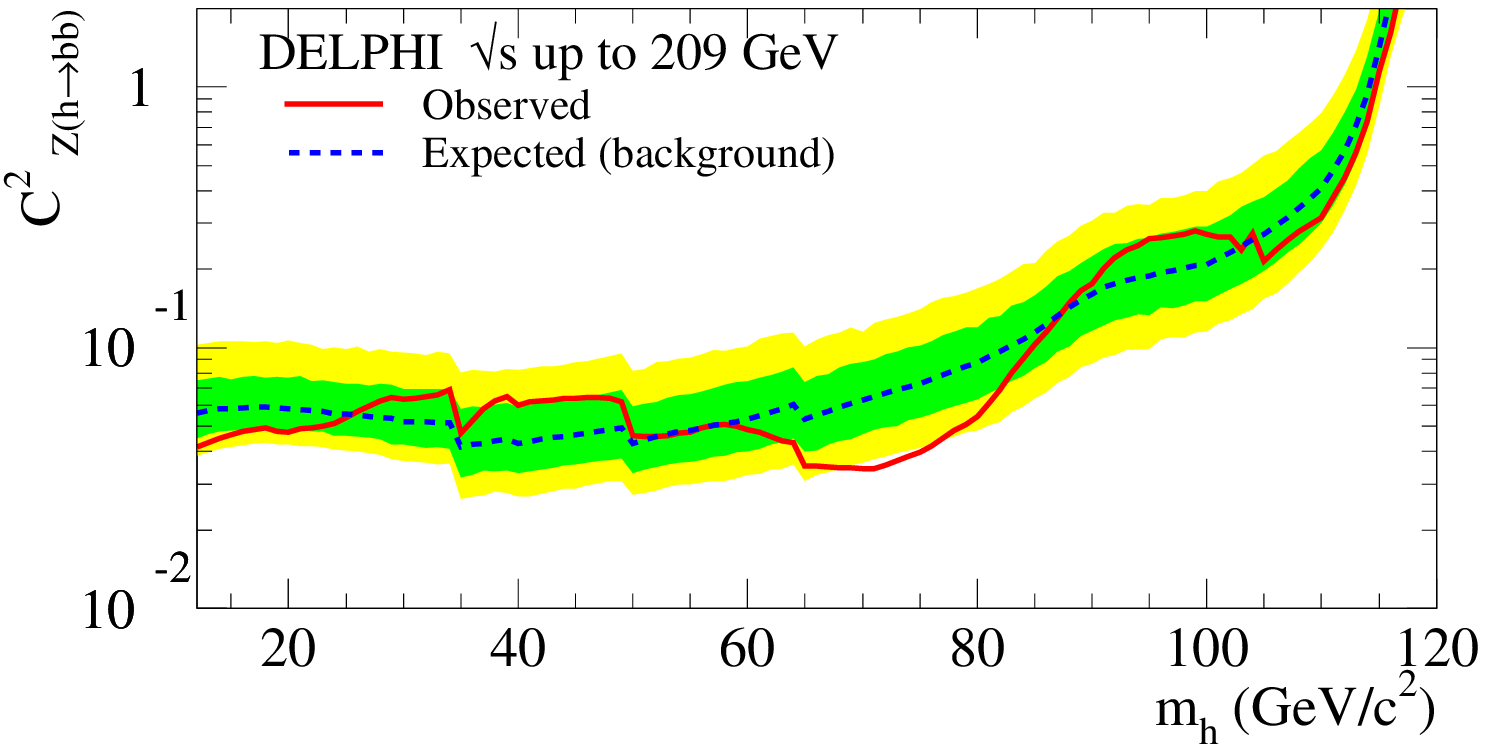,width=.9\textwidth}
\caption[]{
95\% CL upper bounds on the reduction factors 
$\mathrm C^2_{Z(h \mra \tau\tau)}$ and $\mathrm C^2_{Z(h \mra bb)}$, as defined in Section~\ref{higgsprod}.
The limits observed in the data (full curve) are shown together with 
the expected median limits in background process experiments 
(dashed curve). The bands correspond to the 
68.3\% and 95.0\% confidence intervals for background-only experiments.
The shape of the results in the $\tau$ channel is due to the
sensitivity of the LEP2 analyses starting at 40~\GeVcc\ and 
to LEP1 analyses applied on subsets of the data sample only.}
\label{slice_hz}
\end{center}
\end{figure}

The upper limits on the suppression factors for hZ production followed by a direct
decay of the h boson into $\tau$~lepton or b~quark pairs, are shown as a function of
\mh\ in Figure~\ref{slice_hz}. For full strength production and decay, mass limits of
112.4 and 114.6~\GeVcc\ on \mh\ are obtained in the two channels, respectively
(the mass limit in the \tautau\ channel is not absolute, since there is an unexcluded region around
\mh $=$ 40~\GeVcc).
Upper limits on the suppression factors lower than 10\% are obtained
for \mh\ from the \bb\ threshold up to 85~\GeVcc\ in the case of b
decays. The limits are much weaker in the case of $\tau$ decays
with upper bounds of 20\% for \mh\ between 50 and 90~\GeVcc.

Numerical values for the observed exclusions are given in Tables~\ref{cpl_ttZ} and
\ref{cpl_bbZ}.

\subsection{hA and hZ production: cascade decays}
\label{res:cascade}

The analysis described in Section~\ref{4bLEP2} is applied to the search for
Higgs bosons involving cascade decays. Compared to the previous
section, the only differences are the signal selection efficiencies, which are
sensitive to the details of the final state. The primary hA and hZ
production rates are the same as above.

Results on the final state with six b~quarks, originating from hA 
production with intermediary decay of the h boson into two A bosons, 
are displayed in Figure~\ref{contours}. The high number of b~quarks 
in the final state makes the search sensitive even for small 
suppression factors. For full strength production and decay, the limit on \mh\ is 114.5~\GeVcc\ when
\mA $\sim$ \mh/2, and 136.3~\GeVcc\ when \mA $=$ 12~\GeVcc.

Production of four b~quarks in addition to a Z boson
through the process hZ \ra (AA)Z, is constrained as shown in Figure~\ref{contours}. 
The \mh\ range covered is bounded from above because
of the high mass of the associated Z boson. In the case of no
suppression (in other words, if this channel is dominant), the present
analysis constrains the h mass to be above $\sim$95~\GeVcc, for any
\mA\ between the b~quark decay threshold and \mh/2.

The similar hA \ra h(hZ) process is found to be
unconstrained by the present work. The reasons are that the hA
cross-section decreases much faster than the hZ cross-section when
approaching the kinematic limit, leading to reduced
sensitivity. Furthermore, the excess observed in the data taken in
2000a (see Table~\ref{comp4} and the discussion
given in the previous section) is enough to forbid any exclusion in this
channel. This conclusion also applies to the (AZ)A process, as argued in
Section~\ref{higgsprod}. 

Numerical values for the observed exclusions are given in Tables~\ref{cpl_hA6b} and
\ref{cpl_hZ4b2q}.

 
\section{Conclusions}
\label{conclusions}

Searches for Higgs production have been performed in various channels, using the data 
recorded by DELPHI at LEP2, relying extensively on a multi-purpose b-tagging analysis. 
The much studied hA \ra 4b channel has been revisited and extended sensitivity towards large
h and A mass differences was obtained. The decay h \ra AA was also
considered and searched for in hA and hZ production. In these
three cases large portions of the (\mh,\mA) plane are excluded,
depending on a global suppression factor. The decay A \ra hZ was
also studied but was found unconstrained.

Four-b final states were searched for in the LEP1 data, in the hA channel and
in the Yukawa process. The results of the hA channel contribute to the coverage of the 
(\mh,\mA) plane at low masses. The search for the Yukawa process allowed the enhancement
of the h and A coupling to b quarks to be constrained for a large mass range of these bosons. 
The \bb\tautau\ final state was investigated in the context of the Yukawa process, and is constrained
over the same mass range.

Finally, models in which different Higgs doublets couple preferentially to
quarks or to leptons will predict dominant heavy-lepton decays. The four-$\tau$ final state from 
Yukawa production was searched for at LEP1. The hA \ra 4$\tau$ channel was investigated at LEP2, and 
strongly constrained by the present analysis.

The emphasis of this work is on the model-independence of the
results. All results are presented in a form that allows their
reinterpretation in a large class of models of the electroweak scalar sector.

 
\subsection*{Acknowledgements}
\vskip 3 mm
 We are greatly indebted to our technical 
collaborators, to the members of the CERN-SL Division for the excellent 
performance of the LEP collider, and to the funding agencies for their
support in building and operating the DELPHI detector.\\
We acknowledge in particular the support of \\
Austrian Federal Ministry of Education, Science and Culture,
GZ 616.364/2-III/2a/98, \\
FNRS--FWO, Flanders Institute to encourage scientific and technological 
research in the industry (IWT), Belgium,  \\
FINEP, CNPq, CAPES, FUJB and FAPERJ, Brazil, \\
Czech Ministry of Industry and Trade, GA CR 202/99/1362,\\
Commission of the European Communities (DG XII), \\
Direction des Sciences de la Mati$\grave{\mbox{\rm e}}$re, CEA, France, \\
Bundesministerium f$\ddot{\mbox{\rm u}}$r Bildung, Wissenschaft, Forschung 
und Technologie, Germany,\\
General Secretariat for Research and Technology, Greece, \\
National Science Foundation (NWO) and Foundation for Research on Matter (FOM),
The Netherlands, \\
Norwegian Research Council,  \\
State Committee for Scientific Research, Poland, SPUB-M/CERN/PO3/DZ296/2000,
SPUB-M/CERN/PO3/DZ297/2000, 2P03B 104 19 and 2P03B 69 23(2002-2004)\\
FCT - Funda\c{c}\~ao para a Ci\^encia e Tecnologia, Portugal, \\
Vedecka grantova agentura MS SR, Slovakia, Nr. 95/5195/134, \\
Ministry of Science and Technology of the Republic of Slovenia, \\
CICYT, Spain, AEN99-0950 and AEN99-0761,  \\
The Swedish Natural Science Research Council,      \\
Particle Physics and Astronomy Research Council, UK, \\
Department of Energy, USA, DE-FG02-01ER41155. \\
EEC RTN contract HPRN-CT-00292-2002. \\





\label{figures}
 
\appendix

\newpage
\section{Efficiencies}
\label{efficiencies}
Signal efficiencies for all analyses presented in this paper are given below. The quoted uncertainties
are statistical only.
\begin{table}[htbp]
 \caption{Signal efficiencies in the \bb(h \ra \bb)
          and \bb(A \ra \bb) channels (LEP1).}
 \begin{center}  
 \scriptsize
 \begin{tabular}{r|cc|cc}
mass (\GeVcc)
& \multicolumn{2}{c|}{three-jet eff. ($\%$)} 
& \multicolumn{2}{c} {four-jet eff. ($\%$)}\\ 
      & Bin 1 & Bin 2 & Bin 1 & Bin 2 \\
\hline
\mh $= 11$  &$0.5\pm 0.1$& $1.2\pm 0.2$&$0.4\pm 0.1$&$0.5\pm 0.1$\\
      $13$  &$0.6\pm 0.1$& $0.7\pm 0.1$&$0.7\pm 0.1$&$0.4\pm 0.1$\\
      $15$  &$0.4\pm 0.1$& $0.8\pm 0.1$&$0.9\pm 0.1$&$1.1\pm 0.1$\\
      $20$  &$0.5\pm 0.1$& $1.0\pm 0.1$&$1.1\pm 0.2$&$1.5\pm 0.2$\\
      $30$  &$0.7\pm 0.1$& $1.2\pm 0.2$&$1.8\pm 0.2$&$2.1\pm 0.2$\\
      $40$  &$0.6\pm 0.1$& $1.5\pm 0.2$&$1.8\pm 0.2$&$2.4\pm 0.2$\\
      $50$  &$0.6\pm 0.1$& $0.9\pm 0.1$&$1.3\pm 0.2$&$1.8\pm 0.2$\\
\mA $= 11$  &$0.6\pm 0.1$& $1.4\pm 0.2$&$0.9\pm 0.1$&$0.5\pm 0.1$\\
      $13$  &$0.5\pm 0.1$& $1.3\pm 0.2$&$0.8\pm 0.1$&$0.7\pm 0.1$\\
      $15$  &$0.5\pm 0.1$& $1.0\pm 0.1$&$1.1\pm 0.2$&$1.3\pm 0.2$\\
      $20$  &$0.5\pm 0.1$& $1.1\pm 0.2$&$1.4\pm 0.2$&$1.8\pm 0.2$\\
      $30$  &$0.5\pm 0.1$& $1.3\pm 0.2$&$1.7\pm 0.2$&$2.2\pm 0.2$\\
      $40$  &$0.4\pm 0.1$& $1.5\pm 0.2$&$1.8\pm 0.2$&$2.2\pm 0.2$\\
      $50$  &$0.5\pm 0.1$& $1.1\pm 0.1$&$1.7\pm 0.2$&$1.8\pm 0.2$\\
 \end{tabular}
 \end{center}
 \label{bbbb_Y}
\end{table}

\begin{table}[htbp]
\caption{Signal efficiencies in the hA \ra 4b channel (LEP1). The efficiencies are 
symmetric in \mh\ and \mA.}
\begin{center}
\scriptsize
\begin{tabular}{r|cc|cc}
\multicolumn{1}{r|}{mass (\GeVcc)}
& \multicolumn{2}{c|}{three-jet eff. ($\%$)} 
& \multicolumn{2}{c}{four-jet eff. ($\%$)}\\ 
 \mA,\mh & Bin 1 & Bin 2 & Bin 1 & Bin 2 \\
\hline
$12$,$20$  &$0.8\pm 0.1$& $1.7\pm 0.2$&$1.2\pm 0.1$&$1.1\pm 0.1$\\
$12$,$30$  &$0.9\pm 0.1$& $2.0\pm 0.2$&$1.5\pm 0.1$&$1.4\pm 0.1$\\
$12$,$40$  &$1.0\pm 0.1$& $1.8\pm 0.2$&$1.3\pm 0.1$&$1.1\pm 0.1$\\
$12$,$50$  &$1.0\pm 0.1$& $1.9\pm 0.2$&$1.3\pm 0.1$&$1.2\pm 0.1$\\
$12$,$60$  &$0.8\pm 0.1$& $1.5\pm 0.1$&$0.9\pm 0.1$&$0.8\pm 0.1$\\
$12$,$70$  &$0.4\pm 0.1$& $0.7\pm 0.1$&$0.4\pm 0.1$&$0.3\pm 0.1$\\

$20$,$20$  &$0.9\pm 0.1$& $2.0\pm 0.2$&$2.4\pm 0.2$&$3.4\pm 0.2$\\
$20$,$30$  &$0.9\pm 0.1$& $1.6\pm 0.2$&$2.4\pm 0.2$&$3.5\pm 0.3$\\
$20$,$40$  &$0.7\pm 0.1$& $1.3\pm 0.1$&$1.8\pm 0.2$&$2.5\pm 0.2$\\
$20$,$50$  &$0.7\pm 0.1$& $1.3\pm 0.2$&$1.6\pm 0.2$&$2.0\pm 0.2$\\
$20$,$60$  &$0.7\pm 0.1$& $1.1\pm 0.1$&$1.3\pm 0.1$&$1.2\pm 0.1$\\

$30$,$30$  &$0.7\pm 0.1$& $1.8\pm 0.2$&$2.0\pm 0.2$&$2.7\pm 0.2$\\
$30$,$40$  &$0.6\pm 0.1$& $1.2\pm 0.1$&$1.6\pm 0.2$&$2.4\pm 0.2$\\
$30$,$50$  &$0.4\pm 0.1$& $1.3\pm 0.1$&$1.7\pm 0.2$&$2.1\pm 0.2$\\ 

$40$,$40$  &$0.6\pm 0.1$& $1.1\pm 0.1$&$1.8\pm 0.2$&$2.0\pm 0.2$\\

\end{tabular}
\label{ha1-eff}
\end{center}
\end{table}

\begin{table}[htbp]
 \caption{Signal efficiencies in the \bb(h \ra \tautau)
          and \bb(A \ra \tautau) channels (LEP1).}
 \begin{center}  
 \scriptsize 
 \begin{tabular}{r|ccr|c}
 mass (\GeVcc) &  efficiency && mass (\GeVcc) &  efficiency \\ 
            & $(\%)$ &&             & $(\%)$ \\ \hline
  \mh $= 4$ & 0.8 $\pm$ 0.1 &&  \mA $= 4$ & 1.0 $\pm$ 0.1 \\
        $7$ & 1.1 $\pm$ 0.1 &&        $7$ & 1.4 $\pm$ 0.1 \\
        $9$ & 1.3 $\pm$ 0.1 &&        $9$ & 1.8 $\pm$ 0.1 \\
       $10$ & 1.5 $\pm$ 0.1 &&       $10$ & 1.8 $\pm$ 0.1 \\
       $12$ & 1.7 $\pm$ 0.1 &&       $12$ & 1.7 $\pm$ 0.1 \\
       $15$ & 1.9 $\pm$ 0.1 &&       $15$ & 2.0 $\pm$ 0.1 \\
       $20$ & 2.2 $\pm$ 0.2 &&       $20$ & 2.3 $\pm$ 0.2 \\
       $30$ & 3.3 $\pm$ 0.2 &&       $30$ & 3.2 $\pm$ 0.2 \\
       $40$ & 3.8 $\pm$ 0.2 &&       $40$ & 3.8 $\pm$ 0.2 \\
       $50$ & 3.7 $\pm$ 0.2 &&       $50$ & 4.1 $\pm$ 0.2 \\
 \end{tabular}
 \end{center}
 \label{bbtt_Y}
\end{table}

\begin{table}[htbp]
 \caption{Signal efficiencies in the four-prong \tautau(h \ra \tautau)
          and \tautau(A \ra \tautau) channels (LEP1).}
 \begin{center}  
 \scriptsize
 \begin{tabular}{r|ccr|c}
 mass (\GeVcc) &  efficiency && mass (\GeVcc) &  efficiency \\ 
            & $(\%)$ &&             & $(\%)$ \\ \hline
  \mh $= 4$ & 3.0 $\pm$ 0.2 &&  \mA $= 4$ & 3.2 $\pm$ 0.2 \\
        $7$ & 5.3 $\pm$ 0.2 &&        $7$ & 5.6 $\pm$ 0.2 \\
        $9$ & 5.8 $\pm$ 0.2 &&        $9$ & 5.9 $\pm$ 0.2 \\
       $10$ & 6.0 $\pm$ 0.2 &&       $10$ & 5.7 $\pm$ 0.2 \\
       $12$ & 6.3 $\pm$ 0.2 &&       $12$ & 6.2 $\pm$ 0.2 \\
       $15$ & 5.9 $\pm$ 0.2 &&       $15$ & 6.2 $\pm$ 0.2 \\
       $20$ & 6.1 $\pm$ 0.2 &&       $20$ & 5.7 $\pm$ 0.2 \\
       $30$ & 6.2 $\pm$ 0.2 &&       $30$ & 5.8 $\pm$ 0.2 \\
       $40$ & 6.5 $\pm$ 0.2 &&       $40$ & 6.3 $\pm$ 0.2 \\
       $50$ & 6.2 $\pm$ 0.2 &&       $50$ & 5.9 $\pm$ 0.2 \\
 \end{tabular}
 \end{center}
 \label{tttt_Y4p}
\end{table}

\begin{table}[htbp]
 \caption{Signal efficiencies in the six-prong \tautau(h \ra \tautau)
          and \tautau(A \ra \tautau) channels (LEP1).}
 \begin{center}  
 \scriptsize
 \begin{tabular}{r|ccr|c}
 mass (\GeVcc) &  efficiency && mass (\GeVcc) &  efficiency \\ 
            & $(\%)$ &&             & $(\%)$ \\ \hline
  \mh $= 4$ & 2.4 $\pm$ 0.2 &&  \mA $= 4$ & 2.5 $\pm$ 0.2 \\
        $7$ & 3.9 $\pm$ 0.2 &&        $7$ & 4.3 $\pm$ 0.2 \\
        $9$ & 4.5 $\pm$ 0.2 &&        $9$ & 4.6 $\pm$ 0.2 \\
       $10$ & 4.3 $\pm$ 0.2 &&       $10$ & 4.6 $\pm$ 0.2 \\
       $12$ & 4.7 $\pm$ 0.2 &&       $12$ & 4.6 $\pm$ 0.2 \\
       $15$ & 4.7 $\pm$ 0.2 &&       $15$ & 4.8 $\pm$ 0.2 \\
       $20$ & 5.6 $\pm$ 0.2 &&       $20$ & 4.8 $\pm$ 0.2 \\
       $30$ & 5.5 $\pm$ 0.2 &&       $30$ & 5.4 $\pm$ 0.2 \\
       $40$ & 5.5 $\pm$ 0.2 &&       $40$ & 5.3 $\pm$ 0.2 \\
       $50$ & 5.6 $\pm$ 0.2 &&       $50$ & 5.2 $\pm$ 0.2 \\
 \end{tabular}
 \end{center}
 \label{tttt_Y6p}
\end{table}

\begin{table}
\caption{Signal efficiencies in the hA \ra (AA)A \ra 6b channel (LEP2).}
\begin{center}
\scriptsize
\begin{tabular}{r|ccccc}
\multicolumn{1}{r|}{mass (\GeVcc)} & \multicolumn{5}{c}{efficiency (\%)} \\ 
 \mA,\mh & $ \sqrt{s} = 189$ GeV & $192$ GeV & $196$ GeV  & $200$ GeV & $206$ GeV \\
\hline
$12$,$70$  &  $27.1\pm 1.6$ & $26.9\pm 1.6$ & $27.4\pm 1.7$ & $27.3\pm 1.7$ & $26.6\pm 1.6$ \\
$12$,$90$  &  $44.2\pm 2.1$ & $44.0\pm 2.1$ & $44.1\pm 2.1$ & $42.3\pm 2.1$ & $41.8\pm 2.0$ \\
$12$,$110$ &  $47.9\pm 2.2$ & $48.1\pm 2.2$ & $48.8\pm 2.2$ & $49.6\pm 2.2$ & $49.0\pm 2.2$ \\
$12$,$130$ &  $42.8\pm 2.1$ & $43.4\pm 2.1$ & $44.4\pm 2.1$ & $44.1\pm 2.1$ & $44.0\pm 2.1$ \\
$12$,$150$ &  $36.3\pm 1.9$ & $38.1\pm 2.0$ & $39.7\pm 2.0$ & $41.0\pm 2.0$ & $42.4\pm 2.1$ \\
$12$,$170$ &  $ 4.2\pm 0.7$ & $ 4.6\pm 0.7$ & $ 5.9\pm 0.8$ & $11.3\pm 1.1$ & $22.7\pm 1.5$ \\
$30$,$70$  &  $49.1\pm 2.2$ & $49.6\pm 2.3$ & $48.5\pm 2.2$ & $49.0\pm 2.2$ & $48.8\pm 2.2$ \\
$30$,$90$  &  $52.5\pm 2.3$ & $53.2\pm 2.3$ & $53.7\pm 2.3$ & $53.7\pm 2.3$ & $53.7\pm 2.3$ \\
$30$,$110$ &  $54.3\pm 2.3$ & $54.2\pm 2.3$ & $54.4\pm 2.3$ & $54.5\pm 2.3$ & $54.5\pm 2.3$ \\
$30$,$130$ &  $53.2\pm 2.3$ & $53.9\pm 2.3$ & $53.9\pm 2.3$ & $53.7\pm 2.3$ & $53.6\pm 2.3$ \\
$30$,$150$ &  $50.1\pm 2.3$ & $49.8\pm 2.3$ & $50.4\pm 2.3$ & $51.0\pm 2.3$ & $51.0\pm 2.3$ \\
$50$,$110$ &  $56.3\pm 2.4$ & $56.9\pm 2.4$ & $57.9\pm 2.4$ & $57.9\pm 2.4$ & $57.9\pm 2.4$ \\
$50$,$130$ &  $57.0\pm 2.4$ & $57.9\pm 2.4$ & $58.4\pm 2.4$ & $58.5\pm 2.4$ & $58.6\pm 2.4$ \\
\end{tabular}
\label{aaa-eff}
\end{center}
\end{table}

\begin{table}
\caption{Signal efficiencies in the hZ \ra (AA)Z \ra 4b+jets channel (LEP2).}
\begin{center}
\scriptsize
\begin{tabular}{r|ccccc}
\multicolumn{1}{r|}{mass (\GeVcc)} & \multicolumn{5}{c}{efficiency (\%)} \\ 
 \mA,\mh & $ \sqrt{s} = 189$ GeV & $192$ GeV & $196$ GeV  & $200$ GeV & $206$ GeV \\
\hline
$12$,$30$  &  $ 6.9\pm 0.8$ & $ 7.7\pm 0.9$ & $ 8.3\pm 0.9$ & $ 7.6\pm 0.9$ & $ 8.3\pm 0.9$ \\
$12$,$50$  &  $13.8\pm 1.2$ & $13.8\pm 1.2$ & $14.7\pm 1.2$ & $14.8\pm 1.2$ & $14.7\pm 1.2$ \\
$12$,$70$  &  $20.7\pm 1.4$ & $20.3\pm 1.4$ & $19.9\pm 1.4$ & $19.8\pm 1.4$ & $20.2\pm 1.4$ \\
$12$,$90$  &  $20.9\pm 1.4$ & $21.8\pm 1.5$ & $20.9\pm 1.4$ & $21.0\pm 1.4$ & $21.1\pm 1.5$ \\
$12$,$105$ &                &               &               & $23.0\pm 1.5$ & $23.7\pm 1.5$ \\
$20$,$50$  &  $13.0\pm 1.1$ & $12.3\pm 1.1$ & $12.2\pm 1.1$ & $12.3\pm 1.1$ & $12.3\pm 1.1$ \\
$20$,$70$  &  $14.4\pm 1.2$ & $14.4\pm 1.2$ & $13.8\pm 1.2$ & $13.8\pm 1.2$ & $13.7\pm 1.2$ \\
$20$,$90$  &  $19.0\pm 1.4$ & $18.9\pm 1.4$ & $18.4\pm 1.4$ & $18.5\pm 1.4$ & $18.4\pm 1.4$ \\
$20$,$105$ &                &               &               & $19.4\pm 1.4$ & $21.1\pm 1.5$ \\
$30$,$70$  &  $16.8\pm 1.3$ & $17.0\pm 1.3$ & $15.5\pm 1.3$ & $15.6\pm 1.3$ & $15.5\pm 1.3$ \\
$30$,$90$  &  $21.9\pm 1.5$ & $22.3\pm 1.5$ & $22.2\pm 1.5$ & $22.3\pm 1.5$ & $22.3\pm 1.5$ \\
$30$,$105$ &                &               &               & $24.8\pm 1.6$ & $24.8\pm 1.6$ \\
$40$,$90$  &  $22.1\pm 1.6$ & $22.4\pm 1.6$ & $22.2\pm 1.6$ & $22.3\pm 1.6$ & $22.3\pm 1.6$ \\
$40$,$105$ &                 &                &             & $26.1\pm 1.7$ & $25.4\pm 1.7$ \\
\end{tabular}
\label{aaz-eff}
\end{center}
\end{table}

\begin{table}
\caption{Signal efficiencies in the hA \ra h(hZ) \ra 4b+jets channel (LEP2).}
\begin{center}
\scriptsize
\begin{tabular}{r|ccccc}
\multicolumn{1}{r|}{mass (\GeVcc)} & \multicolumn{5}{c}{efficiency (\%)} \\ 
 \mA,\mh & $ \sqrt{s} = 189$ GeV & $192$ GeV & $196$ GeV  & $200$ GeV & $206$ GeV \\
\hline
$12$,$110$ & $10.6\pm 1.0$ & $10.6\pm 1.0$ & $10.5\pm 1.0$ & $10.6\pm 1.0$ & $10.5\pm 1.0$ \\
$12$,$130$ & $14.6\pm 1.2$ & $14.5\pm 1.2$ & $14.8\pm 1.2$ & $14.6\pm 1.2$ & $15.4\pm 1.3$ \\
$12$,$150$ & $14.3\pm 1.2$ & $14.1\pm 1.2$ & $14.3\pm 1.2$ & $14.9\pm 1.2$ & $15.4\pm 1.3$ \\
$12$,$170$ & $10.8\pm 1.1$ & $12.6\pm 1.2$ & $13.5\pm 1.3$ & $13.9\pm 1.3$ & $14.2\pm 1.3$ \\
$30$,$130$ & $15.1\pm 1.3$ & $15.1\pm 1.3$ & $15.4\pm 1.3$ & $15.7\pm 1.3$ & $15.6\pm 1.3$ \\
$30$,$150$ & $15.8\pm 1.3$ & $15.6\pm 1.3$ & $16.0\pm 1.3$ & $16.2\pm 1.3$ & $16.2\pm 1.3$ \\
\end{tabular}
\label{hhz-eff}
\end{center}
\end{table}

\begin{table}
\caption{Signal efficiencies in the hA \ra 4b channel (LEP2). The efficiencies
are symmetric in \mh\ and \mA.}
\begin{center}
\scriptsize
\begin{tabular}{r|ccccc}
\multicolumn{1}{r|}{mass (\GeVcc)} & \multicolumn{5}{c}{efficiency (\%)} \\ 
 \mA,\mh & $ \sqrt{s} = 189$ GeV & $192$ GeV & $196$ GeV  & $200$ GeV & $206$ GeV \\
\hline
$12$,$50$  & $ 2.5\pm 0.5$ & $ 2.0\pm 0.4$ & $ 1.6\pm 0.4$ & $ 1.3\pm 0.4$ & $ 1.3\pm 0.4$ \\
$12$,$70$  & $15.7\pm 1.3$ & $15.3\pm 1.2$ & $15.4\pm 1.2$ & $15.4\pm 1.2$ & $14.5\pm 1.2$ \\
$12$,$90$  & $25.4\pm 1.6$ & $25.0\pm 1.6$ & $24.8\pm 1.6$ & $24.5\pm 1.6$ & $23.6\pm 1.6$ \\
$12$,$110$ & $30.7\pm 1.8$ & $31.8\pm 1.8$ & $31.7\pm 1.8$ & $31.4\pm 1.8$ & $30.9\pm 1.8$ \\ 
$12$,$130$ & $30.5\pm 1.7$ & $31.1\pm 1.8$ & $30.6\pm 1.7$ & $31.8\pm 1.8$ & $31.3\pm 1.8$ \\
$12$,$150$ & $23.1\pm 1.5$ & $23.8\pm 1.5$ & $24.2\pm 1.6$ & $25.2\pm 1.6$ & $26.3\pm 1.6$ \\ 
$12$,$170$ & $ 8.6\pm 0.9$ & $10.0\pm 1.0$ & $11.7\pm 1.1$ & $14.5\pm 1.2$ & $17.1\pm 1.3$ \\
$30$,$30$  & $ 3.0\pm 0.5$ & $ 3.0\pm 0.5$ & $ 2.9\pm 0.5$ & $ 2.8\pm 0.5$ & $ 2.9\pm 0.5$ \\
$30$,$50$  & $16.0\pm 1.3$ & $15.8\pm 1.3$ & $15.2\pm 1.2$ & $14.5\pm 1.2$ & $14.1\pm 1.2$ \\
$30$,$70$  & $30.3\pm 1.7$ & $30.4\pm 1.7$ & $30.8\pm 1.8$ & $30.3\pm 1.7$ & $29.5\pm 1.7$ \\
$30$,$90$  & $35.1\pm 1.9$ & $35.7\pm 1.9$ & $35.0\pm 1.9$ & $35.2\pm 1.9$ & $35.3\pm 1.9$ \\
$30$,$110$ & $35.2\pm 1.9$ & $35.6\pm 1.9$ & $35.4\pm 1.9$ & $34.6\pm 1.9$ & $34.9\pm 1.9$ \\
$30$,$130$ & $31.9\pm 1.8$ & $32.9\pm 1.8$ & $33.7\pm 1.8$ & $33.6\pm 1.8$ & $34.3\pm 1.9$ \\
$30$,$150$ & $24.0\pm 1.5$ & $26.4\pm 1.6$ & $27.4\pm 1.7$ & $27.0\pm 1.6$ & $27.2\pm 1.6$ \\ 
$50$,$50$ & $33.7\pm 1.8$ & $33.8\pm 1.8$ & $33.7\pm 1.8$ & $34.1\pm 1.8$ & $33.5\pm 1.8$ \\
$50$,$70$ & $33.1\pm 1.8$ & $33.7\pm 1.8$ & $32.9\pm 1.8$ & $32.8\pm 1.8$ & $33.4\pm 1.8$ \\
$50$,$90$ & $37.4\pm 1.9$ & $37.9\pm 1.9$ & $38.7\pm 2.0$ & $38.9\pm 2.0$ & $39.2\pm 2.0$ \\
$50$,$110$ & $37.3\pm 1.9$ & $37.3\pm 1.9$ & $37.5\pm 1.9$ & $37.3\pm 1.9$ & $36.8\pm 1.9$ \\ 
$50$,$130$ & $31.8\pm 1.8$ & $33.1\pm 1.8$ & $34.4\pm 1.9$ & $33.8\pm 1.8$ & $34.7\pm 1.9$ \\ 
$70$,$70$ & $36.8\pm 1.9$ & $37.3\pm 1.9$ & $37.4\pm 1.9$ & $37.5\pm 1.9$ & $37.9\pm 1.9$ \\
$70$,$90$ & $41.2\pm 2.0$ & $41.5\pm 2.0$ & $41.7\pm 2.0$ & $42.1\pm 2.1$ & $42.5\pm 2.1$ \\
$70$,$110$ & $37.4\pm 1.9$ & $37.9\pm 1.9$ & $38.9\pm 2.0$ & $38.3\pm 2.0$ & $38.9\pm 2.0$ \\
\end{tabular}
\label{ha2-eff}
\end{center}
\end{table}

\begin{table}
\caption{Signal efficiencies in the hA \ra 4$\tau$ channel (examples given at 
$\sqrt{s}$=200 GeV). The efficiencies are symmetric in \mh\ and \mA.}
\begin{center}
\scriptsize
\begin{tabular}{r|ccc}
\multicolumn{1}{r|}{mass (\GeVcc)} & \multicolumn{3}{c}{efficiency (\%)} \\ 
 \mA,\mh & four-jet & three-jet & two-jet \\
\hline
$  4$,$4$  & $              $ & $              $ & $ 37.0 \pm 2.3 $ \\
$  4$,$15$  & $              $ & $              $ & $ 29.0 \pm 2.2 $ \\
$  4$,$35$  & $              $ & $ 10.6 \pm 2.1 $ & $ 15.8 \pm 2.2 $ \\
$  4$,$50$  & $              $ & $ 11.9 \pm 2.1 $ & $ 11.7 \pm 2.1 $ \\
$  4$,$70$  & $              $ & $ 28.4 \pm 2.2 $ & $  6.3 \pm 2.1 $ \\
$  4$,$90$  & $              $ & $ 43.4 \pm 2.3 $ & $  5.7 \pm 2.1 $ \\
$  4$,$125$  & $              $ & $ 44.7 \pm 2.3 $ & $  3.1 \pm 2.0 $ \\
$  4$,$170$  & $  4.0 \pm 2.0 $ & $ 23.5 \pm 2.2 $ & $  4.7 \pm 2.1 $ \\
$ 15$,$15$  & $  3.5 \pm 2.0 $ & $              $ & $ 19.1 \pm 2.2 $ \\
$ 15$,$35$  & $  5.9 \pm 2.1 $ & $              $ & $ 14.3 \pm 2.2 $ \\
$ 15$,$50$  & $ 10.3 \pm 2.1 $ & $              $ & $ 12.9 \pm 2.1 $ \\
$ 15$,$70$  & $ 26.1 \pm 2.2 $ & $  2.3 \pm 2.0 $ & $ 12.1 \pm 2.1 $ \\
$ 15$,$90$  & $ 32.7 \pm 2.3 $ & $  3.1 \pm 2.0 $ & $ 11.3 \pm 2.1 $ \\
$ 15$,$125$  & $ 32.3 \pm 2.3 $ & $  2.2 \pm 2.0 $ & $  7.8 \pm 2.1 $ \\
$ 15$,$170$  & $ 18.1 \pm 2.2 $ & $  4.1 \pm 2.0 $ & $  8.4 \pm 2.1 $ \\
$ 35$,$35$  & $ 13.3 \pm 2.1 $ & $              $ & $ 12.7 \pm 2.1 $ \\
$ 35$,$50$  & $ 26.4 \pm 2.2 $ & $              $ & $ 11.3 \pm 2.1 $ \\
$ 35$,$70$  & $ 39.0 \pm 2.3 $ & $              $ & $ 10.4 \pm 2.1 $ \\
$ 35$,$90$  & $ 41.1 \pm 2.3 $ & $  2.3 \pm 2.0 $ & $  9.0 \pm 2.1 $ \\
$ 35$,$125$  & $ 38.6 \pm 2.3 $ & $  2.3 \pm 2.0 $ & $  7.6 \pm 2.1 $ \\
$ 35$,$150$  & $ 37.2 \pm 2.3 $ & $  3.0 \pm 2.0 $ & $  6.9 \pm 2.1 $ \\
$ 50$,$50$  & $ 38.7 \pm 2.3 $ & $              $ & $ 10.4 \pm 2.1 $ \\
$ 50$,$70$  & $ 43.5 \pm 2.3 $ & $  2.6 \pm 2.0 $ & $  9.2 \pm 2.1 $ \\
$ 50$,$90$  & $ 42.9 \pm 2.3 $ & $  2.6 \pm 2.0 $ & $  7.6 \pm 2.1 $ \\
$ 50$,$135$  & $ 43.2 \pm 2.3 $ & $  2.6 \pm 2.0 $ & $  6.1 \pm 2.1 $ \\
$ 70$,$70$  & $ 45.5 \pm 2.3 $ & $  2.5 \pm 2.0 $ & $  7.3 \pm 2.1 $ \\
$ 70$,$90$  & $ 49.5 \pm 2.3 $ & $  2.8 \pm 2.0 $ & $  5.4 \pm 2.1 $ \\
$ 70$,$115$  & $ 49.7 \pm 2.3 $ & $  3.5 \pm 2.0 $ & $  5.1 \pm 2.1 $ \\
$ 90$,$90$  & $ 49.4 \pm 2.3 $ & $  4.0 \pm 2.1 $ & $  5.2 \pm 2.1 $ \\
\end{tabular}
\label{eff4tau}
\end{center}
\end{table}

\newpage
\section{Excluded couplings per process}
\label{couplings}
This appendix contains tables of excluded couplings and suppression
factors as functions of the
involved Higgs boson masses, for all processes considered in this
work. The mass granularity has been
reduced in order to limit the size of the tables. {\tt FORTRAN} routines
containing the complete information can be obtained from the
DELPHI collaboration on request.

Note that for the Yukawa process the results are given at the matrix
element level rather than at the cross-section level (i.e. C instead
of $\mathrm C^2$); for all other cases the $\mathrm C^2$ factors are listed. All masses are in GeV$/c^2$.

\begin{table}[h]

\caption{{\bf Yukawa channels:} upper bounds on the Yukawa C factors defined in Section 
         \ref{higgsprod}, as function of \mh\ or \mA~(\GeVcc).}
\begin{center}
\scriptsize
\begin{tabular}{c|ccccccccc}
 \mh,\mA & $4$ & $6$ & $9$ & $12$ & $15$ & $20$ & $30$ & $40$ & $50$ \\
\hline
 $\mathrm C_{bb(h\mra bb)}$             &      &      &      & 17.7 & 18.1 & 20.7 & 29.0 & 48.9 & 108.2 \\
 $\mathrm C_{bb(A\mra bb)}$             &      &      &      & 18.4 & 19.0 & 21.0 & 31.8 & 54.8 & 114.9 \\
 $\mathrm C_{bb(h\mra \tau\tau)}$       & 10.3 & 11.1 & 12.3 & 12.9 & 14.5 & 17.6 & 24.5 & 40.0 &  77.5 \\
 $\mathrm C_{bb(A\mra \tau\tau)}$       & 12.8 & 12.9 & 12.8 & 15.2 & 16.3 & 19.3 & 27.7 & 44.4 &  81.0 \\
 $\mathrm C_{\tau\tau(h\mra \tau\tau)}$ & 27.3 & 27.7 & 30.5 & 35.9 & 44.0 & 57.3 & 120.1 & & \\
 $\mathrm C_{\tau\tau(A\mra \tau\tau)}$ & 29.4 & 28.5 & 31.7 & 37.8 & 44.8 & 62.1 & 128.1 & & \\
\end{tabular}
\label{cpl_Yuk}
\end{center}

\end{table}

\begin{table}[p]
\scriptsize
\caption{{\bf hA \ra 4b:} upper bounds on $\mathrm C^{2}_{hA \mra 4b}$,
combining the analyses presented here and the results of \cite{DELhiggs}.
The results  are given as a function of \mh\ and \mA~(\GeVcc), and are symmetric in \mh\ and \mA.}
\begin{center}
\begin{tabular}{rc|rc|rc|rc}
\mh,\mA & $\mathrm C^2_{hA\mra 4b}$ & \mh,\mA & $\mathrm C^2_{hA\mra 4b}$ & 
\mh,\mA & $\mathrm C^2_{hA\mra 4b}$ & \mh,\mA & $\mathrm C^2_{hA\mra 4b}$ \\
\hline
 12,12 & 0.022 &   90,20 & 0.322 &    90,35 & 0.268 &    65,55 & 0.087 \\
 15,12 & 0.011 &   95,20 & 0.357 &    95,35 & 0.302 &    70,55 & 0.114 \\
 20,12 & 0.005 &  100,20 & 0.409 &   100,35 & 0.264 &    75,55 & 0.137 \\
 25,12 & 0.005 &  105,20 & 0.423 &   105,35 & 0.290 &    80,55 & 0.188 \\
 30,12 & 0.005 &  110,20 & 0.515 &   110,35 & 0.404 &    85,55 & 0.261 \\
 35,12 & 0.007 &  115,20 & 0.628 &   115,35 & 0.525 &    90,55 & 0.260 \\
 40,12 & 0.009 &  120,20 & 0.727 &   120,35 & 0.671 &    95,55 & 0.308 \\
 45,12 & 0.011 &  125,20 & 0.878 &   125,35 & 0.862 &   100,55 & 0.368 \\
 50,12 & 0.015 &  130,20 & $\ge$1 &  130,35 & $\ge$1 &  105,55 & 0.438 \\
 55,12 & 0.025 &   25,25 & 0.003 &    40,40 & 0.022 &   110,55 & 0.582 \\
 60,12 & 0.048 &   30,25 & 0.003 &    45,40 & 0.043 &   115,55 & 0.830 \\
 65,12 & 0.114 &   35,25 & 0.006 &    50,40 & 0.057 &   120,55 & $\ge$1 \\
 70,12 & 0.255 &   40,25 & 0.007 &    55,40 & 0.060 &    60,60 & 0.085 \\
 75,12 & 0.318 &   45,25 & 0.012 &    60,40 & 0.089 &    65,60 & 0.108 \\
 80,12 & 0.335 &   50,25 & 0.017 &    65,40 & 0.084 &    70,60 & 0.123 \\
 85,12 & 0.347 &   55,25 & 0.040 &    70,40 & 0.126 &    75,60 & 0.174 \\      
 90,12 & 0.355 &   60,25 & 0.109 &    75,40 & 0.130 &    80,60 & 0.187 \\      
 95,12 & 0.380 &   65,25 & 0.247 &    80,40 & 0.157 &    85,60 & 0.203 \\      
100,12 & 0.406 &   70,25 & 0.235 &    85,40 & 0.187 &    90,60 & 0.266 \\      
105,12 & 0.445 &   75,25 & 0.253 &    90,40 & 0.188 &    95,60 & 0.327 \\      
110,12 & 0.471 &   80,25 & 0.262 &    95,40 & 0.216 &   100,60 & 0.383 \\      
115,12 & 0.574 &   85,25 & 0.287 &   100,40 & 0.248 &   105,60 & 0.495 \\      
120,12 & 0.671 &   90,25 & 0.316 &   105,40 & 0.363 &   110,60 & 0.666 \\      
125,12 & 0.819 &   95,25 & 0.370 &   110,40 & 0.433 &   115,60 & 0.988 \\      
130,12 & $\ge$1 & 100,25 & 0.387 &   115,40 & 0.554 &   120,60 & $\ge$1 \\      
 15,15 & 0.004 &  105,25 & 0.490 &   120,40 & 0.728 &    65,65 & 0.123 \\      
 20,15 & 0.003 &  110,25 & 0.537 &   125,40 & 0.965 &    70,65 & 0.165 \\      
 25,15 & 0.003 &  115,25 & 0.652 &   130,40 & $\ge$1 &   75,65 & 0.169 \\      
 30,15 & 0.004 &  120,25 & 0.843 &    45,45 & 0.071 &    80,65 & 0.162 \\      
 35,15 & 0.005 &  125,25 & $\ge$1 &   50,45 & 0.065 &    85,65 & 0.208 \\      
 40,15 & 0.007 &   30,30 & 0.005 &    55,45 & 0.063 &    90,65 & 0.234 \\      
 45,15 & 0.010 &   35,30 & 0.006 &    60,45 & 0.072 &    95,65 & 0.353 \\      
 50,15 & 0.013 &   40,30 & 0.010 &    65,45 & 0.083 &   100,65 & 0.417 \\      
 55,15 & 0.025 &   45,30 & 0.015 &    70,45 & 0.082 &   105,65 & 0.598 \\      
 60,15 & 0.048 &   50,30 & 0.023 &    75,45 & 0.149 &   110,65 & 0.947 \\      
 65,15 & 0.120 &   55,30 & 0.049 &    80,45 & 0.209 &   115,65 & $\ge$1 \\     
 70,15 & 0.264 &   60,30 & 0.109 &    85,45 & 0.191 &    70,70 & 0.163 \\      
 75,15 & 0.320 &   65,30 & 0.111 &    90,45 & 0.223 &    75,70 & 0.155 \\      
 80,15 & 0.326 &   70,30 & 0.166 &    95,45 & 0.218 &    80,70 & 0.160 \\      
 85,15 & 0.331 &   75,30 & 0.223 &   100,45 & 0.331 &    85,70 & 0.218 \\      
 90,15 & 0.341 &   803,0 & 0.247 &   105,45 & 0.371 &    90,70 & 0.226 \\      
 95,15 & 0.378 &   85,30 & 0.268 &   110,45 & 0.468 &    95,70 & 0.337 \\      
100,15 & 0.408 &   90,30 & 0.258 &   115,45 & 0.606 &   100,70 & 0.477 \\      
105,15 & 0.447 &   95,30 & 0.299 &   120,45 & 0.812 &   105,70 & 0.722 \\      
110,15 & 0.476 &  100,30 & 0.354 &   125,45 & $\ge$1 &  110,70 & $\ge$1 \\     
115,15 & 0.569 &  105,30 & 0.392 &    50,50 & 0.060 &    75,75 & 0.164 \\      
120,15 & 0.685 &  110,30 & 0.375 &    55,50 & 0.056 &    80,75 & 0.179 \\      
125,15 & 0.841 &  115,30 & 0.444 &    60,50 & 0.054 &    85,75 & 0.228 \\      
130,15 & $\ge$1 & 120,30 & 0.559 &    65,50 & 0.069 &    90,75 & 0.242 \\       
 20,20 & 0.002 &  125,30 & 0.711 &    70,50 & 0.089 &    95,75 & 0.430 \\  
 25,20 & 0.002 &  130,30 & 0.918 &    75,50 & 0.128 &   100,75 & 0.658 \\  
 30,20 & 0.003 &  135,30 & $\ge$1 &   80,50 & 0.229 &   105,75 & $\ge$1 \\ 
 35,20 & 0.004 &   35,35 & 0.009 &    85,50 & 0.239 &    80,80 & 0.171 \\  
 40,20 & 0.006 &   40,35 & 0.014 &    90,50 & 0.267 &    85,80 & 0.237 \\  
 45,20 & 0.008 &   45,35 & 0.024 &    95,50 & 0.285 &    90,80 & 0.306 \\  
 50,20 & 0.013 &   50,35 & 0.045 &   100,50 & 0.372 &    95,80 & 0.482 \\  
 55,20 & 0.025 &   55,35 & 0.088 &   105,50 & 0.444 &   100,80 & 0.913 \\  
 60,20 & 0.059 &   60,35 & 0.092 &   110,50 & 0.496 &   105,80 & $\ge$1 \\ 
 65,20 & 0.162 &   65,35 & 0.139 &   115,50 & 0.668 &    85,85 & 0.273 \\  
 70,20 & 0.273 &   70,35 & 0.119 &   120,50 & 0.927 &    90,85 & 0.415 \\  
 75,20 & 0.288 &   75,35 & 0.209 &   125,50 & $\ge$1 &   95,85 & 0.818 \\  
 80,20 & 0.301 &   80,35 & 0.253 &    55,55 & 0.051 &   100,85 & $\ge$1 \\ 
 85,20 & 0.301 &   85,35 & 0.267 &    60,55 & 0.058 &    90,90 & 0.849 \\  
\end{tabular}
\label{cpl_hA4b}                  
\end{center}
\end{table}

\begin{table}[p]
\caption{{\bf hA \ra 4$\tau$:} upper bounds on $\mathrm C^{2}_{hA\mra 4\tau}$, combining
the two-jet, three-jet and four-jet streams. The results are given as a function of \mh\ and \mA~(\GeVcc), and are symmetric in \mh\ and \mA.}
\begin{center}
\scriptsize
\begin{tabular}{rc|rc|rc|rc}
\mh,\mA & $\mathrm C^2_{hA\mra 4\tau}$ & \mh,\mA & $\mathrm C^2_{hA\mra 4\tau}$ & 
\mh,\mA & $\mathrm C^2_{hA\mra 4\tau}$ & \mh,\mA & $\mathrm C^2_{hA\mra 4\tau}$ \\
\hline
  5,5 & 0.005 &    70,15 & 0.036 &    55,30 & 0.040 &     125,45 & 0.341 \\  
 10,5 & 0.005 &    80,15 & 0.046 &    60,30 & 0.043 &     130,45 & 0.403 \\  	
 15,5 & 0.010 &    90,15 & 0.052 &    65,30 & 0.044 &     135,45 & 0.624 \\  	
 20,5 & 0.012 &   100,15 & 0.067 &    70,30 & 0.043 &     140,45 & $\ge$1 \\ 	
 25,5 & 0.010 &   110,15 & 0.091 &    80,30 & 0.048 &      50,50 & 0.038 \\  	
 30,5 & 0.012 &   115,15 & 0.111 &    90,30 & 0.056 &      55,50 & 0.041 \\  	
 35,5 & 0.017 &   120,15 & 0.134 &   100,30 & 0.075 &      60,50 & 0.043 \\  	
 40,5 & 0.021 &   125,15 & 0.163 &   110,30 & 0.106 &      65,50 & 0.045 \\  	
 45,5 & 0.027 &   130,15 & 0.206 &   115,30 & 0.126 &      70,50 & 0.048 \\  	
 50,5 & 0.032 &   135,15 & 0.270 &   120,30 & 0.156 &      80,50 & 0.061 \\  	
 55,5 & 0.033 &   140,15 & 0.366 &   125,30 & 0.196 &      90,50 & 0.081 \\  	
 60,5 & 0.031 &   145,15 & 0.502 &   130,30 & 0.246 &     100,50 & 0.115 \\  	
 65,5 & 0.034 &   150,15 & 0.711 &   135,30 & 0.335 &     110,50 & 0.175 \\	
 70,5 & 0.037 &   155,15 & $\ge$1 &  140,30 & 0.463 &     115,50 & 0.212 \\  	
 80,5 & 0.050 &    20,20 & 0.007 &   145,30 & 0.665 &     120,50 & 0.292 \\    	
 90,5 & 0.059 &    25,20 & 0.009 &   150,30 & $\ge$1 &    125,50 & 0.402 \\    	
100,5 & 0.074 &    30,20 & 0.010 &    35,35 & 0.021 &     130,50 & 0.637 \\    	
110,5 & 0.104 &    35,20 & 0.017 &    40,35 & 0.023 &     135,50 & 0.884 \\    	
115,5 & 0.125 &    40,20 & 0.016 &    45,35 & 0.031 &     140,50 & $\ge$1 \\   	
120,5 & 0.140 &    45,20 & 0.025 &    50,35 & 0.039 &      55,55 & 0.041 \\    	
125,5 & 0.152 &    50,20 & 0.025 &    55,35 & 0.046 &      60,55 & 0.044 \\    	
130,5 & 0.170 &    55,20 & 0.030 &    60,35 & 0.045 &      65,55 & 0.046 \\    	
135,5 & 0.215 &    60,20 & 0.035 &    65,35 & 0.044 &      70,55 & 0.049 \\    	
140,5 & 0.270 &    65,20 & 0.041 &    70,35 & 0.043 &      80,55 & 0.067 \\    	
145,5 & 0.356 &    70,20 & 0.041 &    80,35 & 0.052 &      90,55 & 0.088 \\    	
150,5 & 0.498 &    80,20 & 0.047 &    90,35 & 0.059 &     100,55 & 0.128 \\    	
155,5 & 0.847 &    90,20 & 0.053 &   100,35 & 0.080 &     110,55 & 0.212 \\    	
160,5 & $\ge$1 &  100,20 & 0.069 &   110,35 & 0.115 &     115,55 & 0.288 \\    	
 10,10 & 0.006 &   110,20 & 0.094 &   115,35 & 0.139 &     120,55 & 0.395 \\    	
 15,10 & 0.011 &   115,20 & 0.122 &   120,35 & 0.172 &     125,55 & 0.536 \\    	
 20,10 & 0.007 &   120,20 & 0.142 &   125,35 & 0.220 &     130,55 & 0.929 \\    	
 25,10 & 0.010 &   125,20 & 0.169 &   130,35 & 0.273 &     135,55 & $\ge$1 \\   	
 30,10 & 0.015 &   130,20 & 0.209 &   135,35 & 0.383 &      60,60 & 0.043 \\    	
 35,10 & 0.016 &   135,20 & 0.308 &   140,35 & 0.534 &      65,60 & 0.048 \\    	
 40,10 & 0.018 &   140,20 & 0.387 &   145,35 & 0.851 &      70,60 & 0.054 \\    	
 45,10 & 0.025 &   145,20 & 0.530 &   150,35 & $\ge$1 &     80,60 & 0.072 \\    	
 50,10 & 0.040 &   150,20 & 0.751 &    40,40 & 0.025 &      90,60 & 0.097 \\    	
 55,10 & 0.043 &   155,20 & $\ge$1 &   45,40 & 0.033 &     100,60 & 0.151 \\    	
 60,10 & 0.044 &    25,25 & 0.008 &    50,40 & 0.040 &     110,60 & 0.263 \\    	
 65,10 & 0.043 &    30,25 & 0.013 &    55,40 & 0.046 &     115,60 & 0.395 \\    	
 70,10 & 0.041 &    35,25 & 0.017 &    60,40 & 0.044 &     120,60 & 0.572 \\    	
 80,10 & 0.052 &    40,25 & 0.018 &    65,40 & 0.043 &     125,60 & 0.773 \\    	
 90,10 & 0.064 &    45,25 & 0.024 &    70,40 & 0.043 &     130,60 & $\ge$1 \\   	
100,10 & 0.068 &    50,25 & 0.029 &    80,40 & 0.055 &      65,65 & 0.053 \\    	
110,10 & 0.089 &    55,25 & 0.035 &    90,40 & 0.067 &      70,65 & 0.060 \\    	
115,10 & 0.110 &    60,25 & 0.041 &   100,40 & 0.093 &      80,65 & 0.079 \\    	
120,10 & 0.123 &    65,25 & 0.044 &   110,40 & 0.129 &      90,65 & 0.111 \\    	
125,10 & 0.135 &    70,25 & 0.043 &   115,40 & 0.150 &     100,65 & 0.175 \\    	
130,10 & 0.177 &    80,25 & 0.052 &   120,40 & 0.184 &     110,65 & 0.333 \\    	
135,10 & 0.214 &    90,25 & 0.054 &   125,40 & 0.250 &     115,65 & 0.555 \\    	
140,10 & 0.265 &   100,25 & 0.062 &   130,40 & 0.330 &     120,65 & $\ge$1 \\   	
145,10 & 0.372 &   110,25 & 0.099 &   135,40 & 0.461 &      70,70 & 0.066 \\    	
150,10 & 0.513 &   115,25 & 0.119 &   140,40 & 0.688 &      80,70 & 0.087 \\    	
155,10 & 0.779 &   120,25 & 0.144 &   155,40 & $\ge$1 &     90,70 & 0.128 \\    	
160,10 & $\ge$1 &  125,25 & 0.180 &    45,45 & 0.038 &     100,70 & 0.212 \\    	
 15,15 & 0.006 &   130,25 & 0.228 &    50,45 & 0.039 &     110,70 & 0.435 \\    	
 20,15 & 0.007 &   135,25 & 0.298 &    55,45 & 0.043 &     115,70 & 0.728 \\    	
 25,15 & 0.009 &   140,25 & 0.420 &    60,45 & 0.043 &     120,70 & $\ge$1 \\   	
 30,15 & 0.011 &   145,25 & 0.582 &    65,45 & 0.043 &      80,80 & 0.117 \\    	
 35,15 & 0.014 &   150,25 & 0.843 &    70,45 & 0.045 &      90,80 & 0.207 \\    	
 40,15 & 0.020 &   155,25 & $\ge$1 &   80,45 & 0.056 &     100,80 & 0.433 \\    	
 45,15 & 0.022 &    30,30 & 0.013 &    90,45 & 0.074 &     115,80 & $\ge$1 \\   	
 50,15 & 0.024 &    35,30 & 0.019 &   100,45 & 0.102 &      90,90 & 0.417 \\    	
 55,15 & 0.029 &    40,30 & 0.020 &   110,45 & 0.150 &     110,90 & $\ge$1 \\   	
 60,15 & 0.033 &    45,30 & 0.025 &   115,45 & 0.186 &     100,100 & $\ge$1 \\   	
 65,15 & 0.036 &    50,30 & 0.028 &   120,45 & 0.237 &             &  \\        	
\end{tabular}
\label{cpl_hA4t}                  
\end{center}
\end{table}

\begin{table}

\caption{{\bf hZ \ra \tautau Z:} upper bounds on $\mathrm C^{2}_{Z(h\mra \tau\tau)}$,
as function of \mh~(\GeVcc), reinterpreting the search for the Standard Model Higgs boson \cite{DELhiggs}.}
\begin{center}
\scriptsize
\begin{tabular}{cc|cc|cc|cc|cc}
\mh & $\mathrm C^2_{Z(h\mra \tau\tau)}$ & \mh & $\mathrm C^2_{Z(h\mra \tau\tau)}$ & 
\mh & $\mathrm C^2_{Z(h\mra \tau\tau)}$ & \mh & $\mathrm C^2_{Z(h\mra \tau\tau)}$ & \mh & $\mathrm C^2_{Z(h\mra \tau\tau)}$ \\
\hline
 12 & 0.285 & 35 & 1.132 & 60 & 0.169 & 85 & 0.088 &110 & 0.590 \\
 15 & 0.316 & 40 & 1.022 & 65 & 0.093 & 90 & 0.102 &115 & $\ge$1\\
 20 & 0.398 & 45 & 0.457 & 70 & 0.082 & 95 & 0.164 & &\\
 25 & 0.530 & 50 & 0.260 & 75 & 0.095 &100 & 0.219 & &\\
 30 & 0.751 & 55 & 0.199 & 80 & 0.067 &105 & 0.297 & &\\
\end{tabular}
\label{cpl_ttZ}
\end{center}

\caption{{\bf hZ \ra \bb Z:} upper bounds on $\mathrm C^{2}_{Z(h\mra bb)}$,
as function of \mh~(\GeVcc), reinterpreting the search for the Standard Model Higgs boson \cite{DELhiggs}.}
\begin{center}
\scriptsize
\begin{tabular}{cc|cc|cc|cc|cc}
\mh & $\mathrm C^2_{Z(h\mra bb)}$ & \mh & $\mathrm C^2_{Z(h\mra bb)}$ & 
\mh & $\mathrm C^2_{Z(h\mra bb)}$ & \mh & $\mathrm C^2_{Z(h\mra bb)}$ & \mh & $\mathrm C^2_{Z(h\mra bb)}$ \\
\hline
 12 & 0.042 & 35 & 0.047 & 60 & 0.049 & 85 & 0.103 &110 & 0.314 \\
 15 & 0.046 & 40 & 0.060 & 65 & 0.035 & 90 & 0.176 &115 & $\ge$1\\
 20 & 0.047 & 45 & 0.064 & 70 & 0.034 & 95 & 0.262 & &\\
 25 & 0.054 & 50 & 0.046 & 75 & 0.040 &100 & 0.273 & &\\
 30 & 0.063 & 55 & 0.047 & 80 & 0.055 &105 & 0.215 & &\\
\end{tabular}
\label{cpl_bbZ}
\end{center}

\end{table}

\begin{table}
\caption{{\bf hA \ra 6b:} upper bounds on $\mathrm C^{2}_{hA\mra 6b}$, as a function of \mh\ and \mA~(\GeVcc).}
\begin{center}
\scriptsize
\begin{tabular}{rc|rc|rc|rc}
\mh,\mA & $\mathrm C^2_{hA\mra 6b}$ & \mh,\mA & $\mathrm C^2_{hA\mra 6b}$ & 
\mh,\mA  &  $\mathrm C^2_{hA\mra 6b}$  & \mh,\mA  &  $\mathrm C^2_{hA\mra 6b}$ \\
\hline
 25,12 & $\ge$1 &  95,15 & 0.242  &  80,25 & 0.176  & 110,35 & 0.396  \\  
 30,12 & $\ge$1 & 100,15 & 0.265  &  85,25 & 0.193  & 115,35 & 0.478  \\          
 35,12 & $\ge$1 & 105,15 & 0.296  &  90,25 & 0.211  & 120,35 & 0.581  \\          
 40,12 & 0.879  & 110,15 & 0.327  &  95,25 & 0.235  & 125,35 & 0.736  \\          
 45,12 & 0.701  & 115,15 & 0.391  & 100,25 & 0.261  & 130,35 & 0.949  \\          
 50,12 & 0.625  & 120,15 & 0.471  & 105,25 & 0.299  & 135,35 & $\ge$1 \\          
 55,12 & 0.256  & 125,15 & 0.571  & 110,25 & 0.339  &  80,40 & 0.195  \\          
 60,12 & 0.189  & 130,15 & 0.733  & 115,25 & 0.410  &  85,40 & 0.216  \\          
 65,12 & 0.183  & 135,15 & 0.898  & 120,25 & 0.503  &  90,40 & 0.299  \\       	 
 70,12 & 0.181  & 140,15 & $\ge$1 & 125,25 & 0.614  &  95,40 & 0.273\\            
 75,12 & 0.209  &  40,20 & 0.547  & 130,25 & 0.764  & 100,40 & 0.320\\            
 80,12 & 0.213  &  45,20 & 0.155  & 135,25 & 0.997  & 105,40 & 0.365\\            
 85,12 & 0.217  &  50,20 & 0.098  & 140,25 & $\ge$1 & 110,40 & 0.440\\            
 90,12 & 0.218  &  55,20 & 0.125  &  60,30 & 0.141  & 115,40 & 0.535\\            
 95,12 & 0.240  &  60,20 & 0.146  &  65,30 & 0.150  & 120,40 & 0.699\\        	 
100,12 & 0.261  &  65,20 & 0.168  &  70,30 & 0.149  & 125,40 & 0.866\\            
105,12 & 0.292  &  70,20 & 0.173  &  75,30 & 0.165  & 130,40 & $\ge$1\\           
110,12 & 0.322  &  75,20 & 0.193  &  80,30 & 0.175  &  90,45 & 0.264\\        	 
115,12 & 0.390  &  80,20 & 0.206  &  85,30 & 0.194  &  95,45 & 0.300\\        	 
120,12 & 0.466  &  85,20 & 0.191  &  90,30 & 0.210  & 100,45 & 0.349\\        	 
125,12 & 0.586  &  90,20 & 0.210  &  95,30 & 0.234  & 105,45 & 0.410\\        	 
130,12 & 0.725  &  95,20 & 0.234  & 100,30 & 0.270  & 110,45 & 0.493\\        	 
135,12 & 0.922  & 100,20 & 0.265  & 105,30 & 0.313  & 115,45 & 0.616\\        	 
140,12 & $\ge$1 & 105,20 & 0.294  & 110,30 & 0.361  & 120,45 & 0.786\\        	 
 30,15 & $\ge$1 & 110,20 & 0.333  & 115,30 & 0.428  & 125,45 & $\ge$1\\       	 
 35,15 & $\ge$1 & 115,20 & 0.390  & 120,30 & 0.524  & 100,50 & 0.391\\        	 
 40,15 & 0.713  & 120,20 & 0.474  & 125,30 & 0.654  & 105,50 & 0.469\\        	 
 45,15 & 0.177  & 125,20 & 0.593  & 130,30 & 0.826  & 110,50 & 0.571\\        	 
 50,15 & 0.195  & 130,20 & 0.723  & 135,30 & $\ge$1 & 115,50 & 0.733\\        	 
 55,15 & 0.202  & 135,20 & 0.938  &  70,35 & 0.159  & 120,50 & 0.956\\        	 
 60,15 & 0.169  & 140,20 & $\ge$1 &  75,35 & 0.168  & 125,50 & $\ge$1\\       	      
 65,15 & 0.179  &  50,25 & 0.111  &  80,35 & 0.189  & 110,55 & 0.688\\        	      
 70,15 & 0.178  &  55,25 & 0.129  &  85,35 & 0.206  & 115,55 & 0.907\\        	      
 75,15 & 0.214  &  60,25 & 0.134  &  90,35 & 0.226  & 120,55 & $\ge$1\\       	      
 80,15 & 0.211  &  65,25 & 0.169  &  95,35 & 0.253  & 120,60 & $\ge$1\\       	      
 85,15 & 0.213  &  70,25 & 0.161  & 100,35 & 0.289  &       & \\       	        	      
 90,15 & 0.215  &  75,25 & 0.178  & 105,35 & 0.335  &       & \\

\end{tabular}
\label{cpl_hA6b}                  
\end{center}
\end{table}

\begin{table}
\caption{{\bf hZ \ra\ 4b+jets:} upper bounds on $\mathrm C^{2}_{Z(AA\mra 4b)}$, as a function of \mh\ and \mA~(\GeVcc).}
\begin{center}
\scriptsize
\begin{tabular}{rc|rc|rc|rc}
\mh,\mA & $\mathrm C^2_{Z(AA\mra 4b)}$ & \mh,\mA & $\mathrm C^2_{Z(AA\mra 4b)}$ & 
\mh,\mA & $\mathrm C^2_{Z(AA\mra 4b)}$ & \mh,\mA & $\mathrm C^2_{Z(AA\mra 4b)}$ \\
\hline
 25,12 & $\ge$1 &  55,15 & 0.244 &    95,20 & 0.696 &    95,30 & 0.579 \\  
 30,12 & 0.324 &   60,15 & 0.252 &   100,20 & 0.947 &   100,30 & 0.776 \\  
 35,12 & 0.281 &   65,15 & 0.240 &   110,20 & $\ge$1 &  110,30 & $\ge$1\\  
 40,12 & 0.250 &   70,15 & 0.262 &    50,25 & 0.253 &    70,35 & 0.273 \\  
 45,12 & 0.230 &   75,15 & 0.273 &    55,25 & 0.262 &    75,35 & 0.287 \\  
 50,12 & 0.218 &   80,15 & 0.302 &    60,25 & 0.273 &    80,35 & 0.296 \\
 55,12 & 0.216 &   85,15 & 0.372 &    65,25 & 0.289 &    85,35 & 0.338 \\  
 60,12 & 0.219 &   90,15 & 0.434 &    70,25 & 0.313 &    90,35 & 0.392 \\  
 65,12 & 0.221 &   95,15 & 0.641 &    75,25 & 0.314 &    95,35 & 0.567 \\  
 70,12 & 0.231 &  100,15 & 0.869 &    80,25 & 0.319 &   100,35 & 0.771 \\  
 75,12 & 0.258 &  110,15 & $\ge$1 &   85,25 & 0.367 &   110,35 & $\ge$1 \\ 
 80,12 & 0.289 &   40,20 & 0.267 &    90,25 & 0.426 &    80,40 & 0.292 \\  
 85,12 & 0.338 &   45,20 & 0.266 &    95,25 & 0.632 &    85,40 & 0.330 \\  
 90,12 & 0.417 &   50,20 & 0.266 &   100,25 & 0.856 &    90,40 & 0.391 \\  
 95,12 & 0.612 &   55,20 & 0.276 &   110,25 & $\ge$1 &   95,40 & 0.570 \\  
100,12 & 0.829 &   60,20 & 0.290 &    60,30 & 0.260 &   100,40 & 0.759 \\  
110,12 & $\ge$1 &  65,20 & 0.311 &    65,30 & 0.276 &   110,40 & $\ge$1 \\ 
 30,15 & 0.303 &   70,20 & 0.333 &    70,30 & 0.292 &    90,45 & 0.503 \\  
 35,15 & 0.295 &   75,20 & 0.344 &    75,30 & 0.296 &    95,45 & 0.586 \\  
 40,15 & 0.276 &   80,20 & 0.363 &    80,30 & 0.314 &   100,45 & $\ge$1 \\ 
 45,15 & 0.250 &   85,20 & 0.401 &    85,30 & 0.340 &   100,50 & $\ge$1 \\ 
 50,15 & 0.233 &   90,20 & 0.467 &    90,30 & 0.393 &   110,55 & $\ge$1 \\ 
\end{tabular}
\label{cpl_hZ4b2q}                
\end{center}
\end{table}

\end{document}